\documentclass{aa}  
\usepackage{graphicx}
\usepackage{txfonts}
\usepackage{lscape}
\usepackage{placeins}
\usepackage{url}
\usepackage{amsmath}
\usepackage{hyperref}
\usepackage{makecell}
\usepackage{multirow}
\usepackage{booktabs}
\usepackage{tabularx}
\usepackage{array}
\newcounter{colnum}
\newcommand{\coln}{\stepcounter{colnum}\arabic{colnum}}
\newcommand{\sersic}{S$\acute{e}$rsic\ }
\newcommand{\software}[1]{\texttt{#1}}

\begin{document} 
    \title{Pushing HST to the limit:}
    \subtitle{Detection completeness and morphology robustness of faint galaxies in the EGS field}

   \author{Limin Lai\inst{1,2,3} \and Zhao-Yu Li\inst{1,2,3}\fnmsep\corrauth{lizy.astro@sjtu.edu.cn} \and Chao Ma\inst{4,5} \and Yang Anpin Li\inst{6,1,2,3}\fnmsep\corrauth{yangli\_anpin@sjtu.edu.cn} \and Luis C. Ho\inst{4,5} \and Mingyang Zhuang\inst{7}}

   \institute{Department of Astronomy, School of Physics and Astronomy, Shanghai Jiao Tong University, 800 Dongchuan Road, Shanghai 200240, People's Republic of China
        \and State Key Laboratory of Dark Matter Physics, School of Physics and Astronomy, Shanghai Jiao Tong University, Shanghai 200240, People's Republic of China
        \and Key Laboratory for Particle Astrophysics and Cosmology (MOE) / Shanghai Key Laboratory for Particle Physics and Cosmology, Shanghai 200240, People’s Republic of China
        \and Kavli Institute for Astronomy and Astrophysics, Peking University, Beijing 100871, People's Republic of China
        \and Department of Astronomy, School of Physics, Peking University, Beijing 100871, People's Republic of China
        \and Shanghai Astronomical Observatory, Chinese Academy of Sciences, 80 Nandan Road, Shanghai 200030, People's Republic of China
        \and Department of Astronomy, University of Illinois Urbana-Champaign, Urbana, IL 61801, USA}
   
   \date{}

  \defcitealias{stefanon2017}{S17}

  \abstract
   {}
   {
   Reliable catalogs at the faint limit are essential for statistical studies of galaxy evolution.
   However, detection and structural measurements of faint galaxies remain challenging. 
   We evaluate detection completeness and structural parameter uncertainties of faint galaxies in the CANDELS/EGS field using mock images and observations from the Hubble Space Telescope (HST) and James Webb Space Telescope (JWST).
   The framework is also applicable to large high-redshift surveys, including Euclid, the upcoming Chinese Space Station Survey Telescope (CSST), and the Nancy Grace Roman Space Telescope.
   }
   {
   We detect sources in the EGS F160W mosaic using a $1.5\sigma$ threshold and quantify detection completeness and false-detection rates with realistic mock images.
   Mock galaxy-pair tests assess blending effects on both source detection and structural measurements in \sersic fitting. 
   In addition, the overlap with CEERS observations is used to derive an empirical criterion for separating true faint galaxies from fake detections using HST photometry.
   }
   {
   We identify 72,467 sources in the CANDELS/EGS field, including 57,444 fainter than 25 mag in F160W. 
   Detection completeness exceeds 95\% for objects brighter than 25 mag but declines to $<50\%$ beyond 27 mag (with false detection rate $\gtrsim50\%$). 
   Structure parameters from single \sersic modeling are reliable down to $m_\mathrm{H}=25$ mag.  
   For galaxies at $25<m_\mathrm{H}<27$ mag, systematic biases in the \sersic index reach $\Delta n\approx5-70\%$, while $\Delta q$ decreases from approximately $-0.03$ to $-0.27$. 
   Close companions ($\sim 0.3~\mathrm{arcsec}$) introduce measurable biases in recovered \sersic magnitudes and shapes, while the impact becomes negligible at larger separations.
   An empirical criterion based on $m_\mathrm{H}$, F814W-F160W color, $R_{50}$, and $z_\mathrm{phot}$ separates true faint galaxies from false detections, reaching 72\% completeness and 84\% purity.
   It yields a final catalog of 59,646 objects consistent with previous CANDELS measurements at bright magnitudes while reaching fainter limits.
   }
   {}


   \maketitle
   \nolinenumbers
\section{Introduction}\label{sec:intro}

The Hubble Space Telescope (HST) has revealed more irregular and faint galaxies and provided multiband photometry and structural measurements, enabling detailed studies of galaxy populations over a wide range of redshifts~\citep{Wuyts2013,van_der_Wel_2014,Shibuya2015}. 
Moreover, with the unprecedented depth and resolution, the James Webb Space Telescope (JWST) has uncovered a surprising population of disk galaxies in the early Universe, some of which appear to be massive, challenging the current theory of galaxy formation and assembly history~\citep{Crespo2024,Nelson2024,Shapley2025,Xiao2025,Barisi2025,Ito2025}. 
The physical processes of the Hubble sequence emergence, such as bulge growth history and bar formation, still remain poorly understood.
Before the JWST era, most previous studies have focused on the luminous galaxies in HST optical/near-infrared (NIR) surveys, while faint galaxies, often limited by observational sensitivity and resolution, have received less attention.
Recently, JWST observations have renewed interest in faint sources due to their unprecedented resolution and sensitivity (e.g.,~\citet{Castellano2022,Endsley2023,Kirkpatrick2023,Gottumukkala2024,Ling2024,Mckay2025,Chemrynska2026}), extending galaxy samples to fainter limits and higher redshifts and making it increasingly important to understand the completeness of source detection and the reliability of structural measurements for objects near the detection threshold.

Although deep surveys routinely reach faint magnitudes, the robustness of faint-source catalogs remains limited by several systematic effects. 
Low signal-to-noise ratios (S/N) can lead to false detections and incompleteness in source catalogs, while blending between nearby sources introduces biases in photometric and structural measurements. 
These effects become more severe in crowded fields or for close pairs with large magnitude contrasts.
Furthermore, the predominantly high-redshift and compact nature of faint galaxies makes traditional visual classification prone to contamination.
Hence, two-dimensional image analysis is essential, such as single \sersic component fitting, bulge/disk decomposition through multiple components fitting, as well as nonparametric methods that characterize galaxy light distributions through statistical morphology indicators.
While structure fitting tools such as \sersic modeling are widely used~\citep{Peng2002galfit,Peng2010galfit}, their performance at the faint limit is not well understood and should be quantified in a systematic way using realistic simulations and independent observations. 
Therefore, uncertainties of the galaxy structure measurements at the faint-end are often difficult to propagate into scientific analyses.

The Extended Groth Strip (EGS) field~\citep{Davis2007}, as part of the Cosmic Assembly Near-infrared Deep Extragalactic Legacy Survey (CANDELS)~\citep{Grogin2011, Koekemoer2011}, provides an ideal dataset to address these issues. 
Its rich panchromatic coverage enables reliable multiband photometry, structural measurements, and robust spectral energy distribution (SED) fitting across a wide redshift range. 
In addition, partial overlap with JWST Cosmic Evolution Early Release Science Survey (CEERS; ERS 1345, PI: S Finkelstein) offers deep and high-resolution near-infrared imaging, allowing independent validation of faint detections. 
In this work, we use HST imaging for source detection and photometric measurements, while JWST data are used for detection verification and to provide additional infrared constraints for sources in the overlapping regions. 
This combination enables a self-consistent assessment of detection completeness and false-detection rates.

Existing source catalogs in the CANDELS/EGS field are primarily optimized for relatively bright galaxies and become increasingly incomplete and uncertain toward the faint end. 
In this paper, we construct a reliable faint-source catalog in the CANDELS/EGS field by pushing the EGS F160W imaging to its detection limits to identify fainter sources. 
We evaluate the detection reliability through extensive simulations and cross-validation with JWST imaging.
We perform realistic mock-image tests to quantify detection completeness and false-detection rates, and measure the impact of blending on both detection and structural parameter measurement.
Using the overlapped CEERS observations, we propose an empirical criterion to distinguish real faint galaxies from spurious detections with HST photometry.
The primary goal of this work is to construct a well-characterized faint-source catalog with quantified detection completeness and structural uncertainties.
The methods developed in this work provide a framework for more robust statistical studies of galaxy evolution using current HST data, and are directly applicable to large imaging surveys such as Euclid~\citep{euclidcollaboration2025euclidquickdatarelease}, the upcoming Chinese Space Station Survey Telescope (CSST)~\citep{csstcollaboration2025introductionchinesespacestation}, and the Nancy Grace Roman Space Telescope~\citep{Roman2015}, where similar challenges in detecting and characterizing faint sources are expected.

This paper is structured as follows: in Sect.~\ref{sec:data}, we introduce the datasets used in our analysis, including HST, JWST, and other ancillary data.
In Sect.~\ref{sec:photometry}, we describe the source detection and photometry measurement methods.
The photometric catalog properties are discussed in Sect.~\ref{sec:catalog_properties}, along with comprehensive tests to quantify detection limit, completeness, and false detection rate, and the parametric structure analysis.
The SED fitting results are shown in Sect.~\ref{sec:sed fitting}.
We provide an empirical criterion to separate the false objects from the true faint galaxies, and also explore the blending effects on the source detection and structure measurements in Sect.~\ref{sec:discussion}.
The main results are summarized in Sect.~\ref{sec:conclusion}.

Throughout our paper, a flat $\Lambda$CDM cosmology model, with $\Omega_m = 0.3$ and a Hubble constant $H_0 = 70\ \mathrm{km\ s^{-1}\ Mpc^{-1}}$, is assumed. 
All magnitudes are expressed in the AB system~\citep{ABmag_oke1983secondary}.
The flux $f_{\nu}$ in our catalog is given in $\mu \mathrm{Jy}$ ($10^{-29} \mathrm{erg\ cm^{-2}s^{-1}Hz^{-1}}$), corresponding to zeropoint (ZP) of $23.9$ mag.

\section{Data}\label{sec:data}

The CANDELS project targets five extragalactic fields, providing optical and near-infrared coverage in ACS/F606W and F814W, as well as WFC3/F125W and F160W bands~\citep{Grogin2011, Koekemoer2011}.
The depths of the F606W and F814W bands were enhanced by the All-wavelength Extended Groth Strip International Survey (AEGIS), reaching $5\sigma$ limiting magnitude for point sources of 28.7 mag and 28.1 mag, respectively.
Moreover, the recent JWST/CEERS survey encompasses about half of the EGS field, providing an independent validation of our source detection and structure measurement methods.
The CEERS images were observed with the Near-Infrared Camera (NIRCam) in the F115W, F150W, F200W, F277W, F356W, F410M, and F444W filters.

To derive photometric redshifts ($z_\mathrm{phot}$) and stellar masses ($M_\bigstar$), we perform SED fitting using the available multiband photometry.
Following~\citet{stefanon2017} (hereafter~\citetalias{stefanon2017}), we incorporate $u^*, g', r', i', z'$ data from the Canada-France-Hawaii Telescope (CFHT)/MegaCam observations~\citep{Gwyn2012CFHT}, near-infrared (NIR) broadband $K_S$ imaging from the Wide-field InfraRed Camera (WIRCam) Deep Survey (WIRDS;~\citep{Bielby2012wircam}), HST WFC3/F140W data from the 3D-HST survey~\citep{Skelton20143dhst}, and the 3.6 and 4.5 $\mu m$ IRAC data from the Spitzer Extended Deep Survey (SEDS;~\citep{Ashby2013}) and S-CANDELS~\citep{Ashby2015spitzer}, thereby extending the wavelength coverage. 
In addition, we include spectroscopic redshifts for 3433 objects spanning $0.01 < z < 11.41$, primarily drawn from the DEEP2 survey~\citep{Coil2004,Willner2006,Cooper2006,Newman2013}, the DEEP3 survey~\citep{Cooper2011,Cooper2012}, HST grism measurements~\citep{Brammer2012grism,Momcheva2016} and the Red Unknowns: Bright Infrared Extragalactic Survey (RUBIES)~\citep{degraaff2024rubies}.
Observations from Gaia~\citep{Gaia2023} and Chandra/ACIS X-ray detections~\citep{Nandra_2005,Nandra2015,Laird_2009} are used to identify the stars and active galactic nuclei (AGNs) from normal galaxies, respectively.
The photometric properties of these datasets, including filters, 5$\sigma$ limiting magnitudes of point sources, and the full width at half maximum (FWHM) of the point spread function (PSF) are summarized in Table~\ref{tab:hst data quality summary}.

\begin{table*}[ht]
    \centering
    \caption{Properties of the panchromatic images used in this work}
    \begin{tabular}{lcccccc}
    \hline 
    Instrument & Filter & $\lambda_\mathrm{c}$  & \thead{PSF FWHM} & \thead{Depth $5\sigma$} & References \\
    &&(\AA)&(arcsec)&(AB)&\\
    \multicolumn{1}{c}{(1)}&(2)&(3)&(4)&(5)&(6)\\
    
    \hline
    HST/ACS&F606W&5959&0.12&28.8&~\citet{Koekemoer2011}\\
    &F814W&8084&0.12&28.2&\\
    HST/WFC3&F125W&12501&0.19&27.6&\\
    &F140W&13971&0.19&26.8&~\citet{Skelton20143dhst,Brammer2012grism}\\
    &F160W&15419&0.20&27.6&~\citet{Koekemoer2011}\\
    \hline
    JWST/NIRCam&F115W&11571&0.056&$\sim 29$\tablefootmark{a}&~\citet{Bagley2023CEERS}\\
    &F150W&15040&0.064& &\\ 
    &F200W&19934&0.077& &\\ 
    &F277W&27670&0.130& &\\ 
    &F356W&35767&0.152& &\\ 
    &F410M&40842&0.159& &\\ 
    &F444W&44154&0.161& &\\
    \hline  
    CFHT/MegaCam&$u^*$&3828&0.95&27.1&~\citet{Gwyn2012CFHT} \\
    &$g'$&4870&0.90&27.3& \\
    &$r'$&6245&0.77&27.2& \\ 
    &$i'$&7676&0.71&27.0& \\
    &$z'$&8872&0.71&26.1& \\ 
    \hline
    CFHT/WIRCam&$K_S$&21574&0.65&24.3&~\citet{Bielby2012wircam} \\
    \hline
    Spitzer/IRAC&3.6$\mu m$&35569&1.80&23.9&~\citet{Ashby2015spitzer}\\
    &4.5$\mu m$&45020&1.82&24.2& \\
    \hline 
    \end{tabular}
    \label{tab:hst data quality summary}
    \tablefoot{Col~(1): Telescope and the instrument; Col~(2): the name of the filter; Col~(3): the central wavelength; Col~(4): PSF FWHM; Col~(5): $5\sigma$ depth of point sources in the image; Col~(6): reference of the corresponding mosaics;\\a): The depths in different points of CEERS are not the same, but vary in the range of 28.3-29.2, please refer to~\citet{Bagley2023CEERS};\\The properties of the non-JWST images listed in this table are adopted from \citetalias{stefanon2017}, while Col~(6) gives the references for the original observations or image mosaics.}
\end{table*}

\subsection{HST data}

In the CANDELS/EGS field, which covers an area of approximately $200\ \mathrm{arcmin^2}$, the nominal exposure times for the WFC3/IR mosaics are approximately $1300\mathrm{s}$ and $2700\mathrm{s}$ in the F125W and F160W filters, respectively. 

The $5 \sigma$ point source detection limits~\footnote{The $5\sigma$ limits are measured in circular apertures with diameters equal to twice the corresponding PSF FWHM.} in the F125W and F160W filters are both $27.6$ mag, with corresponding PSF FWHMs of $0.19''$ and $0.20''$, respectively, as reported by~\citetalias{stefanon2017}.
The ACS mosaics have nominal exposure times of around $6000\mathrm{s}$ and $12000\mathrm{s}$ in the F606W and F814W filters, achieving $5 \sigma$ depths of $28.8$ and $28.2$ mag, respectively. 
The empirical PSF FWHMs are $0.12''$ in both bands (see~\citealp{Grogin2011,Koekemoer2011} for details). 
Additionally, the WFC3/F140W mosaics from the 3D-HST program cover approximately two-thirds of the F160W field, with the PSF FWHM of $0.19''$ and $5 \sigma$ depths of $26.8$ mag.
For our analysis, we directly use the publicly available science images and the corresponding weight maps.
The image mosaics used in this work have a pixel scale of $60~\mathrm{mas~pixel^{-1}}$ and share the common world coordinate system (WCS).
A summary of the data properties is provided in Table~\ref{tab:hst data quality summary}.

\subsection{JWST data}

We utilize publicly available JWST/NIRCam imaging data from the CEERS program, covering an area of $97~\mathrm{arcmin^2}$~\citep{Finkelstein2023CEERS}. 
The mosaics for CEERS pointings 1, 2, 3, and 6 are taken from Release 0.5, while those for pointings 4, 5, 7, 8, 9, and 10 are taken from Release 0.6.
The dataset includes observations obtained with three short-wavelength (SW) filters (F115W, F150W, and F200W) and four long-wavelength (LW) filters (F277W, F356W, F410M, and F444W), achieving typical $5\sigma$ point source depths of $\sim 29$ mag.

The observations and data reduction procedures are described in detail in~\citet{Bagley2023CEERS}. 
For the astrometry calibration, each image is aligned to a reference catalog based on the HST WFC3/F160W mosaic, whose astrometry is tied to Gaia Early Data Release 3 (Gaia-EDR3)~\citep{GaiaEDR32021summary,GaiaEDR32021astrometry}. 
The images are drizzled to a pixel scale of $30~\mathrm{mas~pixel^{-1}}$ and share the common tangent point with the EGS mosaic
\footnote{We notice a small but systematic astrometric offset between the CEERS and EGS mosaics of approximately $0.07''$ in right ascension and $0.04''$ in declination, which is smaller than the PSF FWHM in both the HST and JWST images.}.

\subsection{Additional ancillary data}

To improve the SED fitting, we incorporate additional multiband imaging data from ground- and space-based observations, following~\citetalias{stefanon2017}. 
These datasets extend the wavelength coverage of the SEDs from $0.4\mu m$ to $4.5\mu m$ across 19 bands, including the $u^*, g', r', i', z'$ imaging from CFHT/MegaCam, the $K_S$ bands from WIRDS and the $3.6\mu m$ and $4.5\mu m$ bands from Spitzer/IRAC, which all fully cover the EGS field. 
Further details on the ancillary data can be found in~\citetalias{stefanon2017}.
As summarized in Table~\ref{tab:hst data quality summary}, the angular resolutions of these images are significantly lower than the HST and JWST images.
We also provide flags for potential AGNs identified from Chandra/ACIS X-ray detections and for stars identified using Gaia astrometry, allowing these sources to be treated separately in further scientific analysis.

\subsection{Spectroscopic and 3D-HST grism redshifts}
To assess the quality of our photometric redshift estimates, we compare them with available spectroscopic redshifts in this field, providing an external validation of the SED-based redshift measurements.
Following~\citetalias{stefanon2017}, we incorporate spectroscopic redshift data from the DEEP2 and DEEP3 Galaxy Redshift Surveys~\citep{Coil2004,Cooper2006,Willner2006,Cooper2011,Cooper2012,Newman2013,Zhou2019_DEEP3}, obtained with the Keck/Deep Imaging Multi-Object Spectrograph (DEIMOS) campaign targeting the EGS field. 
The target galaxies in these surveys are generally brighter than $R=24.1$ mag.
We select the sources with high-quality redshifts (ZQUALITY $\geq 3$) extending to $z \sim 1.7$, yielding 1859 objects matched to our catalog.

Furthermore, we include JWST/Near-Infrared Spectrograph (NIRSpec) data from the RUBIES, a $\sim 61$-hour Cycle 2 program utilizing the NIRSpec/MSA instrument.
RUBIES predominantly targets red sources identified in JWST/NIRCam imaging in the Ultra Deep Survey (UDS) and EGS fields, spanning a wide redshift range from $z \sim 0.01$ to $z \sim 12.8$~\citep{degraaff2024rubies}. 
The dataset encompasses reduced low-resolution ($R \sim 100$) PRISM spectra and medium-resolution ($R \sim 1000$) G395M spectra, together with spectroscopic redshifts derived from PRISM spectra following meticulous visual inspection.
We focus on the Grade 3 and Grade 2 galaxies, corresponding to robust and ambiguous (single line detection) redshifts, respectively. 
Objects with visually assigned redshift value of 0 are excluded from our investigation.
This results in 1204 sources in the EGS field with reliable spectroscopic redshifts from RUBIES.

We also include grism redshifts from the 3D-HST survey~\citep{Brammer2012grism, Momcheva2016}.
The grism data provide low-resolution slitless spectra over the EGS field, covering the wavelength range $\sim 1.1-1.7\mu m$, and are particularly useful for identifying emission-line galaxies at intermediate redshifts.
We use the publicly available grism redshift catalog, adopting the sources with $\mathrm{max_{contam}}<0.2$, $\mathrm{JH_\mathrm{mag}}<24$, $\mathrm{f_\mathrm{cover}}>0.9$, and $\mathrm{f_\mathrm{flagged}}<0.1$.
With redshift range from $0.04$ to $1.65$, we finally obtain 1049 grism redshifts from the 3D-HST survey. 

After removing duplicate entries across the input catalogs, the combined spectroscopic and grism-redshift sample contains 3433 objects with reliable redshifts spanning $0.01 < z < 11.41$.

\section{Panchromatic photometry}\label{sec:photometry}

In this section, we describe our procedures for source detection, PSF construction, and multiband photometric measurements. 
Empirical PSFs are constructed from isolated point sources in each band.
We perform forced-aperture photometry, measuring fluxes at fixed source positions across all bands to ensure consistent color measurements, except for the Spitzer/IRAC channels 1 and 2 (CH1 and CH2) images. 
For the Spitzer/IRAC data, we use the software \software{TPHOT}~\citep{Merlin2015tphot,Merlin2016tphot2}, which performs prior-based deblended photometry by constructing source templates from high-resolution HST images and fitting them to the lower-resolution IRAC data after PSF convolution.
This approach is essential due to the significantly lower angular resolution of the Spitzer images compared to HST, which otherwise results in severe source blending.

\subsection{Source detection in HST and JWST images}\label{subsec:source detection}

We perform source detection in the EGS WFC3/F160W mosaic to construct a primary catalog.
The F160W band probes the rest-frame optical light of high-redshift galaxies and yields higher S/N values than the bluer HST filters in the EGS field, making it the most suitable band for identifying faint galaxies without compromising reliable segmentation. 
In addition, source detection is performed in the CEERS NIRCam/F150W and F200W bands to confirm the faint galaxies and assess the false detection rate. 

Source detection and structure fitting are carried out with \software{GALAPAGOS-2}~\citep{Barden2012galapagos,Haussler2022galapagos2}, which integrates \software{SExtractor}~\citep{Bertin1996sextractor} for source detection with \software{GALFITM}~\citep{Haussler2022galapagos2} for two-dimensional \sersic modeling within a unified pipeline. 
We adopted the standard cold and hot detection strategy implemented in \software{GALAPAGOS-2}.
The cold mode is designed to detect bright and compact sources, while the hot mode identifies faint and diffuse sources.
This dual-mode approach is essential for constructing deep catalogs in crowded extragalactic fields, as it balances completeness and reliability. 
To push the detection limit to the fainter magnitudes in the HST images, we set the parameter \texttt{DETECT\_THRESHOLD} to $1.5\sigma$ with \texttt{DETECT\_MINAREA} set to $8$ pixels in hot mode, and \texttt{DETECT\_THRESHOLD} to $5\sigma$ with \texttt{DETECT\_MINAREA} set to $4$ pixels in cold mode.
Sources detected in hot mode that fall within the expanded Kron ellipse of cold-mode objects (semimajor axis $a_\mathrm{Kron,ellip}=1.1~R_\mathrm{Kron}$) are subsequently removed to prevent a single complex galaxy from being segmented into multiple sources.
Other hot-mode sources are retained and merged with cold-mode sources to construct the final catalog and segmentation maps.
The main \software{SExtractor} parameters used in cold and hot modes are summarized in Table~\ref{tab:SE_params}.
With this procedure, we identified 72,467 objects in the F160W mosaic for the EGS field, comprising 25,752 in cold mode and 46,715 in hot mode.

\begin{figure}[htpb]
    \centering
    \includegraphics[width=0.9\linewidth]{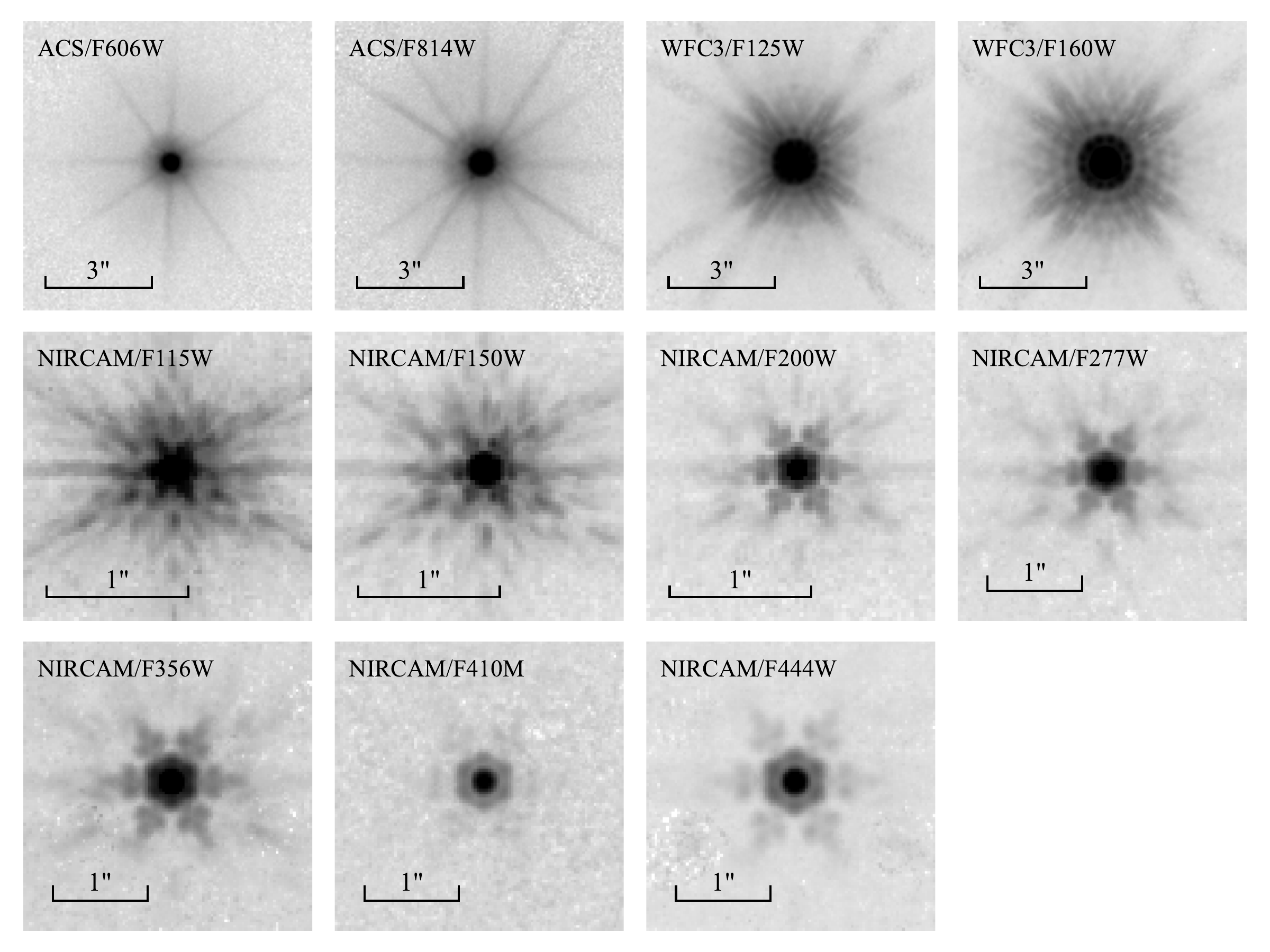}
    \caption{The empirical PSF images of HST and JWST in different filters. The outer structure of PSF is clear, which indicates the high quality of our PSF construction.}
    \label{fig:psf_image}
\end{figure}
\begin{figure}[htpb]
    \centering
    \includegraphics[width=0.9\linewidth]{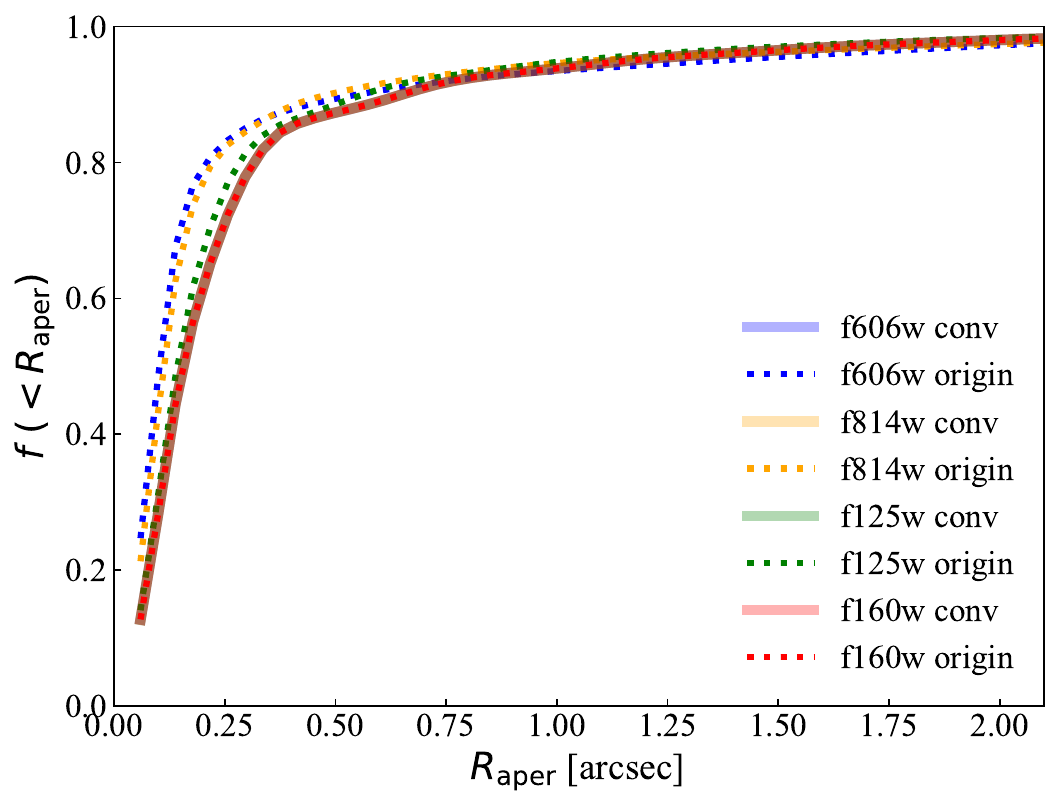}
    \caption{
    The encircled energy growth curves of the PSFs in Fig.~\ref{fig:psf_image} as a function of aperture radius ($R_{\mathrm{aper}}$).
    The dashed and solid lines represent the original PSF and the 
    PSF matched to that of the HST F160W filter, respectively.
    The psf-matching process has good performance as the curves of growth for the matched PSF are consistent.}
    \label{fig:psf_curve}
\end{figure}

Because setting the detection threshold to 1.5$\sigma$ inevitably increases the false detection rate, we used independent observations from the CEERS field to validate faint detections. 
The CEERS F150W and F200W mosaics are used in this work to validate the faint detections in HST images, as their wavelengths are similar to the WFC3/F160W filter.
We therefore performed source detection on the JWST mosaics using similar procedures, adopting slightly higher thresholds appropriate for the deeper JWST data. 
Within the overlapping area of approximately $97$ $\mathrm{arcmin}^2$, we detected 71,604 objects in F150W and 82,084 objects in F200W. 
Cross-matching JWST source catalogs with the HST detections provides a direct estimate of the false detection rate and enables us to derive empirical criteria for identifying the real faint sources.
This cross-validation step is particularly important for a catalog that aims to reach the faint-end limit of the HST images with well-controlled uncertainties.
The final parameter set, listed in Table~\ref{tab:SE_params}, was chosen to maximize the number of true detections while keeping the contamination rate within acceptable limits for the subsequent structure analysis.

\subsection{PSF construction and resolution matching}\label{subsec:psf}

An accurate PSF is essential in galaxy structure analysis.
\citet{Zhuang2024psf} has demonstrated that empirical PSFs from observations provide more accurate structure measurements than theoretical PSF models for both HST and JWST imaging.
To construct the empirical PSF, we select bright point sources from the images in different bands.
Based on the source catalog from \software{SExtractor}, we identify the point sources using the following criteria: FLAGS = 0, CLASS\_STAR > 0.8, and ELLIPTICITY < 0.3. 
Additionally, we adjust the S/N criteria separately for different bands to ensure a sufficient number of point sources with a minimum S/N greater than 50, which helps to better construct the PSF.
To mask other objects near the point source, we expand the mask by smoothing with a kernel of 15 pixels for high-resolution images (from HST and JWST) and 30 pixels for other low-resolution images to exclude the faint envelope of the neighboring bright source, which may not be well identified by the segmentation map (and mask) from \software{SExtractor}. 
We also remove the saturated point sources and other artifacts.
The final empirical PSF is built by median stacking the normalized point source images.
The PSF FWHM is measured as the FWHM of the best-fit 2D Gaussian function.
The PSF FWHM values are summarized in Table~\ref{tab:hst data quality summary}. 
Illustrations of the PSFs for all HST and JWST bands used in our study can be found in Fig.\ref{fig:psf_image}.

To assess the systematic photometric bias across bands that have different PSFs, we construct PSF-matching kernels to homogenize the image resolution.
For the PSF matching process, we utilize \software{Pypher}~\citep{Boucaud_2016_pypher} to obtain the convolution kernels for each band. 
The encircled energy growth curves of HST images before and after the PSF-matching are shown in Fig.~\ref{fig:psf_curve}, where dashed lines represent the original PSF growth curves and solid lines represent the PSF-matched growth curves.
The solid curves are consistent with each other.
The difference between the growth curves is less than 1\% in the outer part of the profile.

\subsection{Nearby bright stars and mosaic boundary}\label{subsec:mask}

\begin{figure}[ht!]
    \centering
    \includegraphics[width=0.4\textwidth]{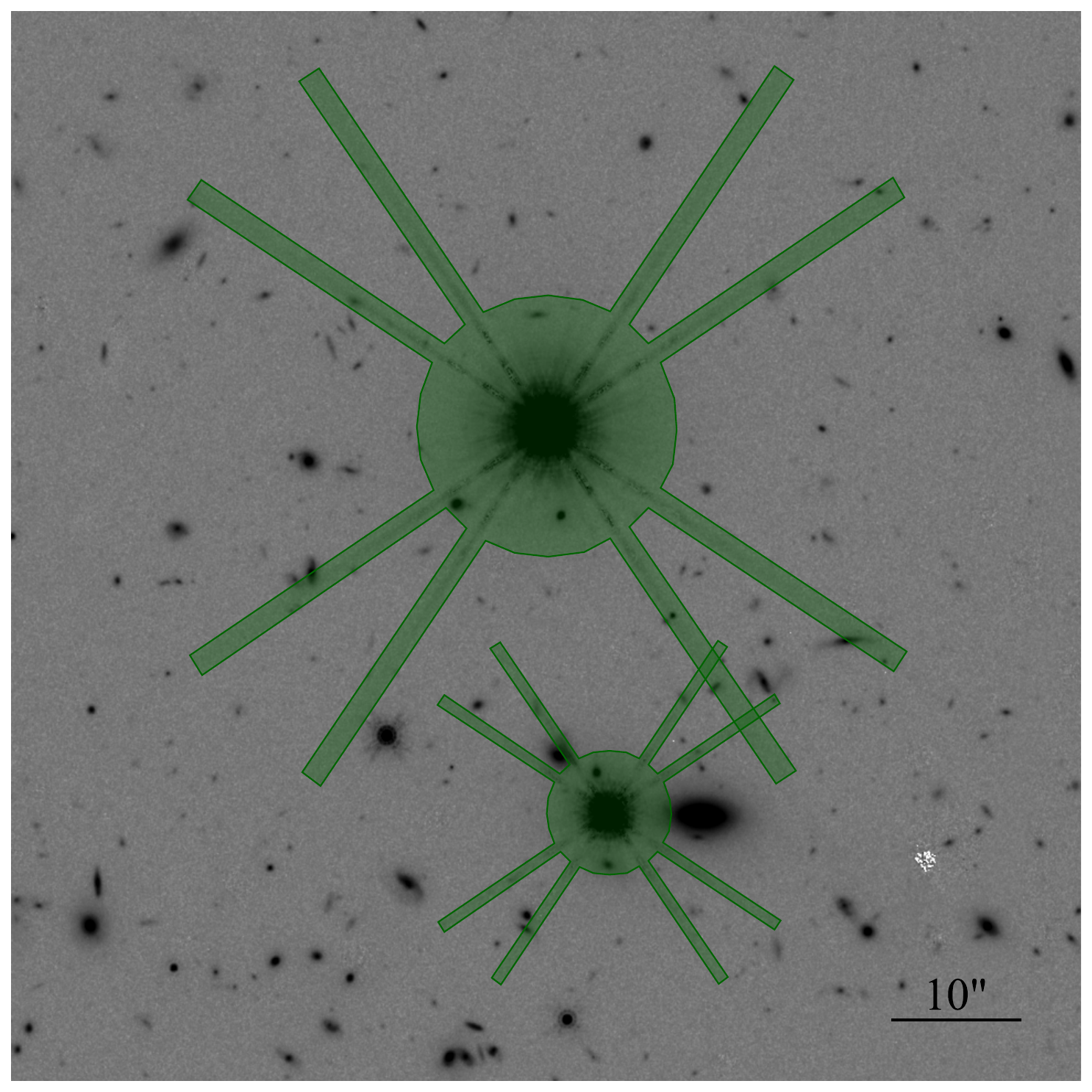}
    \caption{The green polygons represent the mask template of two bright stars in the HST F160W image. }
    \label{fig:mask}
\end{figure}

Galaxies located near bright stars suffer significant contamination in both detection and photometric measurements.
Similar to~\citetalias{stefanon2017}, we visually define the bright star mask templates for both the EGS and CEERS mosaics, encompassing the bright stars and their extended wings.
The resulting mask regions are shown as green regions in Fig.~\ref{fig:mask}.
In these templates, galaxies with an overlap fraction exceeding 10\% with these masks are assigned the flag MASK=2.
In total, approximately $3.6\%$ of galaxies in the EGS field and $\sim 4.8\%$ of galaxies in the CEERS field are affected by nearby bright stars.

The boundaries of the mosaics exhibit lower S/N values due to shorter exposure time.
Galaxies are categorized as near-edge if their distances from their centers to the nearest image edges are less than 200 pixels for HST images, 70 pixels for CEERS pointings 1 and 2, and 40 pixels for other CEERS pointing, which are empirically determined by visually checking the S/N values of image edges.
These galaxies are assigned flag MASK=4. 
In the EGS field, approximately $3.0\%$ of galaxies are near the mosaic edge, while in the CEERS field this fraction is $\sim 11.4\%$.

\subsection{Photometric measurements of galaxies}\label{subsec:photometry measurement}

Accurate photometric redshift estimation and SED fitting require both broad wavelength coverage and consistent flux measurements across different bands.
We therefore conduct photometry measurements on ground-based optical and space-based optical and NIR images, including HST ACS/F606W, F814W, WFC3/F125W, F140W, F160W, JWST NIRCam/F115W, F150W, F200W, F277W, F356W, F410M, F444W, CFHT/$u^*$, $g'$, $r'$, $i'$, $z'$, WIRCam/Ks, and Spitzer IRAC/$3.6\mu m$ and $4.5\mu m$.
Given the large variation in angular resolution among these datasets, a single photometric strategy would not be optimal for all bands. 
We therefore adopted two complementary approaches according to the angular resolution.

For all bands with sufficient resolution to reliably isolate individual sources, including HST, JWST, and CFHT images, we measured fluxes independently in each band using elliptical apertures based on the isophotal shape of each galaxy. 
The aperture shape, orientation, and size are derived from \software{SExtractor} isophotal parameters. 
To minimize biases caused by low S/N detections and blending with nearby bright companions, the semimajor axis of the aperture was defined as
\begin{equation}
    a_\mathrm{ellipse}=
    \begin{cases}
        R_\mathrm{KRON},& \text{default\ value};\\
        R_\mathrm{ISO},& \text{if}\ R_\mathrm{KRON}<R_\mathrm{ISO};\\
        1-1.5\mathrm{FWHM},& \text{if}\ R_\mathrm{ISO}<1.5\mathrm{FWHM}
    \end{cases}
\end{equation}
where $R_\mathrm{KRON}$ is the Kron radius and $R_\mathrm{ISO}=\sqrt{area_\mathrm{seg}}$ is the effective radius derived from the segmentation area.
Since a Kron aperture typically encloses $\sim96\%$ of the total flux~\citep{Kron1980}, we adopt it as the default setting for our measurements.
Visual inspection shows that for objects with $R_\mathrm{KRON}<R_\mathrm{ISO}$, $R_\mathrm{ISO}$ encloses a larger fraction of the flux, so we use $R_\mathrm{ISO}$ for these galaxies.
For objects with $R_\mathrm{ISO}<1.5~\mathrm{FWHM}$, we adopt a fixed aperture size to ensure consistent measurements across images with different PSFs.
Neighboring sources and bad pixels within the aperture were masked and interpolated. 
The local background was estimated in an elliptical annulus between 1.25 and 1.6 times the aperture radius~\citep{Liyang2023}, chosen to provide an area comparable to that of the source aperture.
Because the PSF FWHM varies among the images, 
the smoothing of the object light profile differs from band to band.
Consequently, a fixed aperture for flux measurement would result in different amounts of flux loss.
In our catalog, we applied a PSF growth-curve correction in each band to recover total fluxes. 

This correction assumes that the radial light profile of the source follows the PSF. 
While this assumption is not strictly valid for extended galaxies, it is reasonable for faint and compact objects that are largely PSF-dominated.

\begin{figure*}
    \centering
    \includegraphics[width=0.95\textwidth]{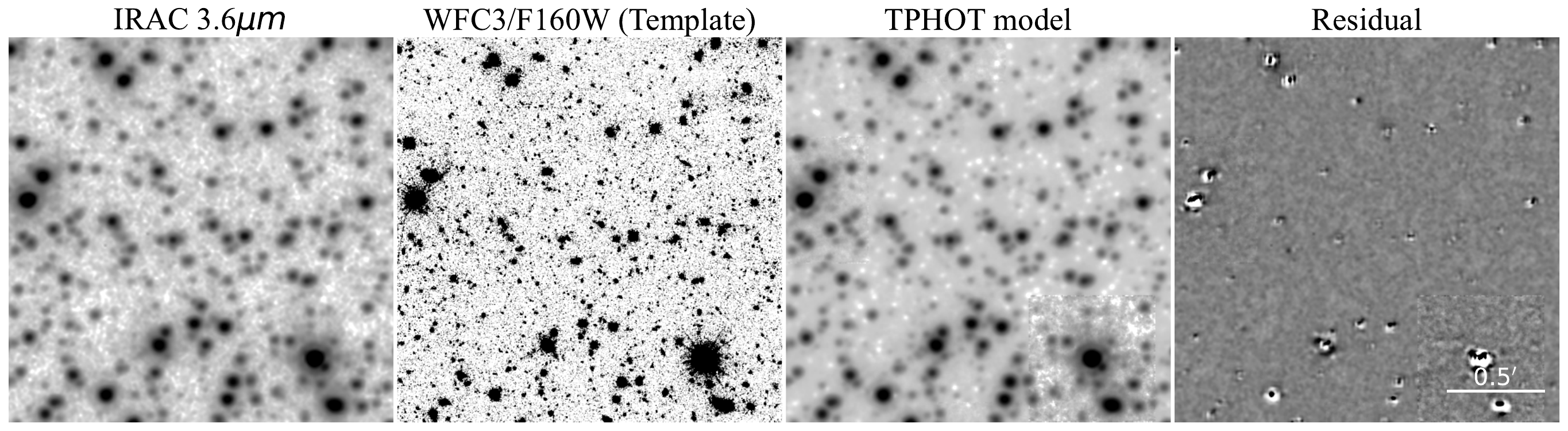}
    \caption{The \software{TPHOT} template fitting examples on Spitzer IRAC/$3.6\mu m$ images. 
    From left to right are the cutout of Spitzer IRAC/$3.6\mu m$ image, corresponding HST WFC3/F160W images (used as fitting template), model image and residual image from \software{TPHOT}.}
    \label{fig:tphot_example}
\end{figure*}

In contrast, for low-resolution images (e.g., Spitzer/IRAC), aperture photometry would require large apertures to capture the total flux, even on PSF-matched images. 
This would significantly reduce the S/N and increase contamination from neighboring sources in crowded regions (as shown in the last two panels in Fig.~\ref{fig:check_aperture_psf_correction}). 
To mitigate these effects, we performed template-fitting photometry using software \software{TPHOT}~\citep{Merlin2015tphot,Merlin2016tphot2}.
HST F160W image, a high-resolution image, serves as the morphological template, which is convolved to match the PSF of the low-resolution images and fitted simultaneously for all objects. 
Following~\citetalias{stefanon2017}, the background level was fixed to zero during the fitting procedure. 
Examples of low-resolution image cutout (IRAC/$3.6\mu m$), the corresponding template from the high-resolution image (WFC3/F160W), the best-fit model, and the residual map are shown in Fig.~\ref{fig:tphot_example}. 
The residuals are generally small, although larger residuals are occasionally observed in central regions for bright sources, likely owing to minor astrometric distortion, color-dependent morphological variations, and uncertainties in the PSF modeling.

In summary, aperture photometry with PSF correction was adopted for high-resolution bands where source morphology can be reliably traced.
On the other hand, template-fitting photometry was used for low-resolution bands to minimize blending contamination and S/N loss. 
Both approaches implicitly assume limited morphological variation with wavelength, which may introduce minor systematic uncertainties but is unlikely to significantly affect the faint, compact galaxies that dominate our sample.

\section{Detection limit, completeness and properties of the photometric catalog}\label{sec:catalog_properties}
In this section, we introduce our tests of the detection limit, completeness, and bias in structure measurement, based on both observations and mock images.

\subsection{Mock image tests on the detection limit and completeness}\label{subsec:detection completeness}

\begin{figure*}
    \centering
    \includegraphics[width=0.9\textwidth]{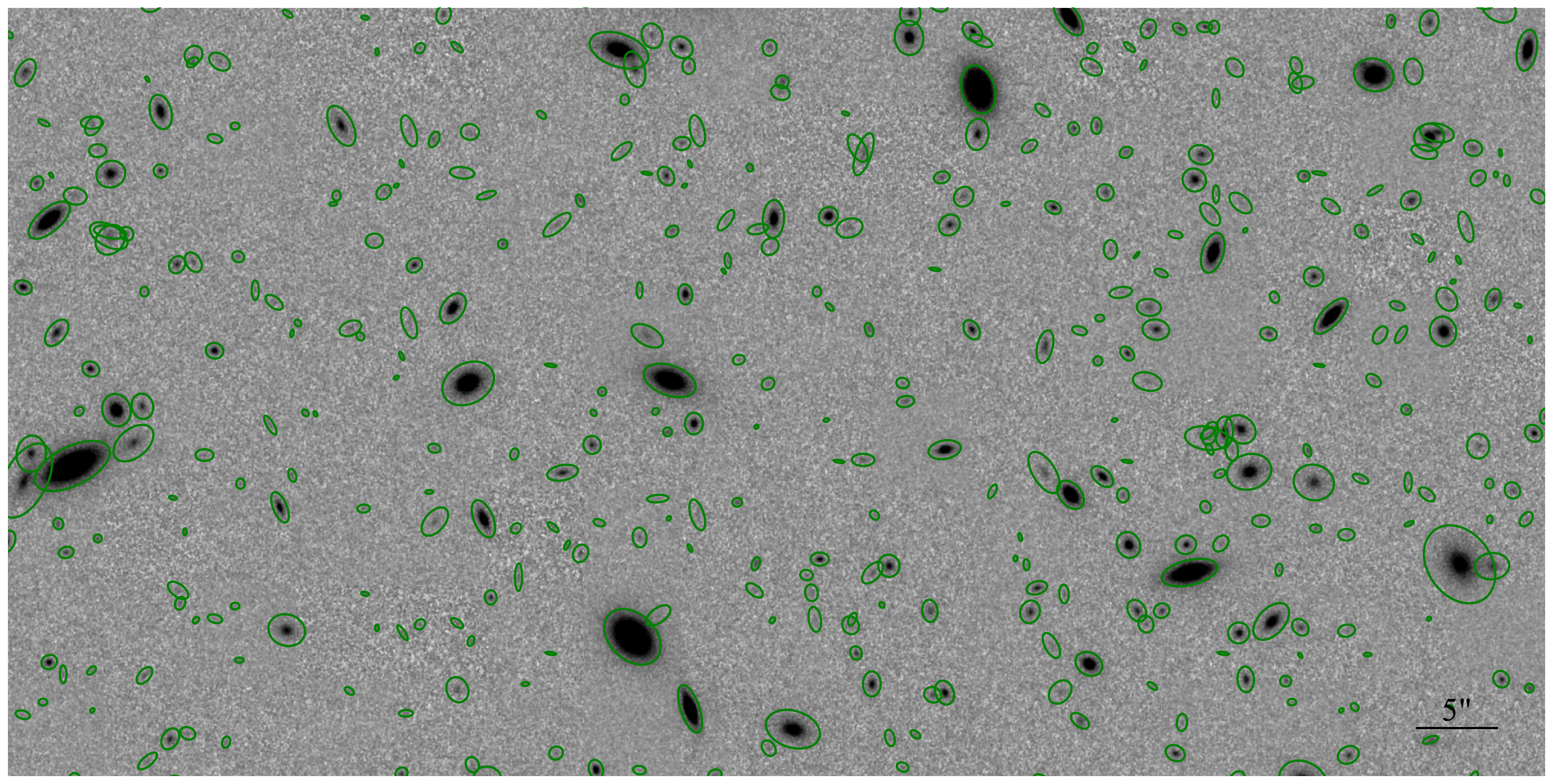}
    \caption{Example of mock image composed of single-\sersic models. The objects are positioned according to the EGG catalog, featuring magnitudes, effective radii, \sersic indices, axis ratios, and position angles that closely resemble those distributions in the HST observations. The green ellipses are Kron ellipses from \software{SExtractor} for each object.}
    \label{fig:mock_image}
\end{figure*}

To understand the detection limit and completeness of our method, we generate the single \sersic mock images.
To mimic the real sky distribution and the structure parameters of galaxies in the HST mosaics, we utilize the Empirical Galaxy Generator (\software{EGG},~\citealt{Schreiber2017egg}) to generate the input mock catalog.
\software{EGG}, specifically its \texttt{egg-gencat} function,
enables us to create mock catalogs similar to the HST observations, offering realistic simulations of high-redshift galaxies across different wavelengths and spatial distributions.
Using \software{EGG}, we produce a mock catalog with a limiting magnitude (equivalent to the peak of the magnitude distribution in \software{EGG}) at 28 mag within an area of $0.01\ \mathrm{degree^2}$, matching the area and depth of the EGS F160W mosaic.
This mock catalog includes parameters such as sky positions, magnitudes, axis ratios, position angles, effective radii for both the bulge and disk components separately, and the bulge-to-disk flux ratio.
The distributions of these parameters are consistent with the HST observations.
The magnitude of the mock galaxies ranges from 15 to 30 mag, and the effective radii range from 0.1 to 200 pixels (0.006 to 12 arcsec).
The \sersic index ($n$) of each mock galaxy ranges from 0.1 to 10.
Notably, the redshift upper limit of galaxies in this catalog is set to $z=8$.
The total number of mock galaxies is 9214.
To create the mock images, we generate the models using \software{GALFIT}~\citep{Peng2002galfit,Peng2010galfit} based on the EGG catalog and combine them with a background, which is carefully extracted from the empty regions in the EGS field.
As shown in Fig.~\ref{fig:mock_image}, the mock image is similar to the real HST F160W observations, which indicates these mock galaxies closely resemble real galaxies.

\begin{figure*}[ht]
    \centering
    \includegraphics[width=0.95\textwidth]{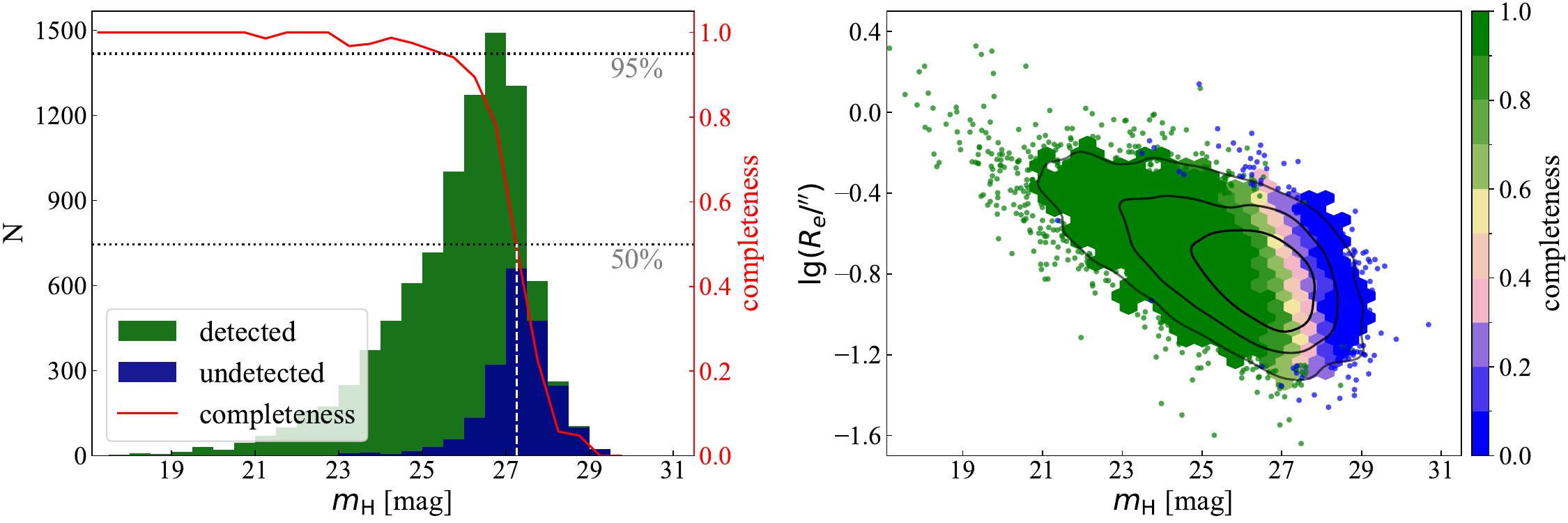}
    \caption{
    Detection completeness analysis based on the mock image. In the left panel, the green and blue histograms show the input F160W magnitude distribution of successfully detected and undetected objects, respectively. The red curve represents the detection completeness (corresponds to the right axis), with the 95\% and 50\% completeness levels (25.7 mag and 27.2 mag, respectively) marked as black dotted lines. The vertical white dashed line indicates the 50\% completeness limit at 27.2 mag. The right panel shows the detection completeness map in the $m_\mathrm{H}-R_\mathrm{e}$ plane. $m_\mathrm{H}$ and $R_\mathrm{e}$ are the input parameters from the EGG catalog. The green-to-blue colormap encodes the detection completeness in each grid. The green and blue points indicate the detected and undetected sources, respectively. The contours represent the morphological distribution of galaxies from the EGG catalog. Clearly, as the galaxy becomes fainter and larger, i.e., with lower surface brightness, the detection completeness drops quickly.}
    \label{fig:completeness}
\end{figure*}

As in Sect.~\ref{subsec:source detection}, we apply \software{GALAPAGOS-2} to the mock images, following the same procedure used for the real F160W mosaic.
Here, the detection completeness is defined as the fraction of input objects that are successfully detected in a given magnitude range. 
The left panel of Fig.~\ref{fig:completeness} shows the magnitude distribution of the detected sources (green) and undetected ones (blue). 
The red curve represents the detection completeness (corresponds to the right axis) as a function of the magnitude.
For magnitudes fainter than 28.7 mag in F160W, no galaxy is detectable, indicating a detection limit of $~28.7$ mag. 
For galaxies brighter than $\thicksim 25$ mag, the detection completeness consistently exceeds 95\%~\footnote{Visual inspection shows that the few undetected objects brighter than 25 mag are strongly blended, making deblending difficult.}.
The detection completeness of galaxies fainter than 25 mag drops with increasing magnitude. 
At magnitude fainter than 27.2 mag, the detection completeness drops below 50\%.

Indeed, the detection completeness depends on the surface brightness for extended objects, which can be described by magnitude ($m_\mathrm{H}$), effective radius ($R_\mathrm{e}$), and \sersic index ($n$).
The right panel of Fig.~\ref{fig:completeness} shows the distribution of the detected sources (green) and undetected ones (blue) in the $m_\mathrm{H}-R_\mathrm{e}$ plane, with the detection completeness represented by the color of hexagons.
The clear color distribution pattern shows that fainter galaxies with larger sizes are less likely to be detected.
In Appendix~\ref{appendix:completeness sersic index}, we further test the detection completeness of galaxies with different \sersic indices ($n=0.5,1,4,8$) in the same $R_\mathrm{e} - m_\mathrm{H}$ parameter space.
The galaxies with higher \sersic indices can be detected more easily, as they have higher central surface brightness.

\begin{figure}[ht]
    \centering
    \includegraphics[width=0.35\textwidth]{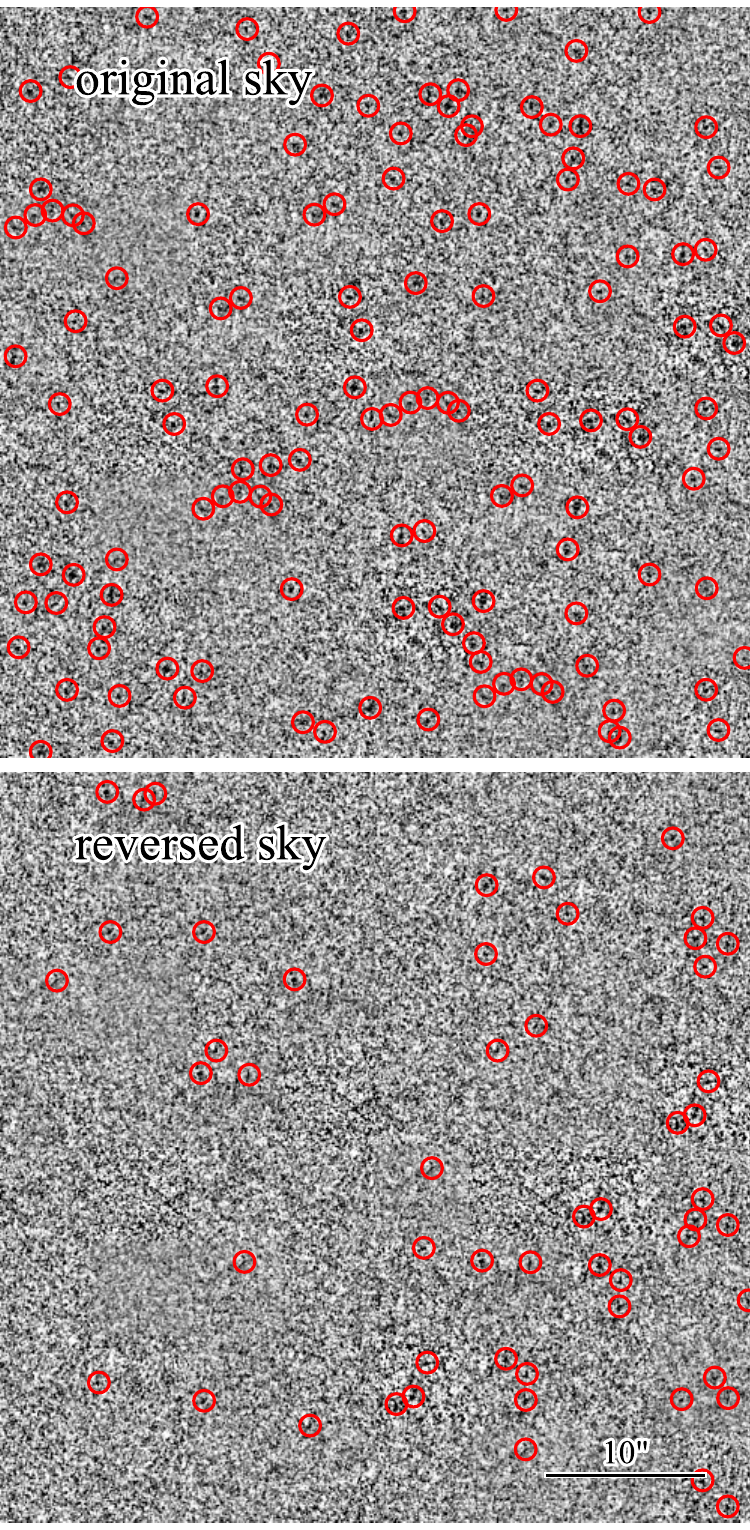}
    \caption{The top and bottom panels show the original and reversed sky images, respectively. The red circles represent the detections on each sky image.}
    \label{fig:sky}
\end{figure}

\begin{figure}[ht]
    \centering
    \includegraphics[width=0.95\linewidth]{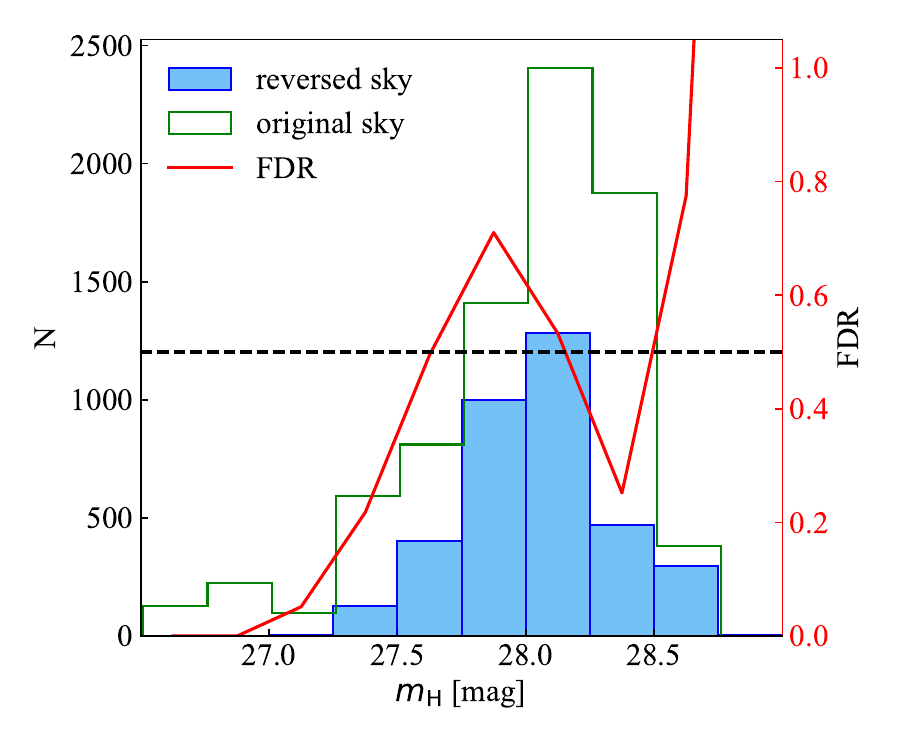}
    \caption{
    F160W magnitude distributions of the detected sources in the original and reversed sky backgrounds, shown as green and blue histograms, respectively.
    The red curve (corresponds to the right axis) indicates the FDR at different magnitudes.
    For sources brighter than $\sim 26.8$ mag, FDR is negligible ($\sim 0$).
    FDR first exceeds 50\% at $\sim 27.6$ mag, suggesting that more than half of the sources at this magnitude value are unreliable.}
    \label{fig:FDR}
\end{figure}

To further assess the detection completeness, we generate three additional sets of mock images: (1) mock images with different fixed \sersic indices, (2) mock galaxies with individual cutouts, ensuring that all of them are isolated, and (3) mock images with a Gaussian background. 
As shown in Fig.~\ref{fig:1d_completeness_diff_n}, surface brightness and blending effects strongly influence the detection completeness, whereas the background pattern has only a slight effect.
More discussions of these tests are presented in Sect.~\ref{sec:discussion} and Appendix~\ref{appendix:completeness}.

\subsection{False detection rate estimation}\label{subsec:FDR}

In the faint end, fake sources can emerge due to the spurious noise in the background.
It is therefore necessary to quantify the false detection rate (FDR) in the HST mosaics and to reduce it empirically.
To assess the FDR, we perform detection tests on both the original and reversed sky images, similar to~\citet{Revalski2023}.
In this work, the reversed sky map is created by multiplying the original sky image by $-1$.
Ideally, if the background noise were purely random and followed a Gaussian distribution, the number of sources detected in the original sky background should be similar to the number of sources detected in the reversed sky images after subtracting the median sky value.
Therefore, sources identified in the two sky images  should all be due to the background fluctuations.
However, in real observations, faint galaxies do exist in the background and will not appear in the reversed sky map. 
In this case, all sources identified in the reversed sky map correspond exclusively to false detections.
Because detections in the reversed image trace spurious sources, whereas the original image contains both real and spurious detections, we estimate the FDR as
\begin{equation}
    \mathrm{FDR} = \frac{N_{\mathrm{reversed}}}{N_{\mathrm{original}}},
\end{equation}
where $N_{\mathrm{reversed}}$ and $N_{\mathrm{original}}$ are the numbers of sources detected in the reversed and original images, respectively.

From the EGS F160W mosaic, we carefully select patches of the sky background without obvious sources and combine these patches to create the sky background and the corresponding reversed sky map, as shown in Fig.~\ref{fig:sky}.
After applying \software{GALAPAGOS-2} with the same detection threshold as used for the observation data, we detect many sources (red circles) in both sky images.
As depicted in Fig.~\ref{fig:FDR}, detections in the reversed sky map are generally fainter than 27.2 mag, suggesting that objects detected in the original image brighter than approximately 27 mag are likely real sources. 
Notably, the number of detections in both the original and reversed sky maps increases with magnitude.
At magnitude fainter than 27.6 mag, FDR surpasses 50\%, suggesting that a substantial proportion of these objects are likely to be false detections rather than real sources.
We notice an apparent decrease in the FDR at magnitudes fainter than 28, which is caused by the asymmetry in the sky pixel distribution, with slightly more positive than negative pixels. 
The presence of potential faint galaxies biases the sky distribution slightly toward positive values. 
As a result, noise fluctuations of intrinsically identical amplitude are preferentially measured as fainter sources in the original sky.
We therefore adopt the magnitude at which the FDR first exceeds 50\% as our final detection magnitude threshold.
In Sect.~\ref{subsec:jwst cross match}, we investigate the structural properties of the false detections in detail to provide an empirical criterion to effectively distinguish real sources from the false ones.

\subsection{Statistical properties of the photometric catalog in the EGS field}\label{subsec:number counts}

\begin{figure}[ht!]
    \centering
    \includegraphics[width=0.45\textwidth]{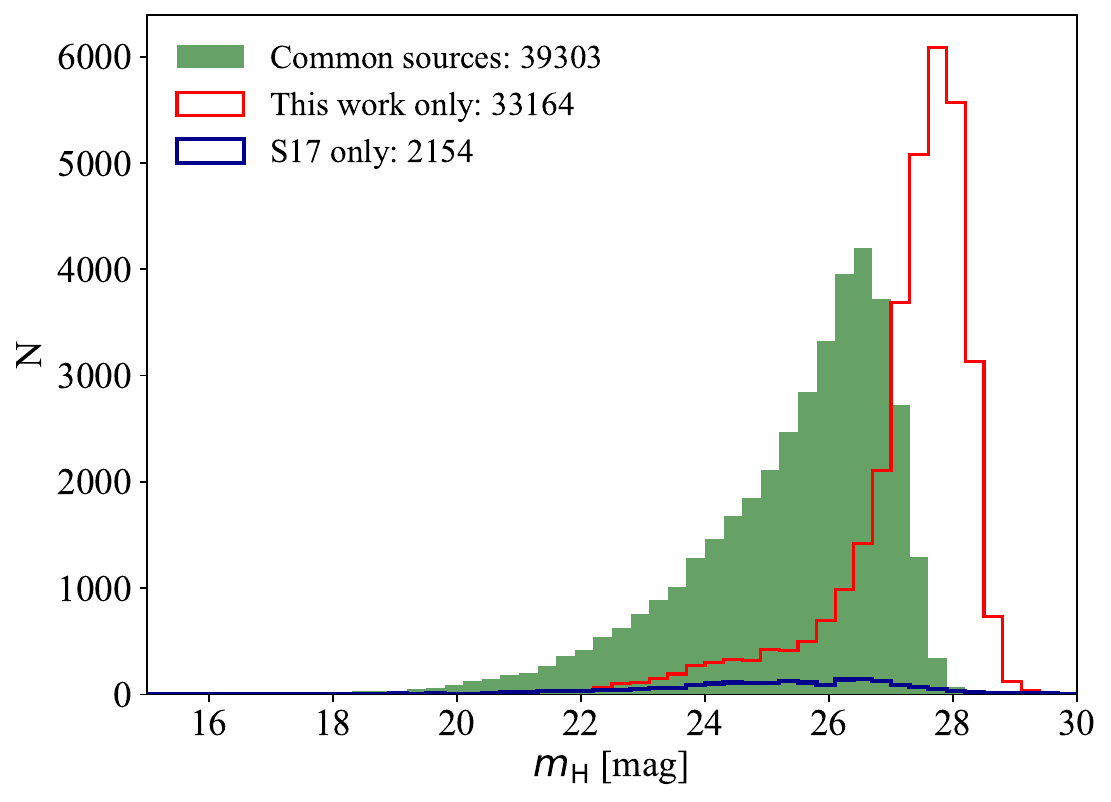}
    \caption{ 
    F160W magnitude distribution of galaxies detected in our work and in~\citetalias{stefanon2017}. 
    The green histogram shows the sources detected in both catalogs (common sources), the red histogram shows objects detected only in our work, and the blue histogram shows objects detected only in \citetalias{stefanon2017}.
    The common sources are generally brighter than 28 mag, peaking at $\sim 26.8$ mag. Most of the objects detected exclusively by our work are fainter than $\sim 26$ mag, with the peak at $\sim 28$ mag.}
    \label{fig:H_hist}
\end{figure}

\begin{figure}[htpb]
    \centering
    \includegraphics[width=0.45\textwidth]{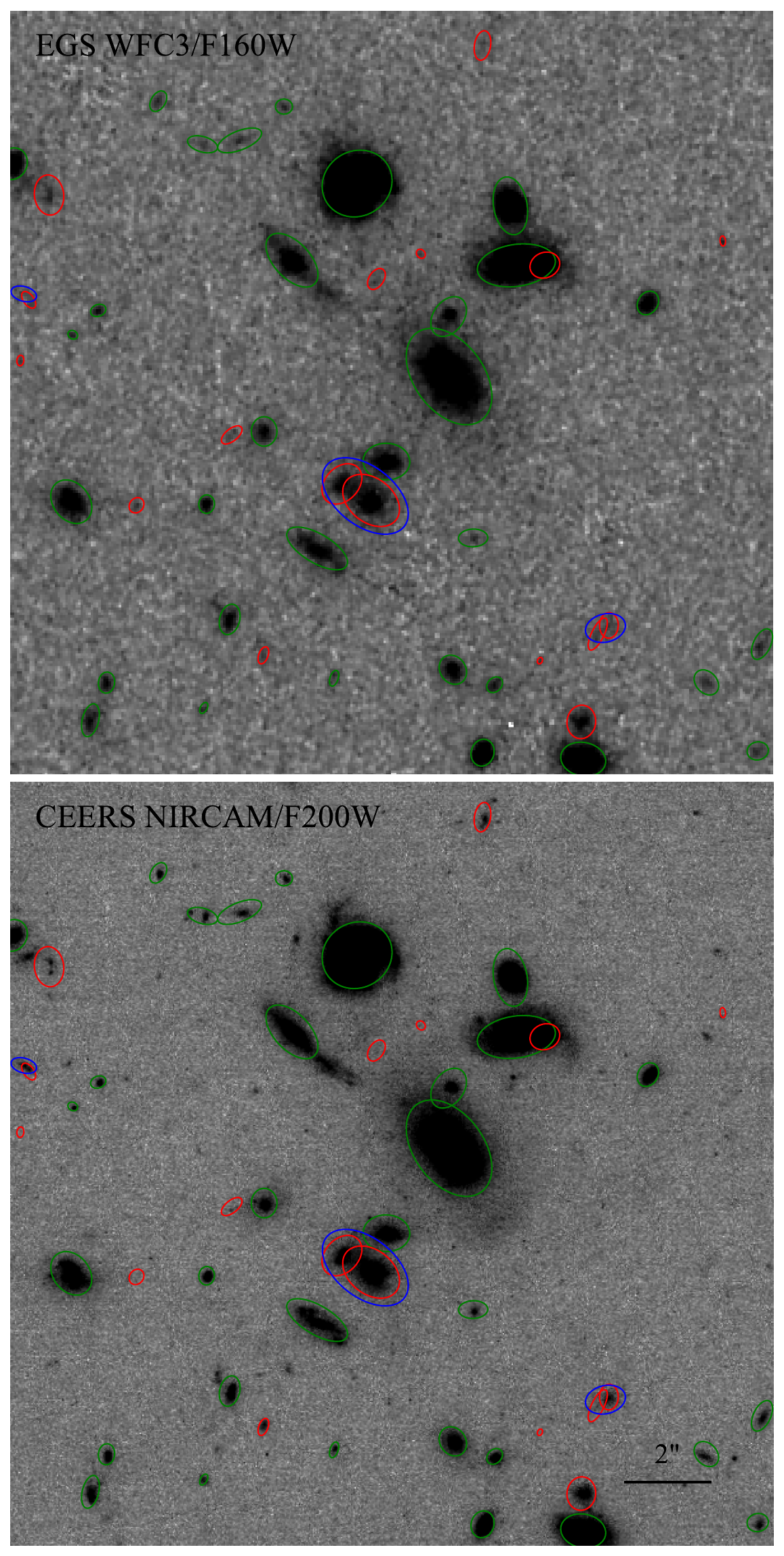}
    \caption{
    Example of source detection in the HST F160W image (top panel) and the JWST F200W image (bottom panel) for the same EGS field.
    Green circles mark the sources detected in both our work and S17, based on the EGS F160W image.
    Red and blue circles indicate objects detected only in our work and only in~\citetalias{stefanon2017}, respectively. 
    Many faint sources in the red circles are clearly visible as real objects in the JWST images, confirming the reliability of our detection method for the faint sources.
    }
    \label{fig:hst_jwst_our_stefanon}
\end{figure}

Using the source detection method described in Sect.~\ref{subsec:source detection}, we successfully identified 72,467 objects in the F160W mosaic of the EGS field. 
This number significantly exceeds the 41,457 sources reported in \citetalias{stefanon2017}.
Fig.~\ref{fig:H_hist} compares the magnitude distribution of our detected objects with that of the sources in~\citetalias{stefanon2017}.
Most sources in~\citetalias{stefanon2017} (39,303) are included in our catalog, while only 2,154 of their sources  remain undetected.
However, we identify 33,164 new objects that were not reported in~\citetalias{stefanon2017}.
These new objects are relatively faint, peaking at $\sim$28 mag but extending to as bright as 22 mag, as shown in Fig.~\ref{fig:hst_jwst_our_stefanon}.
The common sources, on the other hand, mainly populate the bright end.

To further validate our source detection process, Fig.~\ref{fig:hst_jwst_our_stefanon} illustrates a portion of the mosaic in the EGS field (top panel) and the corresponding CEERS mosaic (bottom panel) with detected sources marked by ellipses (green for the common sources; red for the sources only in our catalog; blue for the sources only in~\citetalias{stefanon2017}). 
Upon visual inspection, many of the red-outlined objects appear faint and diffuse in the EGS mosaic.
In the CEERS F200W mosaic (bottom panel), most of these faint sources are likely real galaxies, as shown in the bottom panel.
Conversely, there are also a small number of sources from~\citetalias{stefanon2017} missed by our catalog. 
Visual inspection suggests that some mismatches are caused by the more aggressive source detection strategy in this work, which can better deblend neighboring galaxies\footnote{Some galaxy pairs are not deblended and are identified as a single source in~\citetalias{stefanon2017} catalog, causing mismatches.}.

For the source detection,~\citetalias{stefanon2017} used the \texttt{rms} map, whereas our analysis adopts the \texttt{weight} map, which was typically considered as the inverse of the variance map, and is produced simultaneously with the drizzled image.
The detection threshold in~\citetalias{stefanon2017} was set to better identify relatively bright galaxies and to avoid too many spurious source contamination.
We push the source detection to the faint limit.
As explained in Sect.~\ref{subsec:jwst cross match}, we perform a statistical analysis of the structural parameters of faint sources to empirically remove 12,821 objects, effectively reducing the false detection rate.
Finally, we obtain a final clean catalog of 59,646 objects, with 18,189 objects more than that of~\citetalias{stefanon2017}.

\begin{figure*}[htpb]
    \centering
    \includegraphics[width=0.75\textwidth]{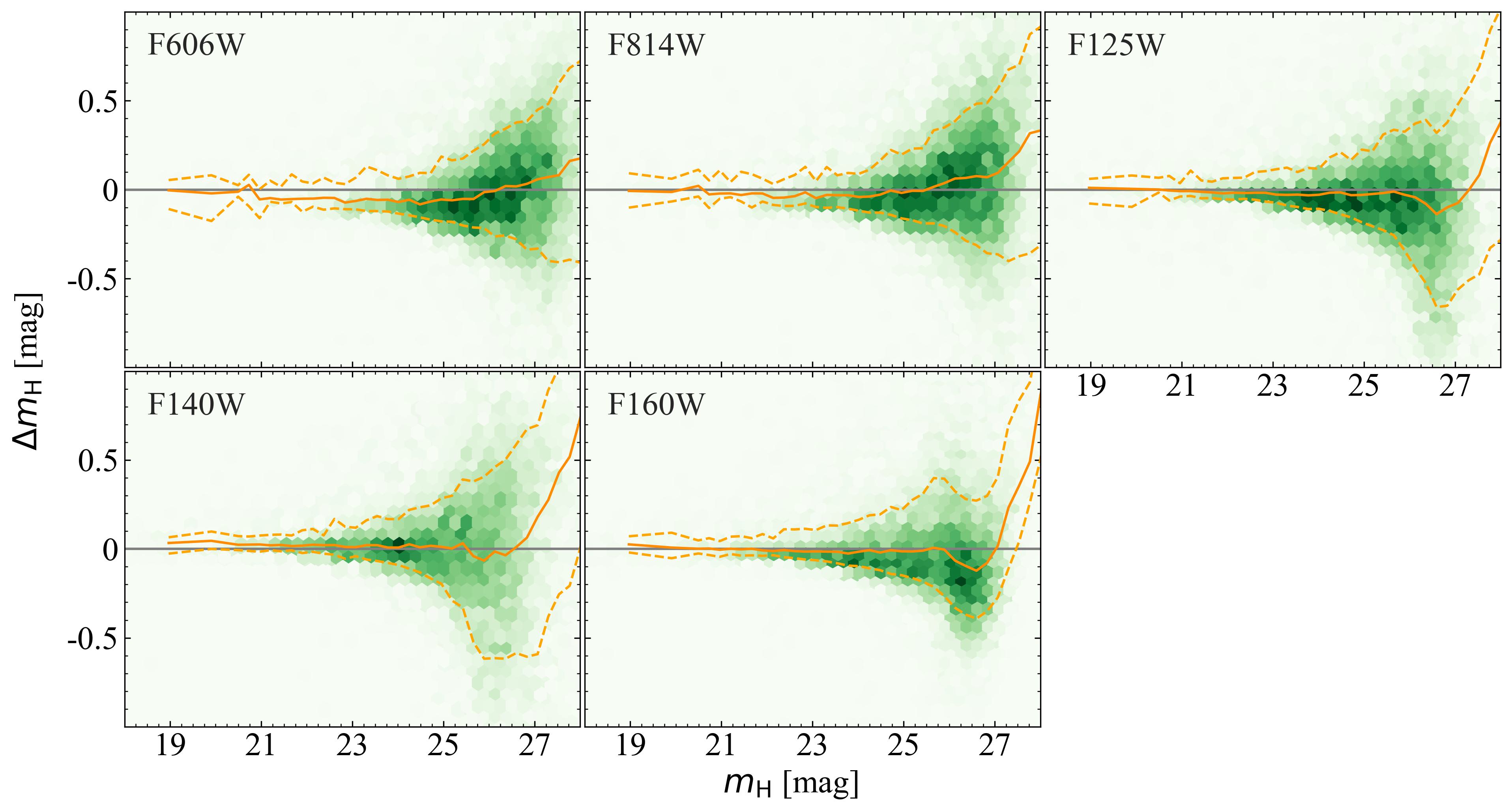}
    \caption{    
    Comparison of the photometry for common sources between this work and \citetalias{stefanon2017} in the EGS field across different bands. The figure shows the magnitude difference ($\Delta m_\mathrm{H}$, this work minus~\citetalias{stefanon2017}) as a function of the flux measured in this work ($m_\mathrm{H}$). Hexagonal bins are color-coded by source density. The yellow solid and dashed lines indicate the median and the $1\sigma$ scatter of the photometric differences within each magnitude bin, respectively. The two measurements agree well for bright sources ($\lesssim 25$ mag), whereas both the scatter and the systematic offset increase significantly for galaxies fainter than 26 mag.
    }
    \label{fig:mag_comp}
\end{figure*}

\begin{figure*}[htpb]
    \centering
    \includegraphics[width=0.95\textwidth]{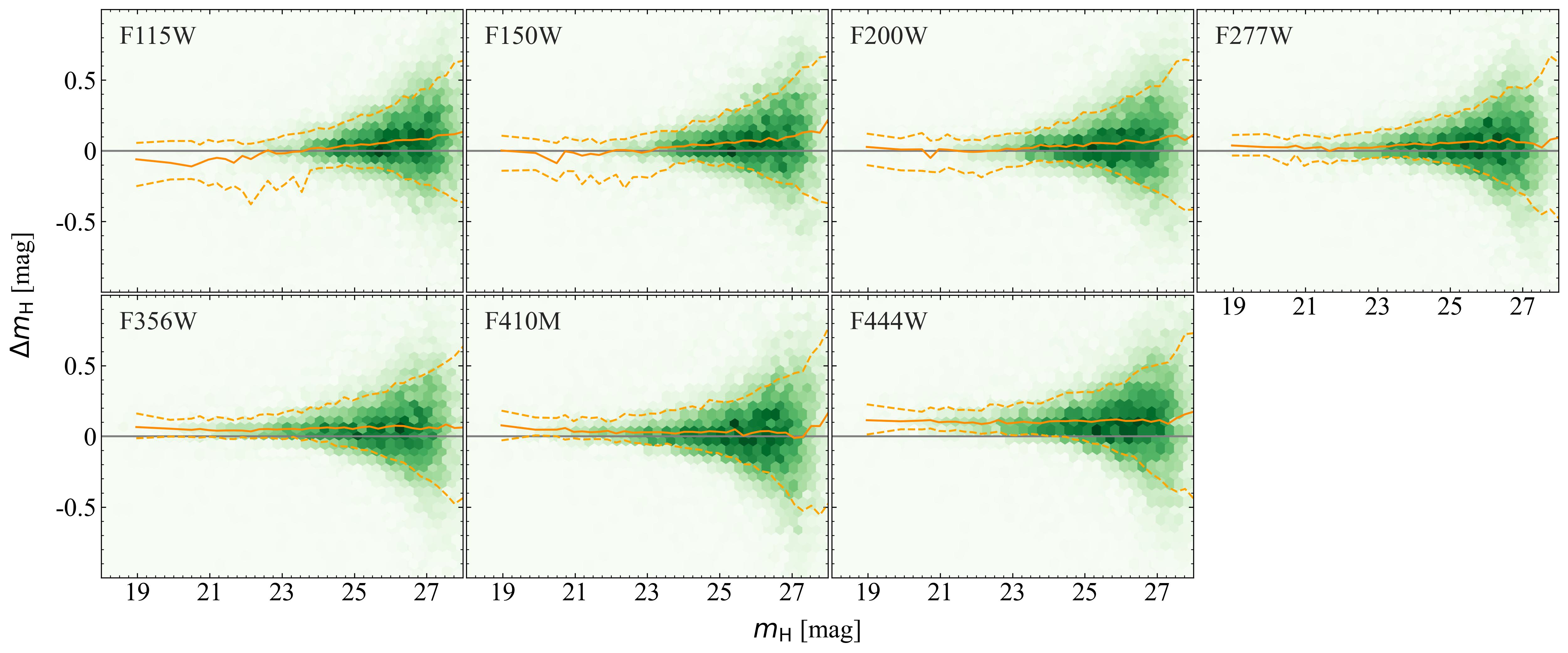}
    \caption{
    Comparison of the photometry for common sources between this work and~\citet{Merlin_2024} for the CEERS images. The magnitude difference ($\Delta m_\mathrm{H}$, this work minus~\citet{Merlin_2024}) is shown as a function of the magnitude measured in this work ($m_\mathrm{H}$). The figure follows the same style as Fig.~\ref{fig:mag_comp}. Overall, given the different aperture choices adopted in this work and by ~\citet{Merlin_2024}, the agreement is good, with only minor systematic offsets and scatter.}
    \label{fig:mag_comp_jwst}
\end{figure*}

We compare the flux values of the common sources measured in our work and~\citetalias{stefanon2017} in Fig.~\ref{fig:mag_comp}.
For the HST bands, the results are generally consistent, with most bands showing minor discrepancies and a median difference typically smaller than 0.1 mag.
On the other hand, for images with lower resolution, particularly at near-infrared wavelengths, the differences are more pronounced. 
Nevertheless, for sufficiently bright sources, the disparities remain under 0.15 mag.

As shown in Fig.~\ref{fig:mag_comp_jwst}, we compare our results with the photometric catalog presented in~\citet{Merlin_2024} for the JWST photometry in the CEERS field. 
Their photometric measurement includes the correction for the Milky Way extinction, which makes their magnitudes slightly brighter than ours. 
Given the high Galactic latitude of the EGS field, this extinction effect is expected to be minor, on the order of 0.01 mag.
In Fig.~\ref{fig:mag_comp_jwst}, for JWST bands, the overall agreement between the two sets of photometric magnitudes is good, with the median magnitude differences close to zero across most bands, except for F444W.
We attribute these significant differences to the different source segmentation methodology used in~\citet{Merlin_2024}. 

\begin{figure*}[htpb]
    \centering
    \includegraphics[width=0.9\textwidth]{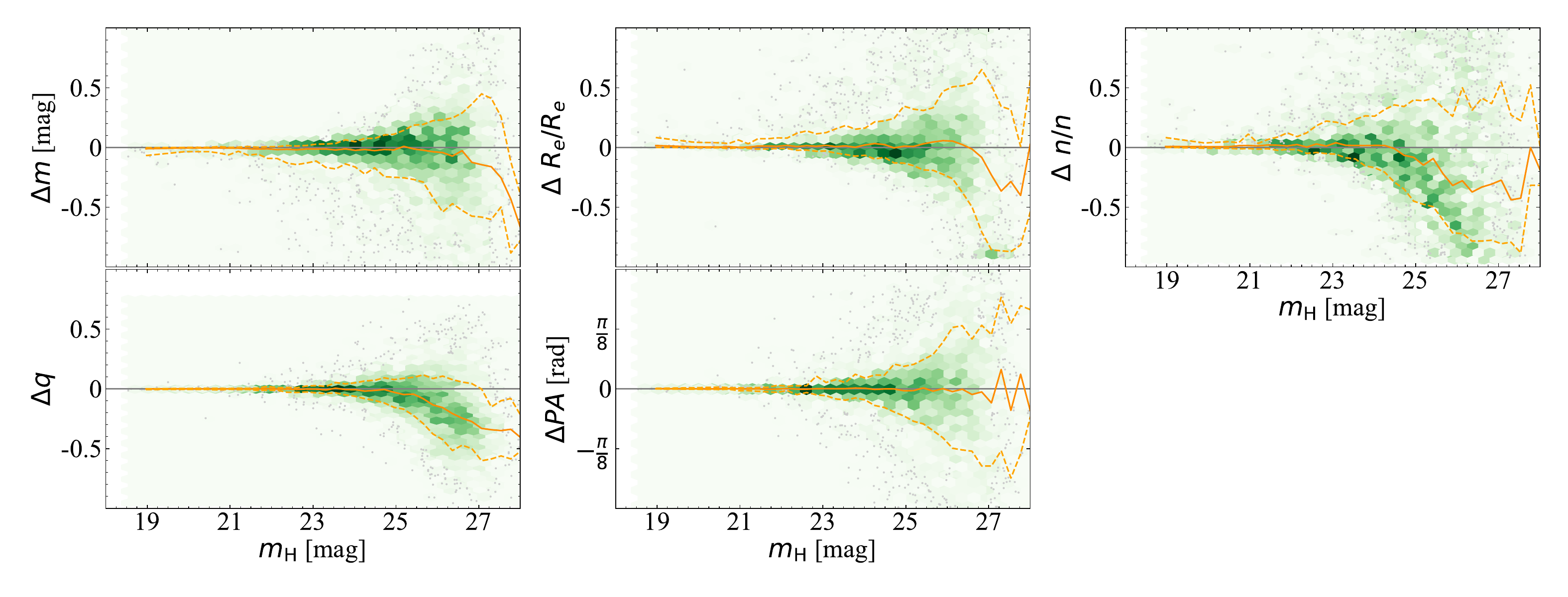}
    \caption{   
    Results of mock image tests used to quantify the uncertainties in the measurement of galaxy structural parameters with \software{GALFITM} for the EGS F160W band. The five panels show the residuals (best-fit minus input values) of the total magnitude, effective radius, \sersic index, axis ratio, and position angle as a function of the input magnitude. 
    The solid and dashed yellow lines indicate the median and the $1\sigma$ dispersion in each magnitude bin, respectively. The tests demonstrate that the structure parameters are reliably recovered for bright sources ($m_\mathrm{H}<25$), whereas both the scatter and systematic bias increase substantially for faint galaxies ($m_\mathrm{H}>26$).
    }
    \label{fig:galfitm_uncertainties}
\end{figure*}

\begin{figure}[htpb]
    \centering
    \includegraphics[width=1\linewidth]{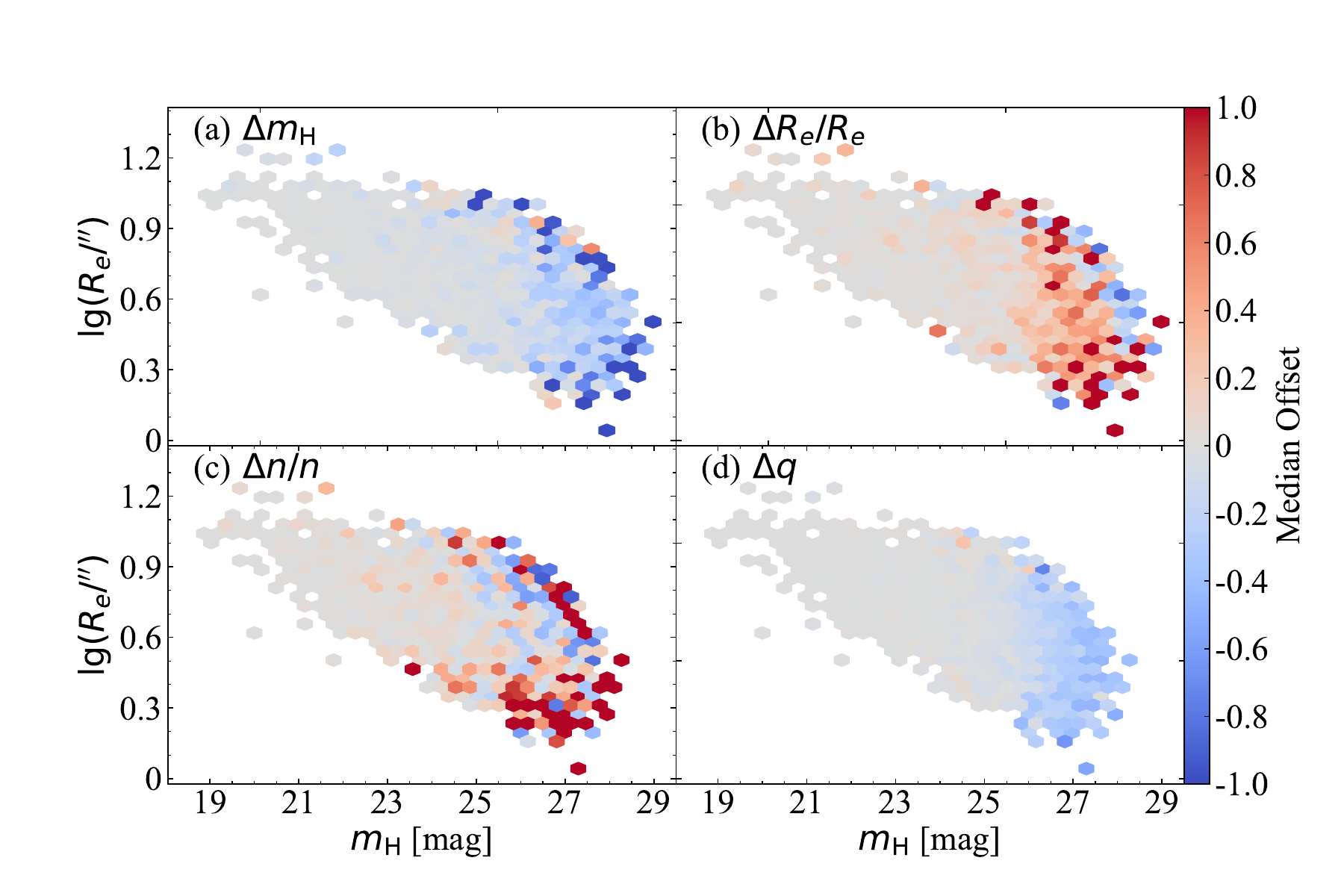}
    \caption{
    Similar to Fig.~\ref{fig:galfitm_uncertainties}, but showing the input $m_\mathrm{H}-R_\mathrm{e}$ distribution color-coded with the median bias of different parameters.
    Each panel shows the median residuals (best-fit minus input) of the $m_\mathrm{H}$, $R_\mathrm{e}$, $n$, and $q$. 
    }
    \label{fig:galfitm_uncertainties_2d}
\end{figure}

\subsection{Uncertainties and bias in the structure parameter measurement}\label{sec:morphology measurement}
Here, we use \software{GALFITM} to perform multiband fitting with the single \sersic model to derive the parametric measurements.
In this subsection, we present the results of the morphology measurements and the tests with mock images, aiming to understand the bias in the measurements of structure parameters, especially for faint galaxies.

We perform single \sersic model fitting on the HST and JWST images using \software{GALFITM}, which allows simultaneous fitting of multiband images by modeling the wavelength dependence of structural parameters with polynomial functions.
Note that \software{GALAPAGOS-2} incorporates \software{GALFITM} for structural fitting and performs automated source fitting in order of increasing magnitude.
However, for galaxies with bright companion(s), the best-fit structure parameters can still be significantly affected even with the fitting strategy implemented in \software{GALAPAGOS-2}.
We therefore visually inspect all galaxy-stamp outputs from \software{GALAPAGOS-2} and rerun \software{GALFITM} with manually adjusted initial parameters for those galaxies with bright companions whose automated fit had clearly failed. 
Of the 72,467 fitted sources, 58 (0.08\% of the fitted sample, or 0.10\% of the 59,646 sources in the final catalog) required this intervention. 
In practice these were all initialization failures, most commonly cases in which the initial effective radius given by \software{GALAPAGOS-2} was substantially larger than the value returned by the fit; for these objects we modified the initial guesses of $R_\mathrm{e}$ and $q$. 
We did not manually refit sources whose structural parameters were poorly constrained primarily because of low S/N, since the fits in that regime are limited by the data rather than by the initial conditions.
For the analysis of galaxy structure parameters, we mainly focus on high-quality images from HST (F606W, F814W, F125W, F160W) and JWST (F115W, F150W, F200W, F277W, F356W, F410M, F444W).

It is challenging to obtain reliable structure parameters for faint galaxies with very low S/N.
As mentioned in Sect.~\ref{sec:data}, we create $\sim 6000$ mock images corresponding to the HST F160W filter (with PSF convolved), combined with the realistic background extracted from EGS F160W mosaic.
We use it to quantify the uncertainties and bias in the structure parameters of the faint galaxies and to determine the limiting magnitude for accurate morphology measurement.
The final results are shown in Fig.~\ref{fig:galfitm_uncertainties}.
We define the bias as the difference between the measured and intrinsic values (measured minus intrinsic). 
To account for the distribution characteristics of the parameters, the bias is quantified separately for each parameter: $\Delta m_\mathrm{H}$, $\Delta R_\mathrm{e}/R_\mathrm{e}$, $\Delta n/n$, $\Delta q$, and the change in position angle (PA), $\Delta \mathrm{PA}$.
For all these morphology parameter distributions, the bias for galaxies brighter than 25 mag is consistent with zero, in agreement with the mock tests of~\citet{Davari_2014}.
We therefore adopt the fitting results directly in the subsequent analysis.
However, for galaxies fainter than 25 mag, systematic bias and large scatter appear, and should be treated with caution.
For the \sersic index $n$, the median bias increases from about 5\% to 70\% of galaxies between 25 mag and 27 mag.
And for the axis ratio $q$, the systematic bias ranges from $-0.03$ to $-0.27$.
For faint and small galaxies, the presence of bright companions or even positive sky-noise fluctuations can bias the fitting: the measured position angle tends to align with that of the contaminating feature, causing the axis ratio $q$ to be smaller (i.e., more elongated).

In Fig.~\ref{fig:galfitm_uncertainties_2d}, we present the bias distribution in the $m_\mathrm{H}-R_\mathrm{e}$ plane, which reflects the surface brightness of galaxies.
The biases become more severe for galaxies of lower surface brightness, with clear systematic underestimation in $m_\mathrm{H}$ and $q$ and overestimation in $R_\mathrm{e}$. 
In addition, the intrinsically small and compact galaxies would be biased to become larger in the best-fitting results.

\begin{figure}[ht]
    \centering
    \includegraphics[width=0.45\textwidth]{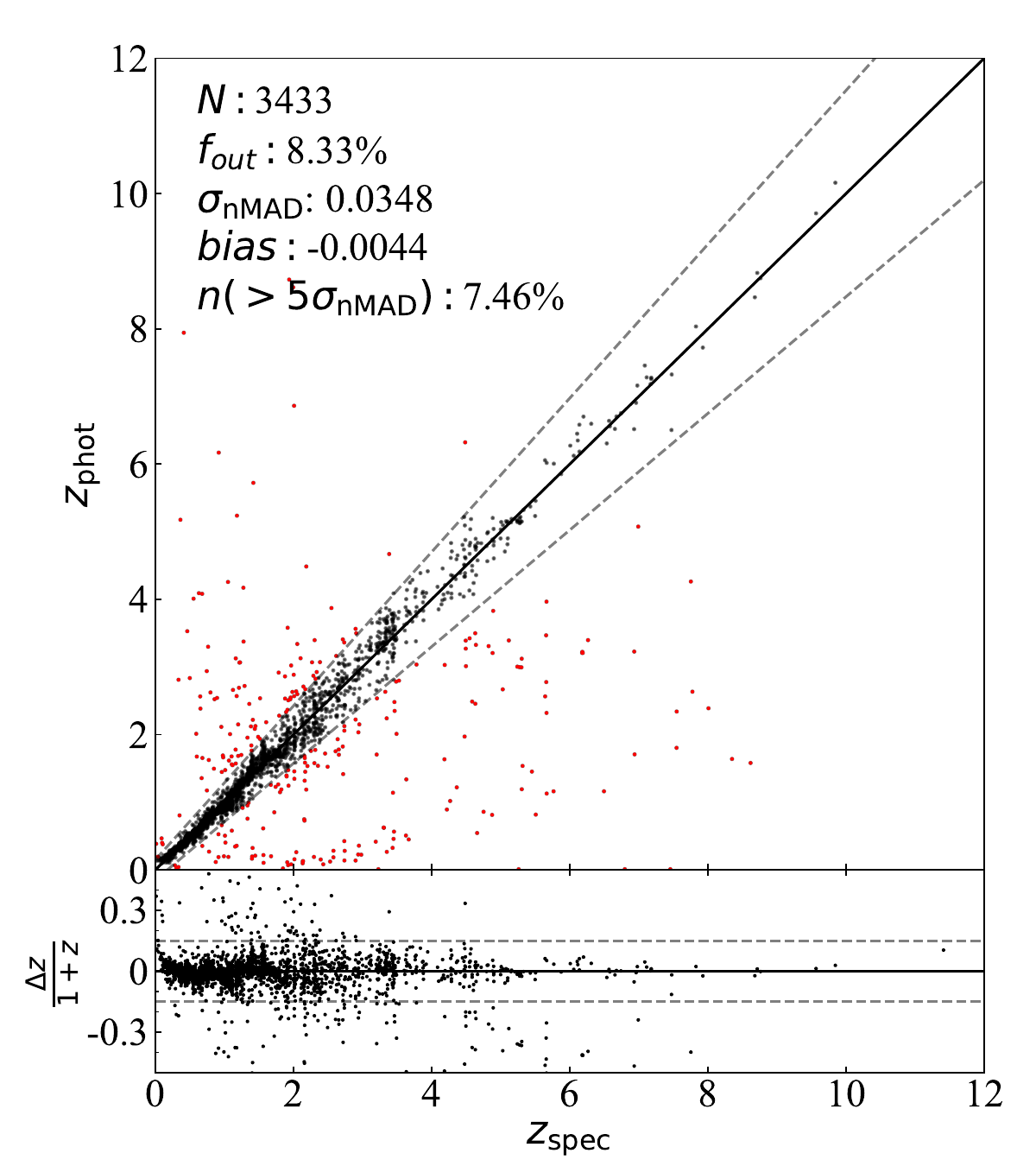}
    \caption{
    Comparison between $z_{\mathrm{phot}}$ and $z_{\mathrm{spec}}$ for 3433 galaxies, derived using 12FSPS templates together with six bluer templates. 
    Outliers are shown as red points for clarity.
    The two redshifts agree well along the one-to-one relation, with an outlier fraction $f_{\mathrm{out}}=8.33\%$, a bias of $-0.0044$, and $\sigma_\mathrm{NMAD}$ = 0.0348.}
    \label{fig:photz_new}
\end{figure}

\begin{figure}[ht]
    \centering
    \includegraphics[width=0.45\textwidth]{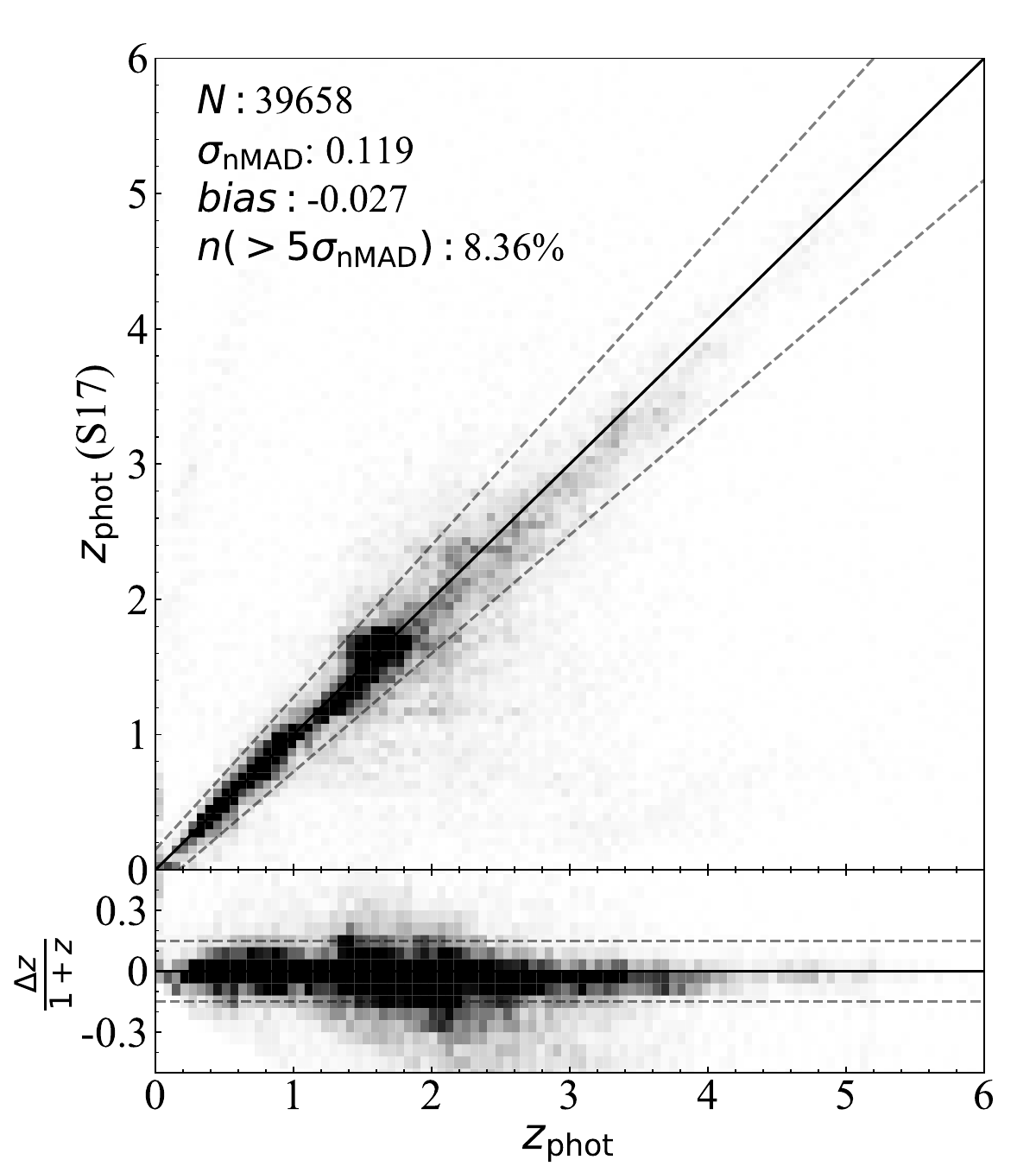}
    \caption{
    Comparison between $z_\mathrm{phot}$ estimates and those of~\citetalias{stefanon2017} for the common sources, showing good consistency.
    The fraction of catastrophic outliers at 8.36\% with $\sigma_\mathrm{NMAD}$ is 0.119.
    The median offset is $-0.027$.
    }
    \label{fig:com_our_ste}
\end{figure}

\section{SED analysis with panchromatic photometry}\label{sec:sed fitting}

Important physical parameters of galaxies, such as the photometric redshift ($z_{\mathrm{phot}}$), stellar mass ($M_\bigstar$), and star formation rate (SFR), can be derived from panchromatic SED analysis. 
In this section, we describe the estimation of photometric redshifts and the stellar masses using the panchromatic dataset presented in Sect.~\ref{sec:data}. 

Objects with CEERS coverage are fitted using 18 bands, while the remaining objects are fitted with 13 bands. 
In both cases, the wavelength coverage spans $0.4-4.5 \mu m$.
Note that in some cases, the lack of mid- and far-infrared data in our panchromatic photometry may introduce significant bias in the estimation of $M_\star$~\citep{Wang_2025}.
Cautions should be taken when interpreting and utilizing these measurements, particularly for high-redshift galaxies.
To minimize contamination by stars, we utilize Gaia data and color-color diagrams to flag point sources (e.g., stars and AGNs) in our catalog, as described in Sect.~\ref{sec:morphology measurement}.

Star formation rates derived from SED fitting are known to be highly sensitive to assumptions about star formation histories, dust attenuation, and template choices, and they can vary significantly among different fitting configurations, especially in the absence of UV to far-infrared wavelength coverage. 
We therefore exclude SFR estimates in the main results of this work, but note that stellar mass estimates from broad-band SED fitting are generally more robust.
Instead, we provide the rest-frame $U-V$ versus $V-J$ (UVJ) diagram to separate star-forming galaxies (SFGs) from quiescent galaxies (QGs), e.g.,~\citealp{williams2009QG},~\citealp{van_der_Wel_2014},~\citealp{Genin_2025}.

\subsection{Photometric redshifts}\label{subsec:photoz}

For the photometric redshift estimation, we adopt the software \software{EAZY}~\citep{Brammer2008_Eazy}, which fits SED with nonnegative linear combinations of stellar population templates to derive the redshift probability distribution functions. 
A significant fraction of galaxies in our catalog are fainter than 26 mag and are likely at high redshift. 
Recent studies have highlighted that some stellar population templates, while performing well at low redshifts, may be unsuitable for high-redshift applications~\citep{Larson_2023,Hainline2024}.
In this work, we adopt the stellar population templates from the "tweak\_fsps\_QSF\_12\_v3" collection~\citep{Conroy2009,Conroy2010} (hereafter 12FSPS) together with six bluer templates specially designed for high-redshift galaxies~\citep{Larson_2023}.

We assess the photometric redshift accuracy using three metrics.
The outlier fraction ($f_{\mathrm{out}}$)~\footnote{The fraction of sources with ${|\Delta z|}/{(1 + z_{\mathrm{spec}})} > 0.15$, where $\Delta z = z_{\mathrm{phot}} - z_{\mathrm{spec}}$.} is adopted to quantify the accuracy of the redshifts.
The systematic bias is characterized by median of $\frac{\Delta z}{1+z_{spec}}$.
Meanwhile, the scatter ($\sigma_{\mathrm{NMAD}}$)~\footnote{Defined as $\sigma_{\mathrm{NMAD}} = 1.48 \times \mathrm{median}(|\frac{\Delta z-\mathrm{median}(\Delta z)}{1 + z_{\mathrm{spec}}}|)$, following~\citet{Brammer2008_Eazy}.} is given by the normalized median absolute deviation (NMAD) to quantify the accuracy of $z_{\mathrm{phot}}$.
We compare the $z_{\mathrm{phot}}$ estimates with and without the six additional bluer templates and find that including these templates reduces $f_{\mathrm{out}}$ from 9.44\% to 8.33\%.

This approach underscores the importance of template selection for optimizing photometric redshift estimates, particularly for faint galaxies (at extreme redshifts).
For the luminosity priors, we specifically focus on the F160W band, which provides a higher S/N than the other HST bands and covers the field homogeneously.
Additionally, we test different systematic error floors~\citep{Dahlen2013} and filter combinations against the available spectroscopic redshifts. 
We find that adopting a systematic error term of 5\% of the observed flux values and excluding the ACS/F606W and F814W bands, which exhibit substantially larger systematic residuals than the other bands, provides the best overall photometric-redshift performance.
We therefore exclude the ACS/F606W and F814W measurements from the \software{EAZY} fitting.
The redshift grid extends from $z=0.01$ to $z=20$, with a step size of 0.005.
After comparing different definitions of photometric redshift provided by \software{EAZY}, we adopt $z_{\mathrm{peak}}$ value as our final $z_{\mathrm{phot}}$, which is the probability-weighted mean redshift around the highest peak of the redshift probability distribution function.

Fig.~\ref{fig:photz_new} compares our photometric redshift estimates with the spectroscopic values for 3433 galaxies fitted with the 12FSPS templates and six additional bluer templates. 
The overall accuracy metrics are $f_\mathrm{out} = 8.33\%$, $bias = -0.0044$, and $\sigma_{\mathrm{NMAD}}=0.0348$.
Dividing the sample into low-redshift ($z_\mathrm{phot}<2$) and high-redshift ($z_\mathrm{phot}\geq 2$) subsamples, we find that the outlier fraction for low-redshift galaxies is $5.61\%$, significantly lower than that of the full sample. 
For high-redshift galaxies, the outlier fraction rises to $19.21\%$.
Among the low-redshift outliers, many are blended with bright companions, which can cause substantial contamination in their photometry. 
Furthermore, uncertainties may arise from possible mismatches between the spectroscopic and photometric catalogs.

In order to assess the robustness of our measurements, we compare our photometric redshifts with those of~\citetalias{stefanon2017} for the common sources, which are based on a similarly broad wavelength coverage.
The photometric redshifts in~\citetalias{stefanon2017} are the median of 10 different measurements.
To evaluate the intrinsic accuracy of two sets of photometric redshifts, we first compare both against the available spectroscopic sample, which contains 3274 galaxies. 
In this work, we obtain an outlier fraction of 7.82\%, $\sigma_{NMAD}=0.0336$, and $bias=-0.0044$.
For the catalog of~\citetalias{stefanon2017}, the corresponding values are $f_\mathrm{out}=6.2\%$, $\sigma_{NMAD}=0.0286$ and $bias=-0.0105$.
Our results are comparable to those of~\citetalias{stefanon2017}, with a slightly smaller bias.
Fig.~\ref{fig:com_our_ste} presents the direct comparison between our photometric redshifts and those from~\citetalias{stefanon2017} for 39,658 common sources. 
The median offset is $-0.027$, with $\sigma_{NMAD}=0.119$.
The fraction of catastrophic outliers between the two catalogs is 8.36\%. 
Differences in CFHT and WIRCam photometry and in the deblending procedures may contribute to the observed offset, especially for faint or deblended sources.
When comparison is restricted to bright galaxies only (e.g., $m<25$ mag), $\sigma_{NMAD}$ decreases to 0.0548 and the bias becomes $-0.0132$, indicating that most discrepancies originate from low-S/N objects.

\subsection{Stellar mass}\label{subsec:stellar mass}

We use \software{CIGALE}~\citep{Boquien2019Cigale} to estimate stellar masses.
The redshift is fixed to the value listed in our final catalog, either to the photometric redshift or spectroscopic redshift (if available).
To avoid overinterpreting the limited panchromatic data, we adopt the simplest template settings, since our wavelength coverage cannot accurately constrain the star formation history, stellar population, or dust attenuation.
For the star formation history, we adopt the \texttt{sfh2exp} module in \software{CIGALE}, which represents a double exponential star formation history and is commonly adopted in the literature~\citep{hassani2025hidden,gandolfi2025ultrahighredshiftcloserbydustobscured,Rutkowski_2025}. 
We choose the BC03~\citep{Bruzal2003} stellar population synthesis (SSP) models with a Chabrier initial mass function (IMF)~\citep{Chabrier2003}. 
Dust attenuation is accounted for with the \texttt{dustatt\_modified\_starburst} model of~\citet{Calzetti2000}. 
As noted in~\citetalias{stefanon2017}, nebular emission can significantly bias stellar mass estimation. 
Therefore, we do not include nebular emission lines in our SED model.
Regarding the possible AGN contribution for galaxies without X-ray detection,~\citet{Ciesla2015AGN} pointed out that constraining the AGN fraction is very challenging when $f_\mathrm{AGN}< 20\%$, and that an unaccounted AGN component can introduce systematic bias in stellar mass and other parameters. 
We perform SED fitting both with and without the AGN component, consistent with the settings described above.
For galaxies with $f_\mathrm{AGN} < 40 \%$, we adopt the results without the AGN component.
The full set of parameter configurations is provided in Table~\ref{tab:cigale_params}.

\begin{figure}[!t]
    \centering
    \includegraphics[width=\columnwidth]{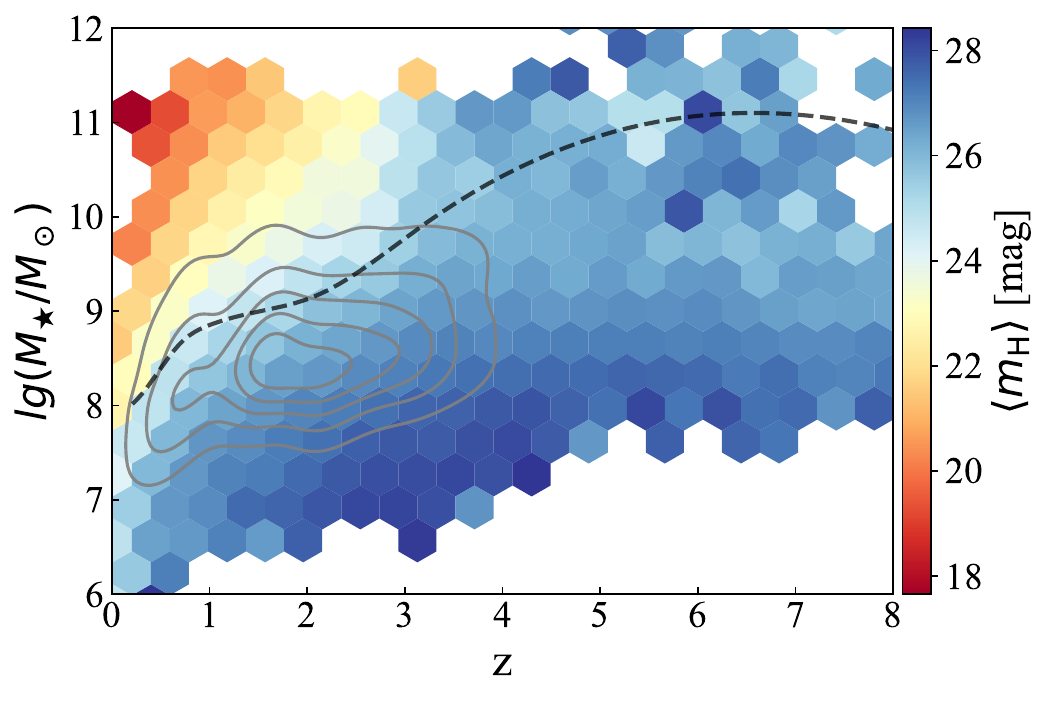}
    \caption{
    Distribution of galaxy stellar mass across different redshift ranges.
    Gray contours show the number density, and the hexagonal bins are color-coded by the mean HST F160W magnitude ($\langle m_\mathrm{H}\rangle$).
    At a given redshift, brighter galaxies tend to have higher stellar masses.
    The black dashed line indicates the 95\% stellar mass completeness.
    }
    \label{fig:z-M_star}
\end{figure}

\begin{figure*}[htpb]
    \centering
    \includegraphics[width=0.9\textwidth]{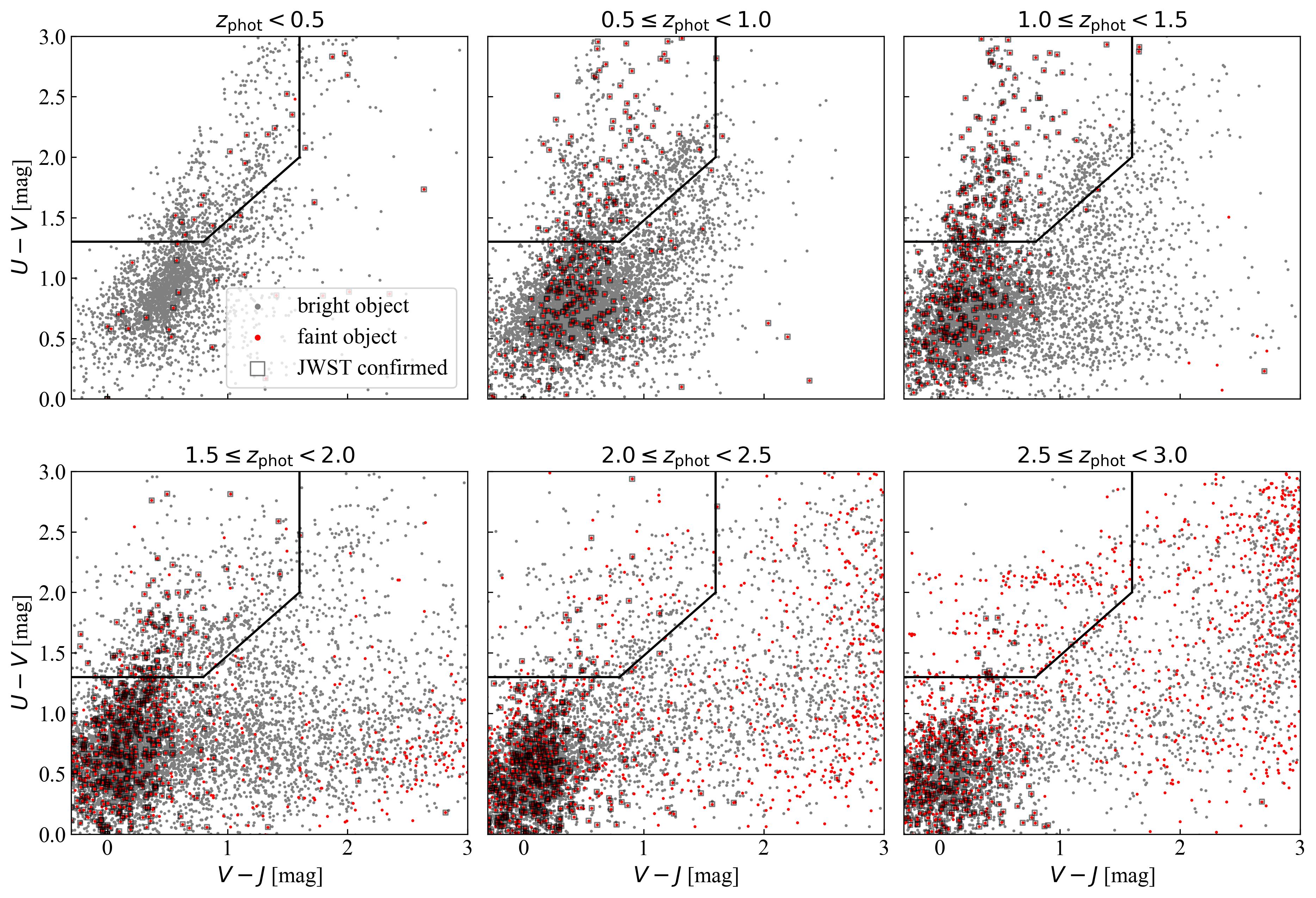}
    \caption{
    Rest-frame UVJ diagram of galaxies detected in the EGS F160W images.
    Each panel corresponds to a redshift bin ranging from $0-0.5$ to $2.5-3.0$.
    Gray points represent the bright galaxies ($m_\mathrm{H}<27$), while the red scatters indicate the faint galaxies($m_\mathrm{H}>27$). Faint galaxies with JWST coverage, which are confirmed as real galaxies, are additionally marked with black boxes.
    The empirical separation between quiescent and star-forming galaxies is shown as the black solid line.
    The faint galaxy population is concentrated in the star-forming region.
    At lower redshifts, a higher fraction of faint galaxies lies in the quiescent region.
    }
    \label{fig:uvj_map}
\end{figure*}

\begin{figure*}[!t]
    \centering
    \includegraphics[width=0.95\linewidth]{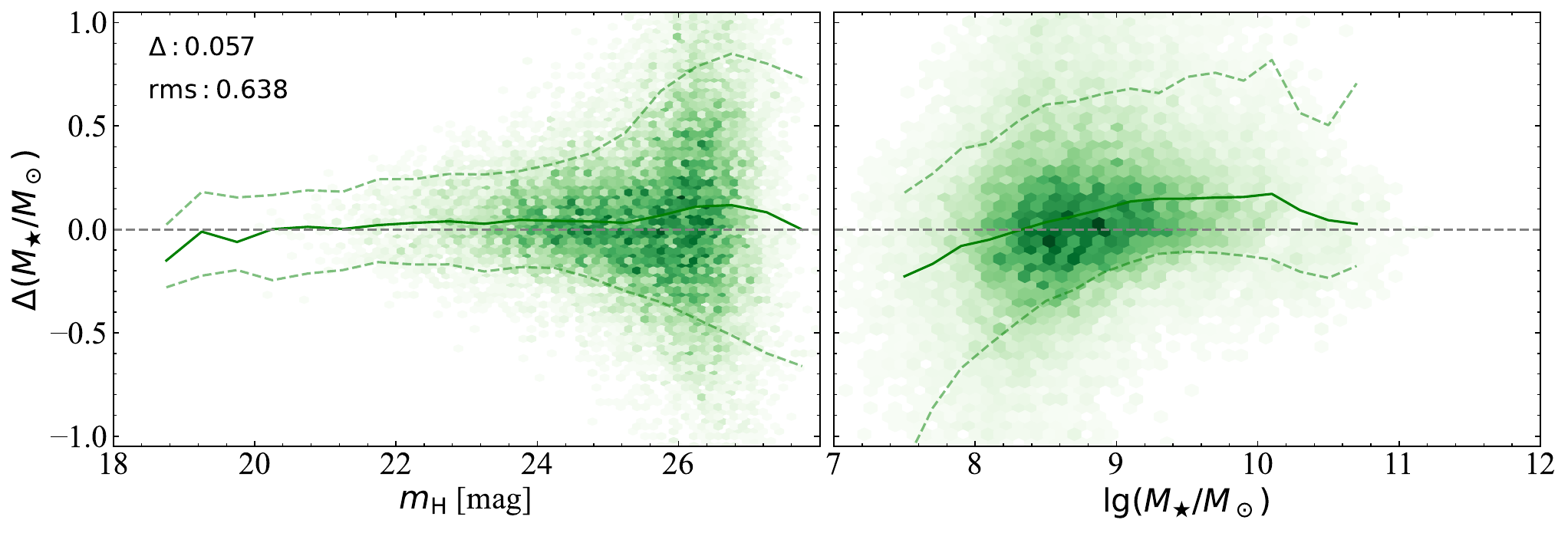}
    \caption{
    Comparison of stellar mass between this work and~\citetalias{stefanon2017}. The left panel shows the stellar mass difference as a function of F160W magnitude, while the right panel presents the difference as a function of stellar mass. The solid line presents the median offset, and the dashed lines show the 16th and 84th percentiles (68\% range). 
    The median stellar mass offset is 0.057, along with an rms of 0.638.
    }
    \label{fig:comapre_M}
\end{figure*}

The distribution of galaxy stellar mass across different redshift ranges is shown in Fig.~\ref{fig:z-M_star}. 
From the stellar mass distribution, it is clear that at high redshift only the most massive galaxies are detected, while lower-mass galaxies remain undetected.
Conversely, at a given redshift, fainter galaxies generally have lower stellar masses.
The 95\% stellar mass completeness limit of full galaxy sample is indicated by the dashed line in Fig.~\ref{fig:z-M_star}.
This completeness limit is derived following the method of~\citet{Song2025}, which estimates the completeness limit as the upper envelope of the stellar mass limit distribution.
The stellar mass limit is calculated as
\begin{equation}
\log(M_{\rm lim}) = \log(M_\star) + 0.4 \times (m_\mathrm{H} - m_{\rm lim}),
\end{equation}
where $m_{\rm lim}=25.7$ mag is the limiting magnitude corresponding to a 95\% detection completeness.
For our catalog, the completeness exceeds 95\% for galaxies with stellar masses above $10^{8.9} M_{\odot}$ at $z<1$.

Here, we point out the large uncertainty in SFR estimation and provide rest-frame UVJ colors as an empirical diagnostic for separating SFGs and QGs.
The UVJ diagram can serve as a practical alternative classification tool for users of the present catalog.
As shown in Fig.~\ref{fig:uvj_map}, the black solid line presents the separation of these two populations.
Faint galaxies with JWST coverage are highlighted by black boxes. 
Their distribution is concentrated on the left side of the UVJ diagram, which is likely a consequence of the more accurate SED measurements and rest-frame colors afforded by the deeper JWST data. 
Galaxies without JWST coverage tend to scatter more widely, and some may extend into the right region of the diagram partly due to larger photometric uncertainties. 
This suggests that the classification of objects in the right portion of the UVJ diagram should be interpreted with caution.

\begin{table}[ht]
    \centering
    \caption{The main parameters in CIGALE}
    \begin{tabular}{ lp{5cm} }
    \hline
    \hline
    \multicolumn{2}{c}{\texttt{sfh2exp}} \\
    $\tau$ (main) [Gyr] & 0.1, 1, 5, 12, 18 \\
    Age (main) [Gyr] & 0.5, 1, 5, 12 \\
    $f_{\mathrm{burst}}$&0, 0.001, 0.01, 0.05, 0.1, 0.15\\
    burst age [Myr]&10, 100, 200\\
    \hline
    \multicolumn{2}{c}{\texttt{bc03}}\\
    IMF&Chabrier\\
    Metallicity & 0.0004, 0.004, 0.02\\
    \hline
    \multicolumn{2}{c}{\texttt{dustatt\_modified\_starburst}}\\
    $\mathrm{E_{BV}}$ lines[mag] & 0.3\\
    $\mathrm{E_{BV}}$ factor & 0.44\\
    $\mathrm{R_{V}}$ & 3.1\\
    \hline
    \multicolumn{2}{c}{\texttt{redshifting}}\\
    Redshift & fixed\\
    \hline
    \end{tabular}
    \label{tab:cigale_params}
\end{table}

Systematic variations in stellar mass estimates can arise from several processes, including photometric measurements, photometric redshift estimation, SED model assumptions, and the choice of software fitting algorithm.
Before comparing with previous works, we first verified that different SED fitting codes, such as \software{LePhare}~\citep{Arnouts1999Lephare,Ilbert2006Lephare}, \software{Bagpipes}~\citep{Carnall_2018,Carnall_2019}, and \software{FAST++}~\citep{Kriek2009FAST}, yield consistent stellar masses when provided with the same input photometry, redshifts, and model assumptions. 
We then performed controlled comparisons with the catalog of~\citetalias{stefanon2017} to quantify the relative impact of photometry, photometric redshift, and model assumptions. 
To disentangle the effects of different inputs, we recomputed the stellar masses in three steps: 
(1) using the literature photometry and redshifts, but substituting only the SED modeling with our assumptions; 
(2) further replacing the literature redshifts with our photometric redshifts; 
and (3) using our full set of photometry and redshifts.
This progressive approach allows us to isolate the impact of each factor in the final results.

Fig.~\ref{fig:comapre_M} compares our final stellar masses with those from~\citetalias{stefanon2017}. 
For galaxies brighter than 26 mag, the median offset is less than 0.1 dex, and the scatter is below 0.2 dex, indicating good agreement for high-S/N sources. 
For fainter galaxies, by contrast, the scatter increases to $\sim0.4$ dex. 
This behavior is expected, as uncertainties in flux measurements propagate nonlinearly into stellar mass estimates via SED fitting, especially when the photometry is noise-dominated or affected by blending.
We find that differences in photometry are the dominant contributor to the stellar mass offsets, while variations in SED modeling and photometric redshift estimates play a secondary role.
In addition, differences in aperture definition, PSF correction, and the treatment of blended sources between the two catalogs can introduce color-dependent biases that further affect stellar mass estimates.
Overall, our analysis shows that discrepancies in stellar mass are primarily driven by differences in photometric measurements, especially for faint and low-mass galaxies. 
This result highlights the importance of accurate and homogeneous photometry for robust stellar mass estimates, whereas variations in the fitting software or SED modeling assumptions introduce comparatively smaller systematic effects.

\section{Discussion}\label{sec:discussion}

Based on the source detection in the F160W image of the EGS field, supplemented by photometric information from the CEERS field, we discuss the detection and verification of the faint objects in our catalog, as well as the statistical properties of their structure parameters. 

\begin{figure}[!t]
    \centering
    \includegraphics[width=0.48\textwidth]{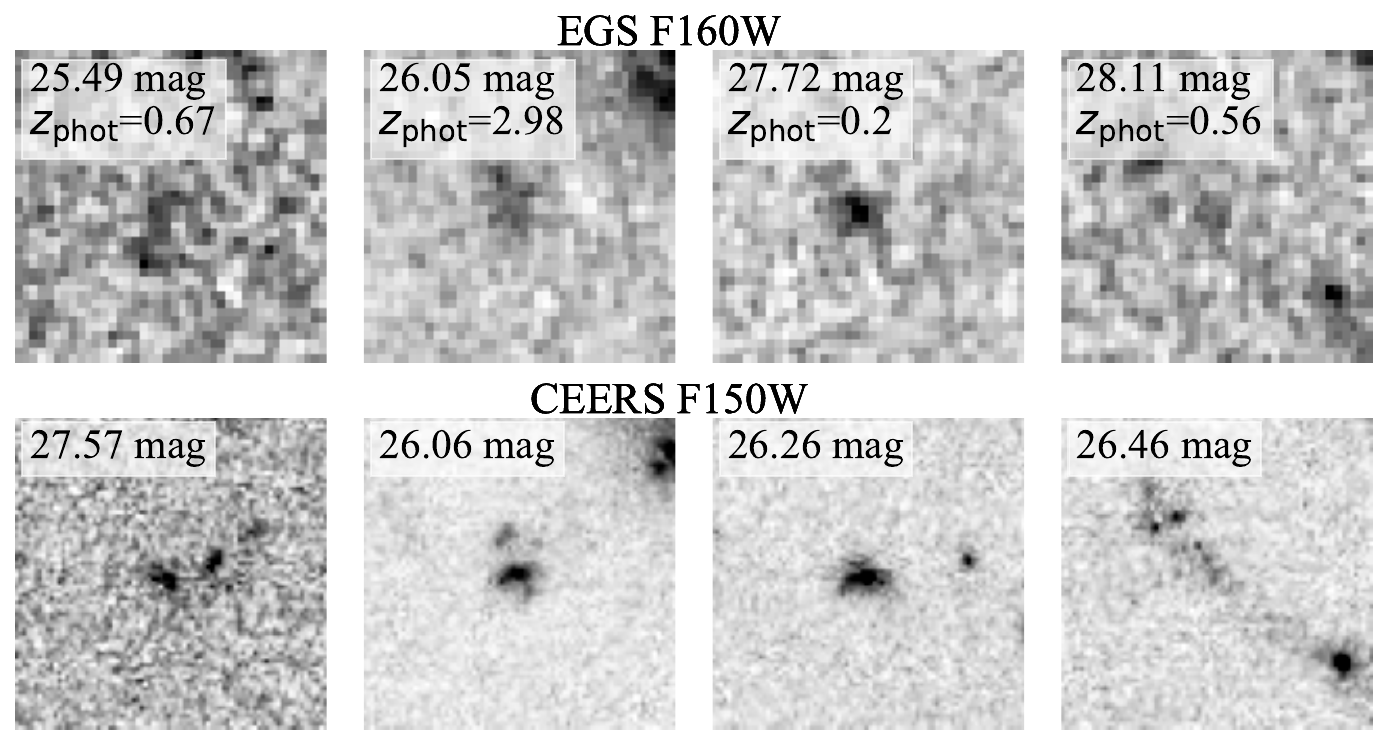}
    \caption{
    Examples of faint objects in EGS F160W image (top row) and CEERS F150W images (bottom row), with redshift and magnitude information in each panel. Each column represents the same objects.}
    \label{fig:cross_check}
\end{figure}

\begin{figure}[!t]
    \centering
    \includegraphics[width=0.45\textwidth]{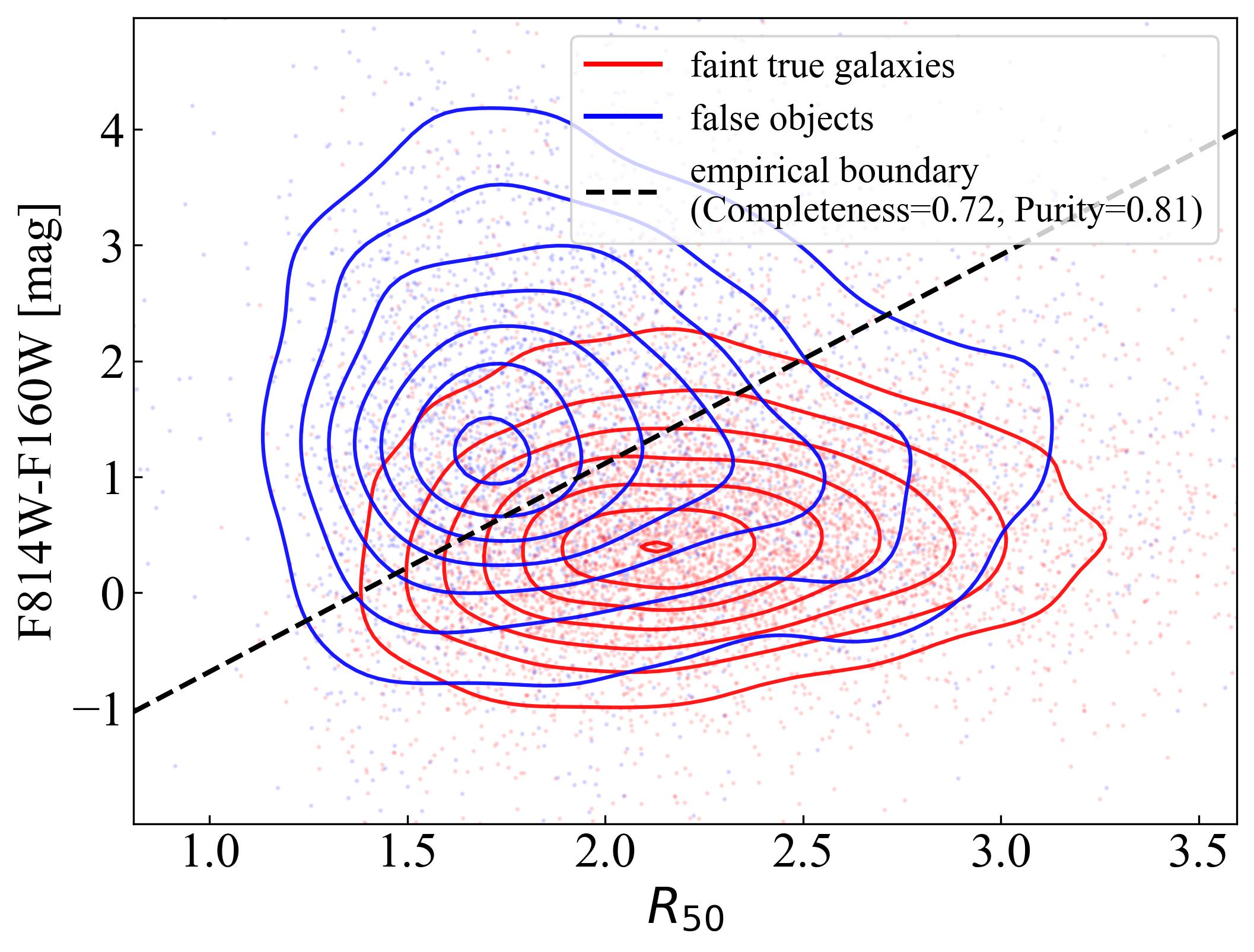}
    \caption{
    Distribution of faint sources ($m_\mathrm{H}>27$) in the space of half-light radius (FLUX\_RADIUS from \software{SExtractor}) and $F814W - F160W$ color. Red points and contours represent true galaxies, whereas blue points and contours correspond to false detections. The black dashed line marks the empirical separation criterion, which achieves 72\% completeness and 81\% purity after removing the sources above this line.
    }
    \label{fig:true_galaxy_selection_contour}
\end{figure}

\begin{figure}[ht]
    \centering
    \includegraphics[width=0.42\textwidth]{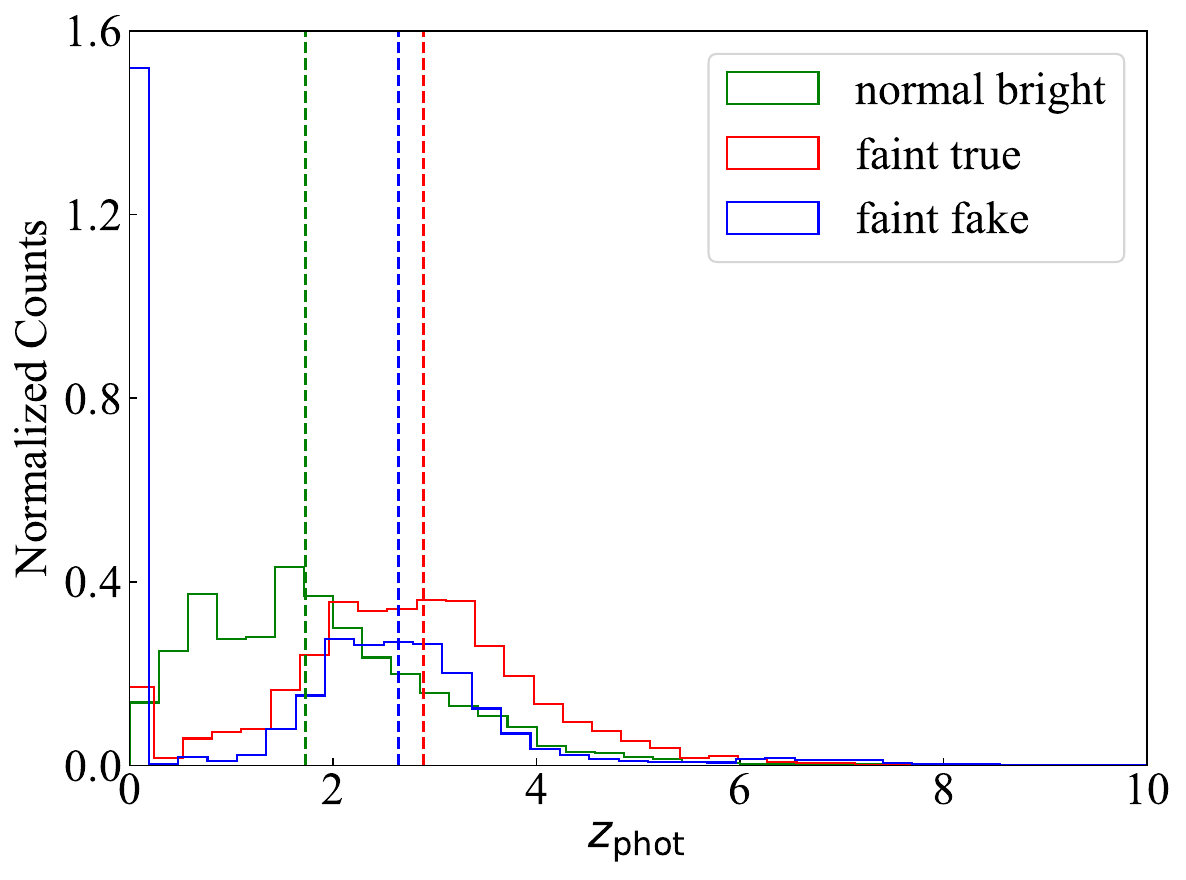}
    \caption{
    Distribution of $z_{\mathrm{phot}}$ for normal bright galaxies (green), faint true galaxies (red), and false detections (blue). Vertical dashed lines mark the median $z_{\mathrm{phot}}$ values of the three populations: 1.73, 2.89, and 2.65, respectively. Notably, approximately 31\% of false detections have $z_{\mathrm{phot}} \lesssim 0.01$.
    }
    \label{fig:z_phot_hist}
\end{figure}

\subsection{Detection and verification of the HST faint objects}\label{subsec:jwst cross match}

We cross-match the EGS F160W catalog with the CEERS catalogs in the F150W and F200W bands, using a minimum separation of $0.3''$. 
As depicted in Fig.~\ref{fig:cross_check}, many faint objects in the EGS F160W image are resolved as extended objects in the CEERS image, suggesting that they are real galaxies.
Statistically, within the EGS field covered by CEERS, approximately $60\%$ of the objects in our catalog at $m_\mathrm{H}\sim 27$ mag are also confirmed as real sources in the CEERS mosaic, while the remaining $\sim 40\%$ are attributed to background noise.
For objects fainter than 28 mag, the fraction of false detections rises to 72\%, consistent with the estimation in Sect.~\ref{subsec:FDR}.

To establish an empirical criterion to separate the real sources from false detections, we also perform detailed comparisons of the structural parameters of the two populations.
For detections fainter than 27 mag, we find that the size$-$color space can effectively distinguish real galaxies from spurious ones, as shown in Fig.~\ref{fig:true_galaxy_selection_contour}. 
We define the half-light radius using FLUX\_RADIUS from \software{SExtractor} and denote it as $R_\mathrm{50}$.
The distributions of faint true galaxies and false detections show a clear offset in this parameter space.
We also examine other structure parameters, including the \sersic index, axis ratio, and additional nonparametric metrics (such as concentration, asymmetry, and clumpiness). 
However, no clear distinction between real and false detections could be made.
To retain a large number of faint galaxies in our catalog with high purity, we adopt the following empirical relation in size$-$color space to statistically remove the spurious objects, 
\begin{equation}
    F814W-F160W >1.8\times (R_{50} - 1.38),
\end{equation}
which can achieve 72\% completeness and 81\% purity.

We further compare the redshift distributions of the three populations, i.e., normal bright galaxies, faint true galaxies, and faint false detections, as shown in Fig.~\ref{fig:z_phot_hist}. 
The faint true galaxies exhibit systematically higher $z_{\mathrm{phot}}$ values than normal bright galaxies. 
Approximately 31\% of the false detections have $z_{\mathrm{phot}} \lesssim 0.01$, indicating that $z_{\mathrm{phot}}$ can also serve as an additional criterion for separating faint true galaxies from false sources.
Combining these two criteria yields a selected sample with 72\% completeness and 84\% purity.
In total, 12,821 sources are excluded from the final catalog by these criteria.
Because the empirical criterion is calibrated in the CEERS overlap region, we compare the F160W imaging and faint-source properties inside and outside this region. 
The similar image depths, PSF FWHMs, and distributions of $m_\mathrm{H}$ and $R_{50}$ indicate that the CEERS overlap is representative of the EGS field after regions near mosaic edges and bright stars are excluded. 
The overall calibration framework can be transferred to other observations, but the validation region should have an imaging depth, PSF, and pixel scale comparable to those of the target data; otherwise, the empirical criterion should be recalibrated.

\subsection{Blending effects in source detection and structure parameter measurement}\label{subsec:completeness blending}

\begin{figure}
    \centering
    \includegraphics[width=0.42\textwidth]{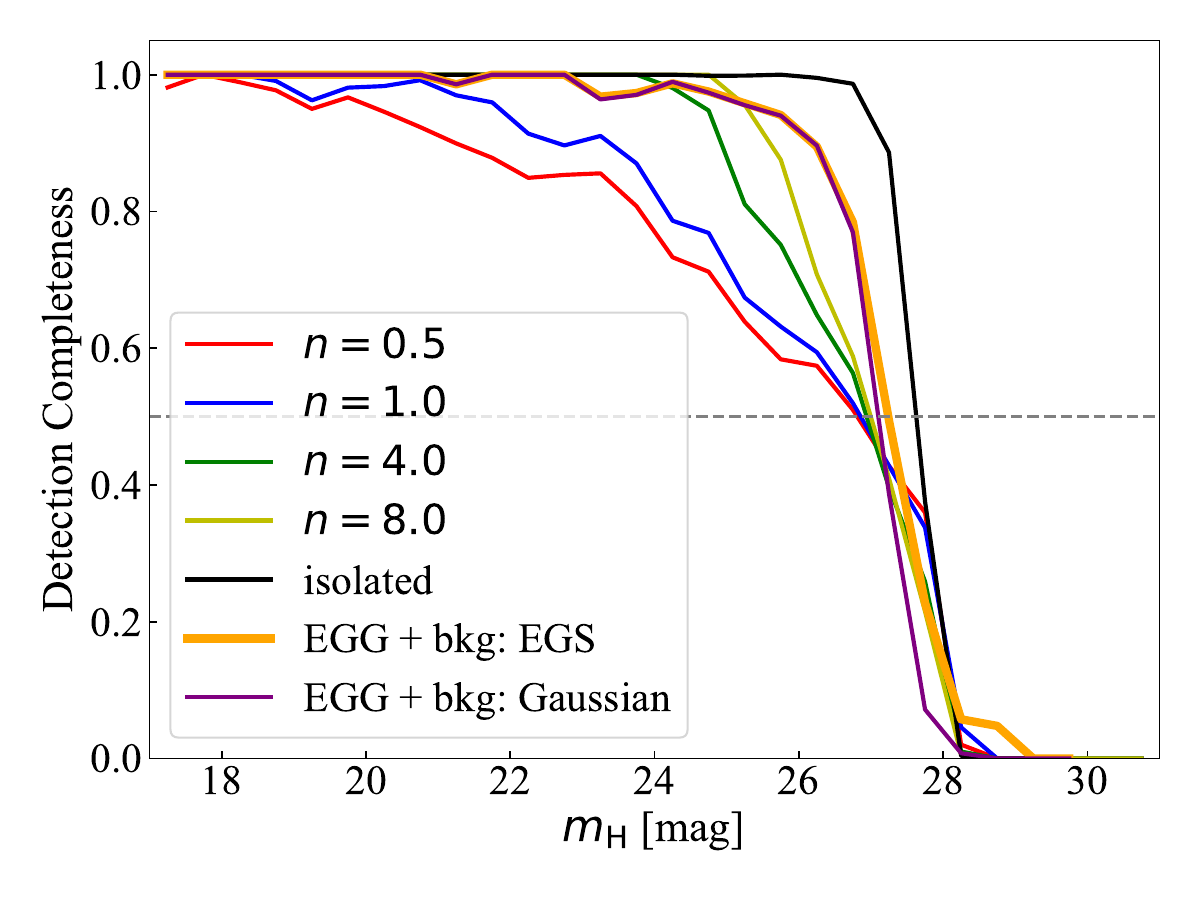}
    \caption{
    Comparison of the detection completeness distributions for different mock tests: galaxies with fixed \sersic index n = 0.5, 1.0, 4.0, 8.0, the isolated mock galaxies, and mock galaxies with EGS background and Gaussian background.}
    \label{fig:1d_completeness_diff_n}
\end{figure}

In real observations, source blending is unavoidable owing to the finite angular resolution, line-of-sight projection, and extended wings of the PSF. 
These effects cause neighboring galaxies to overlap on the image, reducing detection completeness and biasing source identification, particularly at faint magnitudes.
To better understand the impact of blending within the real galaxy population, we compare the detection completeness measured from mock images of isolated galaxies with that obtained from the full mock catalog described in Sect.~\ref{subsec:detection completeness}.
The two samples share the same magnitude and structural parameter distributions.
The detection completeness curves for the different tests are shown as color-coded solid lines in Fig.~\ref{fig:1d_completeness_diff_n}.
At all magnitudes, the detection completeness is systematically higher for isolated galaxies. 
The difference becomes particularly significant for galaxies fainter than 26 mag.
Even for galaxies brighter than 26 mag, the completeness is clearly reduced when companions are present. 
This demonstrates that blending affects not only the faintest sources but also relatively bright galaxies when the projected separations are small (typically for $d\lesssim10\ \mathrm{pixel}$ in F160W image, $\sim 0.6~\mathrm{arcsec}$).
Unlike photometric or structural biases that can be calibrated statistically, blending-induced incompleteness cannot be corrected in a straightforward way at the detection stage.
The probability of missed detection depends simultaneously on projected separation, relative brightness, intrinsic galaxy size, PSF shape, and local background fluctuations. 
These combined effects prevent a simple analytic correction. 
Instead, realistic simulations that incorporate the observed distribution of galaxy sizes and separations are required to quantify completeness losses.
Overall, our results indicate that blending is a nonnegligible contributor to detection incompleteness in deep HST imaging and must be taken into account in statistical studies of faint galaxies.

\begin{figure*}
    \centering
    \includegraphics[width=0.95\textwidth]{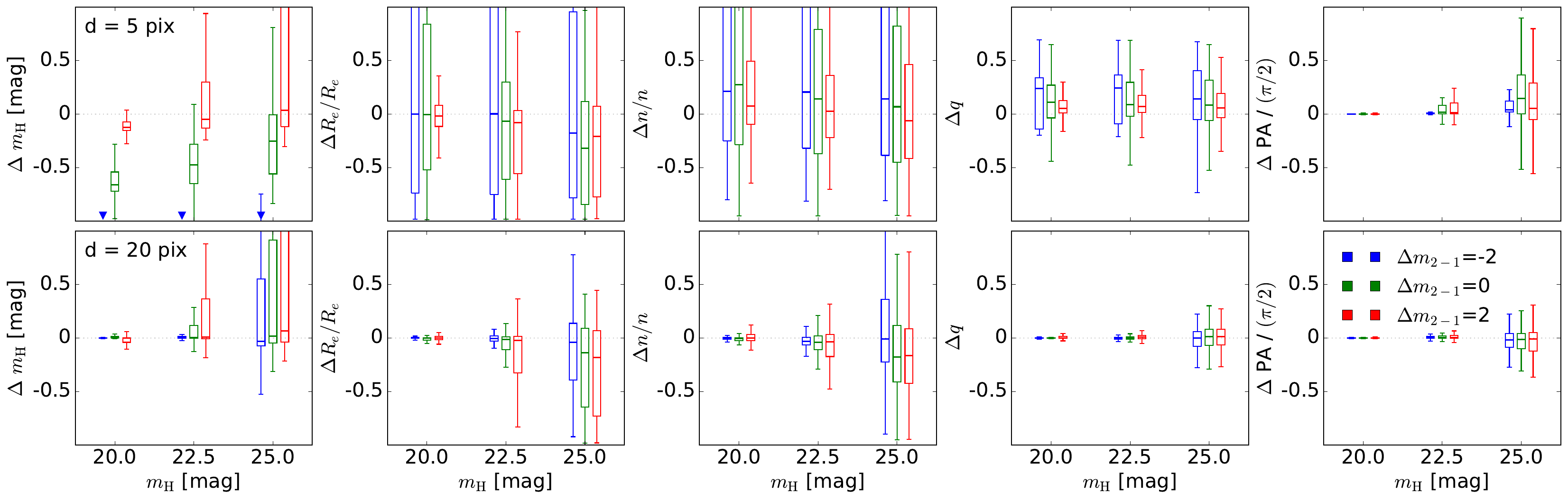}
    \caption{
    Fitting biases as functions of magnitude difference ($\Delta m_{2-1}$) and primary galaxy brightness ($m_1$) for two separations ($d$): $5$ pixels (upper panels) and $20$ pixels (bottom panels).
    Boxes show the interquartile range and are color-coded by magnitude difference $\Delta m_{2-1}$, with blue for brighter companion ($\Delta m_{2-1}=-2$), green for equally bright companion ($\Delta m_{2-1}=0$) and red for fainter one ($\Delta m_{2-1}=2$).
    The horizontal dashed line marks zero bias. 
    For clarity, the x-axis positions are constructed by mapping the discrete input $m_1$ values to equally spaced integers and adding a uniform offset to avoid overlapping.
    Triangles represent median value lower than $-1$ mag.
    }
    \label{fig:structure_bias_deblending}
\end{figure*}

To quantify the impact of blended companions on the structural parameter measurements of galaxies, we generate controlled mock images of galaxy pairs. 
We construct a sample of 10,368 galaxy pairs, each consisting of a primary (target) galaxy and a companion. 
The primary galaxies have magnitudes $m_1=20.0,22.5,25.0$, \sersic indices $n=1.0,2.0,4.0$, effective radius $R_\mathrm{e,1}=1,5,20,50$ pixels (pixel scale is $0.06~\mathrm{arcsec/pixel}$), and axis ratios $q_1=0.3,0.8$. 
We consider magnitude differences $\Delta m_{2-1} = -2,0,2$ and projected separations between the centers of two mock galaxies of $d=5,10,20,50$ pixels. 
The structure parameters of companion galaxies ($n_2, R_\mathrm{e,2}$) are drawn from the same sets of discrete values as those of the primary, while the position angles are held fixed to isolate the effects of size, brightness, and separation.
The axis ratio of companion is fixed to $q_2=0.6$. 
Similar to Sect.~\ref{subsec:detection completeness}, all mock images are convolved with an empirical PSF derived from HST WFC3/F160W observations, and sky background and Poisson noise are added to match the typical depth of the EGS mosaic. 

The mock images are also processed through the full \software{GALAPAGOS-2} pipeline, in the same way as the EGS mosaics. 
We compare the best-fit parameters of the primary galaxies with their input values to quantify the measurement biases.
Fig.~\ref{fig:structure_bias_deblending} shows the fitting biases as functions of separation, companion brightness, and primary galaxy magnitude. 
The upper and bottom panels represent separations of $5$ and $20$ pixels, respectively. 
Blue, green, and red colors correspond to magnitude differences $\Delta m_{2-1}$ of $-2$, $0$, and $2$. 
At the smallest separation in our tests ($d=5$ pixels$=0.3^{\prime\prime}$), deblending frequently fails and the pair is fitted as a single object, especially when the magnitude difference is large.
This leads to systematic brightening of the primary galaxies, with a median $\Delta m_\mathrm{H}$ more negative than $-1$ for $\Delta m_{2-1} = -2$ (blue), and to large scatter in all structural parameters. 
At $d=10$ pixels ($0.6^{\prime\prime}$), most systems are correctly deblended, but a bright companion ($\Delta m_{2-1} = -2$) still induces a noticeable bias ($\Delta mag \approx -0.5$ mag) with large scatter in $R_\mathrm{e}$ and $n$. 
For the larger separations ($d\gtrsim 20$ pixels), the median biases for primary galaxies brighter than $25$ mag are consistent with zero, and the scatter is small ($\lesssim 0.1$), indicating that blending effects become negligible in this regime.
Galaxies of different magnitudes exhibit varying degrees of measurement bias, even at fixed separation and magnitude difference.

This complexity indicates that the observed biases cannot be attributed to a single dominant parameter, but instead result from the interplay of multiple galaxy properties.
We employ a random forest regression model (scikit-learn implementation) to quantify the relative importance of galaxy properties in driving the measured biases.
The model uses as input the photometric and structural parameters of both galaxies together with their projected separation, and predicts the biases in magnitude and structural parameters. 
The results indicate that for magnitude bias ($\Delta mag$), the most important features are the magnitude difference ($\Delta m_{2-1}$) and separation ($d$). 
For structural parameters such as the effective radius ($R_{\rm e}$) and \sersic index ($n$), the biases are primarily driven by the companion effective radius ($R_{\rm e,2}$) and the primary galaxy magnitude ($m_1$). 
Furthermore, when the normalized separation $d/(R_{\rm e,1}+R_{\rm e,2})$ is introduced as a feature, its predictive power becomes comparable to that of the individual structural parameters. 
This confirms that the degree of profile overlap, rather than separation alone, determines the onset of significant bias.

Overall, our tests demonstrate that single \sersic fitting remains reliable for primary galaxies brighter than $25$ mag, provided that the projected separation exceeds 10 pixels ($\sim 0.6\mathrm{arcsec}$).
At smaller separations, however, systematic biases in flux and structural parameters become significant, particularly when a bright or extended companion is present.

\section{Summary and conclusion}\label{sec:conclusion}

We present a multiwavelength source detection and analysis framework applied to HST EGS and JWST CEERS imaging data. 
We identify 72,467 sources in the CANDELS/EGS field, of which 12,821 are removed by our empirical criteria.
Completeness tests based on mock images show a detection completeness exceeding 95\% for sources brighter than 25 mag, which drops below 50\% at fainter magnitudes. 
Detection completeness depends not only on total magnitude but also on surface brightness; we therefore provide completeness maps in the $m_\mathrm{H}-R_\mathrm{e}$ plane. 
Through cross-matching detections between HST F160W and JWST F150W and F200W bands, we quantify the false detection rate, finding that over 50\% of objects fainter than 27 mag in F160W are true galaxies. 
False detection rates derived from negative sky tests are consistent with cross-matching results.
To further assess the impact of blending and deblending on source detection, we perform dedicated mock tests in which galaxies are inserted both in isolation and in realistic projected environments. 
By comparing these two scenarios, we quantify the completeness loss induced by blending. 
The detection completeness of isolated galaxies is systematically higher than that for galaxies in blended environments, particularly at faint magnitudes, demonstrating that blending is a dominant factor preventing the recovery of low-surface-brightness systems that would otherwise be detectable.
Even for galaxies brighter than 26 mag, incompleteness remains nonnegligible in crowded regions. 
These tests show that blending effects cannot be fully corrected at the detection stage, because they depend simultaneously on galaxy separation, projection effect, and the PSF. 
Our deblending experiments provide an empirical estimate of the practical detection limit under realistic observing conditions.

Morphological analysis is conducted using single \sersic modeling with \software{GALFITM}. 
We quantify biases and uncertainty in faint galaxy fits and establish reliable magnitude limits for structure parameters. 
Galaxies with $25<m_\mathrm{H} <27$ mag exhibit significant systematic biases: $\Delta n$ ranges from 5\% to 70\%, and $\Delta q$ ranges from -0.03 to -0.27.
For extremely faint sources ($m_\mathrm{H} > 27$ mag), the fitting structural parameters are largely unreliable.
We also evaluate the impact of deblending on the structure measurements. 
When a companion lies close to the target, imperfect deblending biases the fitted magnitude and structure parameters, especially if the companion is bright or extended. 
The bias becomes negligible at larger separations, defining the practical limits within which single-component fits remain reliable.
Faint true galaxies exhibit smaller axis ratios than false detections and follow the empirical relation derived from bright galaxies. 

We perform forced-aperture photometry across the EGS and CEERS fields and validate our measurements against existing catalogs from~\citetalias{stefanon2017} and~\citet{Merlin_2024}, confirming good agreement. 
Using panchromatic data, we derive photometric redshifts ($z_\mathrm{phot}$) and stellar masses ($M_\bigstar$). 
Comparison with spectroscopic redshifts from DEEP2+3 and RUBIES and grism spectroscopic redshifts yields an outlier fraction of 8.33\%, a systematic bias of $-0.0044$, and a dispersion of $\sigma_\mathrm{NMAD} = 0.0348$ for 3433 objects.
For galaxies in the CEERS regions, our $z_\mathrm{phot}$ estimates yield an outlier fraction of 5.57\%, a systematic bias of $0.00$, and a dispersion of $\sigma_\mathrm{NMAD} = 0.0232$.
Our stellar mass estimates for faint galaxies are systematically higher than those of~\citet{stefanon2017}, likely owing to systematic differences in the underlying photometry.

The methodologies developed and validated in this work, particularly the quantification of completeness as a function of both magnitude and size, the characterization of structural parameter biases at the faint end, and the empirical separation of real sources from false detections using multiwavelength cross-validation, offer valuable lessons for upcoming large-area surveys. 
For Euclid, upcoming facilities CSST and Roman, our results underscore the necessity of incorporating realistic galaxy simulations (including blending and clustering effects) into survey design and data processing pipelines. 
Relying solely on point-source detection limits is insufficient for galaxy evolution studies.  
We recommend adopting validation frameworks that are both simulation-driven and paired with real-data validation, in order to mitigate systematic biases in future surveys.

\section*{Data availability}
The catalog (Table~\ref{tab:source_catalog}) is available in electronic form at the CDS via anonymous ftp to \url{cdsarc.u-strasbg.fr} (130.79.128.5) or via \url{https://cdsweb.u-strasbg.fr/cgi-bin/qcat?J/A+A/}.

\begin{acknowledgements}
We thank the anonymous referee for constructive comments that improved the manuscript.
We thank Jinyi Shangguan, Pinsong Zhao, Ruancun Li, Wen Sun, Changhao Chen, Xinyang Li, Yixiao Lyu, and Jiazhao Li for valuable discussions and constructive suggestions, which have helped improve the quality of this work.
This work is supported by the National Natural Science Foundation of China under grant Nos. 12233001, 12533004, by  the National Key R\&D Program of China under grant No. 2024YFA1611602, by a Shanghai Natural Science Research Grant (24ZR1491200), by the ``111'' project of the Ministry of Education under grant No. B20019, and by the China Manned Space Program with grant Nos. CMS-CSST-2025-A08, CMS-CSST-2025-A09 and CMS-CSST-2025-A11. This work made use of the Gravity Supercomputer at the Department of Astronomy, Shanghai Jiao Tong University.
This project is supported in part by Office of Science and Technology, Shanghai Municipal Government (grant Nos. 24DX1400100, ZJ2023-ZD-001).

\end{acknowledgements}

\bibliographystyle{aa}
\bibliography{aa60867-26}

\begin{thebibliography}{86}
\expandafter\ifx\csname natexlab\endcsname\relax\def\natexlab#1{#1}\fi

\bibitem[{{Arnouts} {et~al.}(1999){Arnouts}, {Cristiani}, {Moscardini}, {Matarrese}, {Lucchin}, {Fontana}, \& {Giallongo}}]{Arnouts1999Lephare}
{Arnouts}, S., {Cristiani}, S., {Moscardini}, L., {et~al.} 1999, \mnras, 310, 540

\bibitem[{{Ashby} {et~al.}(2015){Ashby}, {Willner}, {Fazio}, {Dunlop}, {Egami}, {Faber}, {Ferguson}, {Grogin}, {Hora}, {Huang}, {Koekemoer}, {Labb{\'e}}, \& {Wang}}]{Ashby2015spitzer}
{Ashby}, M.~L.~N., {Willner}, S.~P., {Fazio}, G.~G., {et~al.} 2015, \apjs, 218, 33

\bibitem[{{Ashby} {et~al.}(2013){Ashby}, {Willner}, {Fazio}, {Huang}, {Arendt}, {Barmby}, {Barro}, {Bell}, {Bouwens}, {Cattaneo}, {Croton}, {Dav{\'e}}, {Dunlop}, {Egami}, {Faber}, {Finlator}, {Grogin}, {Guhathakurta}, {Hernquist}, {Hora}, {Illingworth}, {Kashlinsky}, {Koekemoer}, {Koo}, {Labb{\'e}}, {Li}, {Lin}, {Moseley}, {Nandra}, {Newman}, {Noeske}, {Ouchi}, {Peth}, {Rigopoulou}, {Robertson}, {Sarajedini}, {Simard}, {Smith}, {Wang}, {Wechsler}, {Weiner}, {Wilson}, {Wuyts}, {Yamada}, \& {Yan}}]{Ashby2013}
{Ashby}, M.~L.~N., {Willner}, S.~P., {Fazio}, G.~G., {et~al.} 2013, \apj, 769, 80

\bibitem[{{Bagley} {et~al.}(2023){Bagley}, {Finkelstein}, {Koekemoer}, {Ferguson}, {Arrabal Haro}, {Dickinson}, {Kartaltepe}, {Papovich}, {P{\'e}rez-Gonz{\'a}lez}, {Pirzkal}, {Somerville}, {Willmer}, {Yang}, {Yung}, {Fontana}, {Grazian}, {Grogin}, {Hirschmann}, {Kewley}, {Kirkpatrick}, {Kocevski}, {Lotz}, {Medrano}, {Morales}, {Pentericci}, {Ravindranath}, {Trump}, {Wilkins}, {Calabr{\`o}}, {Cooper}, {Costantin}, {de la Vega}, {Hilbert}, {Hutchison}, {Larson}, {Lucas}, {McGrath}, {Ryan}, {Wang}, \& {Wuyts}}]{Bagley2023CEERS}
{Bagley}, M.~B., {Finkelstein}, S.~L., {Koekemoer}, A.~M., {et~al.} 2023, \apjl, 946, L12

\bibitem[{{Barden} {et~al.}(2012){Barden}, {H{\"a}u{\ss}ler}, {Peng}, {McIntosh}, \& {Guo}}]{Barden2012galapagos}
{Barden}, M., {H{\"a}u{\ss}ler}, B., {Peng}, C.~Y., {McIntosh}, D.~H., \& {Guo}, Y. 2012, \mnras, 422, 449

\bibitem[{{Bari{\v{s}}i{\'c}} {et~al.}(2025){Bari{\v{s}}i{\'c}}, {Jones}, {Mortensen}, {Nanayakkara}, {Chen}, {Sanders}, {Bullock}, {Bundy}, {Faucher-Gigu{\`e}re}, {Glazebrook}, {Henry}, {Ju}, {Malkan}, {Morishita}, {Obreschkow}, {Roy}, {Espejo Salcedo}, {Shapley}, {Treu}, {Wang}, \& {Westfall}}]{Barisi2025}
{Bari{\v{s}}i{\'c}}, I., {Jones}, T., {Mortensen}, K., {et~al.} 2025, \apj, 983, 139

\bibitem[{{Bertin} \& {Arnouts}(1996)}]{Bertin1996sextractor}
{Bertin}, E. \& {Arnouts}, S. 1996, \aaps, 117, 393

\bibitem[{{Bielby} {et~al.}(2012){Bielby}, {Hudelot}, {McCracken}, {Ilbert}, {Daddi}, {Le F{\`e}vre}, {Gonzalez-Perez}, {Kneib}, {Marmo}, {Mellier}, {Salvato}, {Sanders}, \& {Willott}}]{Bielby2012wircam}
{Bielby}, R., {Hudelot}, P., {McCracken}, H.~J., {et~al.} 2012, \aap, 545, A23

\bibitem[{{Boquien} {et~al.}(2019){Boquien}, {Burgarella}, {Roehlly}, {Buat}, {Ciesla}, {Corre}, {Inoue}, \& {Salas}}]{Boquien2019Cigale}
{Boquien}, M., {Burgarella}, D., {Roehlly}, Y., {et~al.} 2019, \aap, 622, A103

\bibitem[{{Boucaud} {et~al.}(2016){Boucaud}, {Bocchio}, {Abergel}, {Orieux}, {Dole}, \& {Hadj-Youcef}}]{Boucaud_2016_pypher}
{Boucaud}, A., {Bocchio}, M., {Abergel}, A., {et~al.} 2016, \aap, 596, A63

\bibitem[{{Brammer} {et~al.}(2008){Brammer}, {van Dokkum}, \& {Coppi}}]{Brammer2008_Eazy}
{Brammer}, G.~B., {van Dokkum}, P.~G., \& {Coppi}, P. 2008, \apj, 686, 1503

\bibitem[{{Brammer} {et~al.}(2012){Brammer}, {van Dokkum}, {Franx}, {Fumagalli}, {Patel}, {Rix}, {Skelton}, {Kriek}, {Nelson}, {Schmidt}, {Bezanson}, {da Cunha}, {Erb}, {Fan}, {F{\"o}rster Schreiber}, {Illingworth}, {Labb{\'e}}, {Leja}, {Lundgren}, {Magee}, {Marchesini}, {McCarthy}, {Momcheva}, {Muzzin}, {Quadri}, {Steidel}, {Tal}, {Wake}, {Whitaker}, \& {Williams}}]{Brammer2012grism}
{Brammer}, G.~B., {van Dokkum}, P.~G., {Franx}, M., {et~al.} 2012, \apjs, 200, 13

\bibitem[{{Bruzual} \& {Charlot}(2003)}]{Bruzal2003}
{Bruzual}, G. \& {Charlot}, S. 2003, \mnras, 344, 1000

\bibitem[{{Calzetti} {et~al.}(2000){Calzetti}, {Armus}, {Bohlin}, {Kinney}, {Koornneef}, \& {Storchi-Bergmann}}]{Calzetti2000}
{Calzetti}, D., {Armus}, L., {Bohlin}, R.~C., {et~al.} 2000, \apj, 533, 682

\bibitem[{{Carnall} {et~al.}(2019){Carnall}, {McLure}, {Dunlop}, {Cullen}, {McLeod}, {Wild}, {Johnson}, {Appleby}, {Dav{\'e}}, {Amorin}, {Bolzonella}, {Castellano}, {Cimatti}, {Cucciati}, {Gargiulo}, {Garilli}, {Marchi}, {Pentericci}, {Pozzetti}, {Schreiber}, {Talia}, \& {Zamorani}}]{Carnall_2019}
{Carnall}, A.~C., {McLure}, R.~J., {Dunlop}, J.~S., {et~al.} 2019, \mnras, 490, 417

\bibitem[{{Carnall} {et~al.}(2018){Carnall}, {McLure}, {Dunlop}, \& {Dav{\'e}}}]{Carnall_2018}
{Carnall}, A.~C., {McLure}, R.~J., {Dunlop}, J.~S., \& {Dav{\'e}}, R. 2018, \mnras, 480, 4379

\bibitem[{{Castellano} {et~al.}(2022){Castellano}, {Fontana}, {Treu}, {Santini}, {Merlin}, {Leethochawalit}, {Trenti}, {Vanzella}, {Mestric}, {Bonchi}, {Belfiori}, {Nonino}, {Paris}, {Polenta}, {Roberts-Borsani}, {Boyett}, {Brada{\v{c}}}, {Calabr{\`o}}, {Glazebrook}, {Grillo}, {Mascia}, {Mason}, {Mercurio}, {Morishita}, {Nanayakkara}, {Pentericci}, {Rosati}, {Vulcani}, {Wang}, \& {Yang}}]{Castellano2022}
{Castellano}, M., {Fontana}, A., {Treu}, T., {et~al.} 2022, \apjl, 938, L15

\bibitem[{{Chabrier}(2003)}]{Chabrier2003}
{Chabrier}, G. 2003, \apjl, 586, L133

\bibitem[{{Chemerynska} {et~al.}(2026){Chemerynska}, {Atek}, {Furtak}, {Chisholm}, {Endsley}, {Kokorev}, {Rosdahl}, {Blaizot}, {Adamo}, {Bouwens}, {Fujimoto}, {Korber}, {Mason}, {McQuinn}, {Mu{\~n}oz}, {Natarajan}, {Nelson}, {Oesch}, {Pan}, {Richard}, {Saldana-Lopez}, {Schaerer}, {Volonteri}, {Zitrin}, {Berg}, {Claeyssens}, {Dessauges-Zavadsky}, {Jecmen}, {Labb{\'e}}, {Naidu}, \& {Trebitsch}}]{Chemrynska2026}
{Chemerynska}, I., {Atek}, H., {Furtak}, L.~J., {et~al.} 2026, \mnras, 546, staf2267

\bibitem[{{Ciesla} {et~al.}(2015){Ciesla}, {Charmandaris}, {Georgakakis}, {Bernhard}, {Mitchell}, {Buat}, {Elbaz}, {LeFloc'h}, {Lacey}, {Magdis}, \& {Xilouris}}]{Ciesla2015AGN}
{Ciesla}, L., {Charmandaris}, V., {Georgakakis}, A., {et~al.} 2015, \aap, 576, A10

\bibitem[{{Coil} {et~al.}(2004){Coil}, {Davis}, {Madgwick}, {Newman}, {Conselice}, {Cooper}, {Ellis}, {Faber}, {Finkbeiner}, {Guhathakurta}, {Kaiser}, {Koo}, {Phillips}, {Steidel}, {Weiner}, {Willmer}, \& {Yan}}]{Coil2004}
{Coil}, A.~L., {Davis}, M., {Madgwick}, D.~S., {et~al.} 2004, \apj, 609, 525

\bibitem[{{Conroy} \& {Gunn}(2010)}]{Conroy2010}
{Conroy}, C. \& {Gunn}, J.~E. 2010, \apj, 712, 833

\bibitem[{{Conroy} {et~al.}(2009){Conroy}, {Gunn}, \& {White}}]{Conroy2009}
{Conroy}, C., {Gunn}, J.~E., \& {White}, M. 2009, \apj, 699, 486

\bibitem[{{Cooper} {et~al.}(2011){Cooper}, {Aird}, {Coil}, {Davis}, {Faber}, {Juneau}, {Lotz}, {Nandra}, {Newman}, {Willmer}, \& {Yan}}]{Cooper2011}
{Cooper}, M.~C., {Aird}, J.~A., {Coil}, A.~L., {et~al.} 2011, \apjs, 193, 14

\bibitem[{{Cooper} {et~al.}(2012){Cooper}, {Griffith}, {Newman}, {Coil}, {Davis}, {Dutton}, {Faber}, {Guhathakurta}, {Koo}, {Lotz}, {Weiner}, {Willmer}, \& {Yan}}]{Cooper2012}
{Cooper}, M.~C., {Griffith}, R.~L., {Newman}, J.~A., {et~al.} 2012, \mnras, 419, 3018

\bibitem[{{Cooper} {et~al.}(2006){Cooper}, {Newman}, {Croton}, {Weiner}, {Willmer}, {Gerke}, {Madgwick}, {Faber}, {Davis}, {Coil}, {Finkbeiner}, {Guhathakurta}, \& {Koo}}]{Cooper2006}
{Cooper}, M.~C., {Newman}, J.~A., {Croton}, D.~J., {et~al.} 2006, \mnras, 370, 198

\bibitem[{{Crespo G{\'o}mez} {et~al.}(2024){Crespo G{\'o}mez}, {Colina}, {{\'A}lvarez-M{\'a}rquez}, {Bik}, {Boogaard}, {{\"O}stlin}, {Pei{\ss}ker}, {Walter}, {Labiano}, {P{\'e}rez-Gonz{\'a}lez}, {Greve}, {Wright}, {Alonso-Herrero}, {Caputi}, {Costantin}, {Eckart}, {Garc{\'\i}a-Mar{\'\i}n}, {Gillman}, {Hjorth}, {Iani}, {Langeroodi}, {Pye}, {Rinaldi}, {Tikkanen}, {van der Werf}, {Lagage}, \& {van Dishoeck}}]{Crespo2024}
{Crespo G{\'o}mez}, A., {Colina}, L., {{\'A}lvarez-M{\'a}rquez}, J., {et~al.} 2024, \aap, 691, A325

\bibitem[{{CSST Collaboration} {et~al.}(2026){CSST Collaboration}, {Gong}, {Miao}, {Zhan}, {Li}, {Shangguan}, {Li}, {Liu}, {Chen}, {Yuan}, {Zhou}, {Liu}, {Yu}, {Ji}, {Qi}, {Liu}, {Dai}, {Wang}, {Zheng}, {Hao}, {Dou}, {Ao}, {Lin}, {Zhang}, {Wang}, {Sun}, {Li}, {Li}, {Xu}, {Li}, {Li}, {Wu}, {Zhang}, {Wang}, {Bai}, {Cai}, {Cai}, {Cao}, {Chan}, {Chang}, {Chen}, {Chen}, {Chen}, {Chen}, {Cui}, {Dong}, {Du}, {Duan}, {Fan}, {Fan}, {Fan}, {Fan}, {Fang}, {Fu}, {Fu}, {Fu}, {Gao}, {Gu}, {Gu}, {Guo}, {Han}, {Hu}, {Huang}, {Ho}, {Jiang}, {Jiang}, {Jing}, {Kang}, {Kong}, {Li}, {Li}, {Li}, {Li}, {Li}, {Li}, {Liao}, {Lin}, {Liu}, {Liu}, {Liu}, {Liu}, {Mao}, {Mao}, {Meng}, {Pang}, {Peng}, {Peng}, {Shan}, {Shen}, {Shen}, {Shen}, {Shi}, {Shi}, {Tan}, {Tian}, {Wang}, {Wang}, {Wang}, {Wang}, {Wu}, {Wu}, {Wu}, {Xu}, {Xue}, {Xue}, {Yang}, {Yang}, {Yao}, {Yuan}, {Yuan}, {Zhang}, {Zhang}, {Zhang}, {Zhang}, {Zhang}, {Zhao}, {Zhao}, {Zhong}, {Zhong}, {Zhou}, {Zhu}, \& {Zu}}]{csstcollaboration2025introductionchinesespacestation}
{CSST Collaboration}, {Gong}, Y., {Miao}, H., {et~al.} 2026, Science China Physics, Mechanics, and Astronomy, 69, 239501

\bibitem[{{Dahlen} {et~al.}(2013){Dahlen}, {Mobasher}, {Faber}, {Ferguson}, {Barro}, {Finkelstein}, {Finlator}, {Fontana}, {Gruetzbauch}, {Johnson}, {Pforr}, {Salvato}, {Wiklind}, {Wuyts}, {Acquaviva}, {Dickinson}, {Guo}, {Huang}, {Huang}, {Newman}, {Bell}, {Conselice}, {Galametz}, {Gawiser}, {Giavalisco}, {Grogin}, {Hathi}, {Kocevski}, {Koekemoer}, {Koo}, {Lee}, {McGrath}, {Papovich}, {Peth}, {Ryan}, {Somerville}, {Weiner}, \& {Wilson}}]{Dahlen2013}
{Dahlen}, T., {Mobasher}, B., {Faber}, S.~M., {et~al.} 2013, \apj, 775, 93

\bibitem[{{Davari} {et~al.}(2014){Davari}, {Ho}, {Peng}, \& {Huang}}]{Davari_2014}
{Davari}, R., {Ho}, L.~C., {Peng}, C.~Y., \& {Huang}, S. 2014, \apj, 787, 69

\bibitem[{{Davis} {et~al.}(2007){Davis}, {Guhathakurta}, {Konidaris}, {Newman}, {Ashby}, {Biggs}, {Barmby}, {Bundy}, {Chapman}, {Coil}, {Conselice}, {Cooper}, {Croton}, {Eisenhardt}, {Ellis}, {Faber}, {Fang}, {Fazio}, {Georgakakis}, {Gerke}, {Goss}, {Gwyn}, {Harker}, {Hopkins}, {Huang}, {Ivison}, {Kassin}, {Kirby}, {Koekemoer}, {Koo}, {Laird}, {Le Floc'h}, {Lin}, {Lotz}, {Marshall}, {Martin}, {Metevier}, {Moustakas}, {Nandra}, {Noeske}, {Papovich}, {Phillips}, {Rich}, {Rieke}, {Rigopoulou}, {Salim}, {Schiminovich}, {Simard}, {Smail}, {Small}, {Weiner}, {Willmer}, {Willner}, {Wilson}, {Wright}, \& {Yan}}]{Davis2007}
{Davis}, M., {Guhathakurta}, P., {Konidaris}, N.~P., {et~al.} 2007, \apjl, 660, L1

\bibitem[{{de Graaff} {et~al.}(2025){de Graaff}, {Brammer}, {Weibel}, {Lewis}, {Maseda}, {Oesch}, {Bezanson}, {Boogaard}, {Cleri}, {Cooper}, {Gottumukkala}, {Greene}, {Hirschmann}, {Hviding}, {Katz}, {Labb{\'e}}, {Leja}, {Matthee}, {McConachie}, {Miller}, {Naidu}, {Price}, {Rix}, {Setton}, {Suess}, {Wang}, {Whitaker}, \& {Williams}}]{degraaff2024rubies}
{de Graaff}, A., {Brammer}, G., {Weibel}, A., {et~al.} 2025, \aap, 697, A189

\bibitem[{{Endsley} {et~al.}(2023){Endsley}, {Stark}, {Whitler}, {Topping}, {Chen}, {Plat}, {Chisholm}, \& {Charlot}}]{Endsley2023}
{Endsley}, R., {Stark}, D.~P., {Whitler}, L., {et~al.} 2023, \mnras, 524, 2312

\bibitem[{{Euclid Collaboration} {et~al.}(2026){Euclid Collaboration}, {Aussel}, {Tereno}, {Schirmer}, {Alguero}, {Altieri}, {Balbinot}, {de Boer}, {Casenove}, {Corcho-Caballero}, {Furusawa}, {Furusawa}, {Hudson}, {Jahnke}, {Libet}, {Macias-Perez}, {Masoumzadeh}, {Mohr}, {Odier}, {Scott}, {Vassallo}, {Verdoes Kleijn}, {Zacchei}, {Aghanim}, {Amara}, {Andreon}, {Auricchio}, {Awan}, {Azzollini}, {Baccigalupi}, {Baldi}, {Balestra}, {Bardelli}, {Basset}, {Battaglia}, {Belikov}, {Bender}, {Biviano}, {Bonchi}, {Bonino}, {Branchini}, {Brescia}, {Brinchmann}, {Camera}, {Ca{\~n}as-Herrera}, {Capobianco}, {Carbone}, {Cardone}, {Carretero}, {Casas}, {Castander}, {Castellano}, {Castignani}, {Cavuoti}, {Chambers}, {Cimatti}, {Colodro-Conde}, {Congedo}, {Conselice}, {Conversi}, {Copin}, {Courbin}, {Courtois}, {Cropper}, {Cuby}, {Da Silva}, {da Silva}, {Degaudenzi}, {de Jong}, {De Lucia}, {Di Giorgio}, {Dinis}, {Dolding}, {Dole}, {Douspis}, {Dubath}, {Duncan}, {Dupac}, {Dusini}, {Ealet}, {Escoffier}, {Fabricius}, {Farina},
  {Farinelli}, {Faustini}, {Ferriol}, {Fotopoulou}, {Fourmanoit}, {Frailis}, {Franceschi}, {Franzetti}, {Galeotta}, {George}, {Gillard}, {Gillis}, {Giocoli}, {G{\'o}mez-Alvarez}, {Gracia-Carpio}, {Granett}, {Grazian}, {Grupp}, {Guzzo}, {Gwyn}, {Haugan}, {Herent}, {Hoar}, {Hoekstra}, {Holliman}, {Holmes}, {Hook}, {Hormuth}, {Hornstrup}, {Hudelot}, {Ili{\'c}}, {Jhabvala}, {Joachimi}, {Keih{\"a}nen}, {Kermiche}, {Kiessling}, {Kubik}, {Kuijken}, {K{\"u}mmel}, {Kunz}, {Kurki-Suonio}, {Lahav}, {Le Boulc'h}, {Le Brun}, {Le Mignant}, {Liebing}, {Ligori}, {Lilje}, {Lindholm}, {Lloro}, {Mainetti}, {Maino}, {Maiorano}, {Mansutti}, {Marcin}, {Marggraf}, {Markovic}, {Martinelli}, {Martinet}, {Marulli}, {Massey}, {Maurogordato}, {McCracken}, {Medinaceli}, {Mei}, {Melchior}, {Mellier}, {Meneghetti}, {Merlin}, {Meylan}, {Mora}, {Moresco}, {Morris}, {Moscardini}, {Mourre}, {Nakajima}, {Neissner}, {Nichol}, {Niemi}, {Nightingale}, {Nutma}, {Padilla}, {Paltani}, {Pasian}, {Peacock}, {Pedersen}, {Percival}, {Pettorino}, {Pires},
  {Polenta}, {Pollack}, {Poncet}, {Popa}, {Pozzetti}, {Racca}, {Raison}, {Rebolo}, {Renzi}, {Rhodes}, {Riccio}, {Rix}, {Romelli}, {Roncarelli}, {Rossetti}, {Rusholme}, {Saglia}, {Sakr}, {S{\'a}nchez}, {Sapone}, {Sartoris}, {Sauvage}, {Schewtschenko}, {Schneider}, {Scodeggio}, {Secroun}, {Sefusatti}, \& {Seidel}}]{euclidcollaboration2025euclidquickdatarelease}
{Euclid Collaboration}, {Aussel}, H., {Tereno}, I., {et~al.} 2026, \aap, 711, A1

\bibitem[{{Euclid Collaboration} {et~al.}(2023){Euclid Collaboration}, {Merlin}, {Castellano}, {Bretonni{\`e}re}, {Huertas-Company}, {Kuchner}, {Tuccillo}, {Buitrago}, {Peterson}, {Conselice}, {Caro}, {Dimauro}, {Nemani}, {Fontana}, {K{\"u}mmel}, {H{\"a}u{\ss}ler}, {Hartley}, {Alvarez Ayllon}, {Bertin}, {Dubath}, {Ferrari}, {Ferreira}, {Gavazzi}, {Hern{\'a}ndez-Lang}, {Lucatelli}, {Robotham}, {Schefer}, {Tortora}, {Aghanim}, {Amara}, {Amendola}, {Auricchio}, {Baldi}, {Bender}, {Bodendorf}, {Branchini}, {Brescia}, {Camera}, {Capobianco}, {Carbone}, {Carretero}, {Castander}, {Cavuoti}, {Cimatti}, {Cledassou}, {Congedo}, {Conversi}, {Copin}, {Corcione}, {Courbin}, {Cropper}, {Da Silva}, {Degaudenzi}, {Dinis}, {Douspis}, {Dubath}, {Duncan}, {Dupac}, {Dusini}, {Farrens}, {Ferriol}, {Frailis}, {Franceschi}, {Franzetti}, {Galeotta}, {Garilli}, {Gillis}, {Giocoli}, {Grazian}, {Grupp}, {Haugan}, {Hoekstra}, {Holmes}, {Hormuth}, {Hornstrup}, {Hudelot}, {Jahnke}, {Kermiche}, {Kiessling}, {Kitching}, {Kohley}, {Kunz},
  {Kurki-Suonio}, {Ligori}, {Lilje}, {Lloro}, {Mansutti}, {Marggraf}, {Markovic}, {Marulli}, {Massey}, {McCracken}, {Medinaceli}, {Melchior}, {Meneghetti}, {Meylan}, {Moresco}, {Moscardini}, {Munari}, {Niemi}, {Padilla}, {Paltani}, {Pasian}, {Pedersen}, {Percival}, {Polenta}, {Poncet}, {Popa}, {Pozzetti}, {Raison}, {Rebolo}, {Renzi}, {Rhodes}, {Riccio}, {Romelli}, {Rossetti}, {Saglia}, {Sapone}, {Sartoris}, {Schneider}, {Secroun}, {Seidel}, {Sirignano}, {Sirri}, {Skottfelt}, {Starck}, {Tallada-Cresp{\'\i}}, {Taylor}, {Tereno}, {Toledo-Moreo}, {Tutusaus}, {Valenziano}, {Vassallo}, {Wang}, {Weller}, {Zacchei}, {Zamorani}, {Zoubian}, {Andreon}, {Bardelli}, {Boucaud}, {Colodro-Conde}, {Di Ferdinando}, {Graci{\'a}-Carpio}, {Lindholm}, {Mauri}, {Mei}, {Neissner}, {Scottez}, {Tramacere}, {Zucca}, {Baccigalupi}, {Balaguera-Antol{\'\i}nez}, {Ballardini}, {Bernardeau}, {Biviano}, {Borgani}, {Borlaff}, {Burigana}, {Cabanac}, {Cappi}, {Carvalho}, {Casas}, {Castignani}, {Cooray}, {Coupon}, {Courtois}, {Cucciati},
  {Davini}, {De Lucia}, {Desprez}, {Escartin}, {Escoffier}, {Farina}, {Ganga}, {Garcia-Bellido}, {George}, {Gozaliasl}, {Hildebrandt}, {Hook}, {Ilbert}, {Ili{\'c}}, {Joachimi}, {Kansal}, {Keihanen}, {Kirkpatrick}, {Loureiro}, {Macias-Perez}, {Magliocchetti}, {Mainetti}, {Maoli}, {Marcin}, {Martinelli}, {Martinet}, {Matthew}, {Maturi}, {Metcalf}, {Monaco}, {Morgante}, \& {Nadathur}}]{Euclid2023}
{Euclid Collaboration}, {Merlin}, E., {Castellano}, M., {et~al.} 2023, \aap, 671, A101

\bibitem[{{Finkelstein} {et~al.}(2023){Finkelstein}, {Bagley}, {Ferguson}, {Wilkins}, {Kartaltepe}, {Papovich}, {Yung}, {Arrabal Haro}, {Behroozi}, {Dickinson}, {Kocevski}, {Koekemoer}, {Larson}, {Le Bail}, {Morales}, {P{\'e}rez-Gonz{\'a}lez}, {Burgarella}, {Dav{\'e}}, {Hirschmann}, {Somerville}, {Wuyts}, {Bromm}, {Casey}, {Fontana}, {Fujimoto}, {Gardner}, {Giavalisco}, {Grazian}, {Grogin}, {Hathi}, {Hutchison}, {Jha}, {Jogee}, {Kewley}, {Kirkpatrick}, {Long}, {Lotz}, {Pentericci}, {Pierel}, {Pirzkal}, {Ravindranath}, {Ryan}, {Trump}, {Yang}, {Bhatawdekar}, {Bisigello}, {Buat}, {Calabr{\`o}}, {Castellano}, {Cleri}, {Cooper}, {Croton}, {Daddi}, {Dekel}, {Elbaz}, {Franco}, {Gawiser}, {Holwerda}, {Huertas-Company}, {Jaskot}, {Leung}, {Lucas}, {Mobasher}, {Pandya}, {Tacchella}, {Weiner}, \& {Zavala}}]{Finkelstein2023CEERS}
{Finkelstein}, S.~L., {Bagley}, M.~B., {Ferguson}, H.~C., {et~al.} 2023, \apjl, 946, L13

\bibitem[{{Gaia Collaboration} {et~al.}(2021){Gaia Collaboration}, {Brown}, {Vallenari}, {Prusti}, {de Bruijne}, {Babusiaux}, {Biermann}, {Creevey}, {Evans}, {Eyer}, {Hutton}, {Jansen}, {Jordi}, {Klioner}, {Lammers}, {Lindegren}, {Luri}, {Mignard}, {Panem}, {Pourbaix}, {Randich}, {Sartoretti}, {Soubiran}, {Walton}, {Arenou}, {Bailer-Jones}, {Bastian}, {Cropper}, {Drimmel}, {Katz}, {Lattanzi}, {van Leeuwen}, {Bakker}, {Cacciari}, {Casta{\~n}eda}, {De Angeli}, {Ducourant}, {Fabricius}, {Fouesneau}, {Fr{\'e}mat}, {Guerra}, {Guerrier}, {Guiraud}, {Jean-Antoine Piccolo}, {Masana}, {Messineo}, {Mowlavi}, {Nicolas}, {Nienartowicz}, {Pailler}, {Panuzzo}, {Riclet}, {Roux}, {Seabroke}, {Sordo}, {Tanga}, {Th{\'e}venin}, {Gracia-Abril}, {Portell}, {Teyssier}, {Altmann}, {Andrae}, {Bellas-Velidis}, {Benson}, {Berthier}, {Blomme}, {Brugaletta}, {Burgess}, {Busso}, {Carry}, {Cellino}, {Cheek}, {Clementini}, {Damerdji}, {Davidson}, {Delchambre}, {Dell'Oro}, {Fern{\'a}ndez-Hern{\'a}ndez}, {Galluccio}, {Garc{\'\i}a-Lario},
  {Garcia-Reinaldos}, {Gonz{\'a}lez-N{\'u}{\~n}ez}, {Gosset}, {Haigron}, {Halbwachs}, {Hambly}, {Harrison}, {Hatzidimitriou}, {Heiter}, {Hern{\'a}ndez}, {Hestroffer}, {Hodgkin}, {Holl}, {Jan{\ss}en}, {Jevardat de Fombelle}, {Jordan}, {Krone-Martins}, {Lanzafame}, {L{\"o}ffler}, {Lorca}, {Manteiga}, {Marchal}, {Marrese}, {Moitinho}, {Mora}, {Muinonen}, {Osborne}, {Pancino}, {Pauwels}, {Petit}, {Recio-Blanco}, {Richards}, {Riello}, {Rimoldini}, {Robin}, {Roegiers}, {Rybizki}, {Sarro}, {Siopis}, {Smith}, {Sozzetti}, {Ulla}, {Utrilla}, {van Leeuwen}, {van Reeven}, {Abbas}, {Abreu Aramburu}, {Accart}, {Aerts}, {Aguado}, {Ajaj}, {Altavilla}, {{\'A}lvarez}, {{\'A}lvarez Cid-Fuentes}, {Alves}, {Anderson}, {Anglada Varela}, {Antoja}, {Audard}, {Baines}, {Baker}, {Balaguer-N{\'u}{\~n}ez}, {Balbinot}, {Balog}, {Barache}, {Barbato}, {Barros}, {Barstow}, {Bartolom{\'e}}, {Bassilana}, {Bauchet}, {Baudesson-Stella}, {Becciani}, {Bellazzini}, {Bernet}, {Bertone}, {Bianchi}, {Blanco-Cuaresma}, {Boch}, {Bombrun}, {Bossini},
  {Bouquillon}, {Bragaglia}, {Bramante}, {Breedt}, {Bressan}, {Brouillet}, {Bucciarelli}, {Burlacu}, {Busonero}, {Butkevich}, {Buzzi}, {Caffau}, {Cancelliere}, {C{\'a}novas}, {Cantat-Gaudin}, {Carballo}, {Carlucci}, {Carnerero}, {Carrasco}, {Casamiquela}, {Castellani}, {Castro-Ginard}, {Castro Sampol}, {Chaoul}, {Charlot}, {Chemin}, {Chiavassa}, {Cioni}, {Comoretto}, {Cooper}, {Cornez}, {Cowell}, {Crifo}, {Crosta}, {Crowley}, {Dafonte}, {Dapergolas}, {David}, \& {David}}]{GaiaEDR32021summary}
{Gaia Collaboration}, {Brown}, A.~G.~A., {Vallenari}, A., {et~al.} 2021, \aap, 649, A1

\bibitem[{{Gaia Collaboration} {et~al.}(2023){Gaia Collaboration}, {Vallenari}, {Brown}, {Prusti}, {de Bruijne}, {Arenou}, {Babusiaux}, {Biermann}, {Creevey}, {Ducourant}, {Evans}, {Eyer}, {Guerra}, {Hutton}, {Jordi}, {Klioner}, {Lammers}, {Lindegren}, {Luri}, {Mignard}, {Panem}, {Pourbaix}, {Randich}, {Sartoretti}, {Soubiran}, {Tanga}, {Walton}, {Bailer-Jones}, {Bastian}, {Drimmel}, {Jansen}, {Katz}, {Lattanzi}, {van Leeuwen}, {Bakker}, {Cacciari}, {Casta{\~n}eda}, {De Angeli}, {Fabricius}, {Fouesneau}, {Fr{\'e}mat}, {Galluccio}, {Guerrier}, {Heiter}, {Masana}, {Messineo}, {Mowlavi}, {Nicolas}, {Nienartowicz}, {Pailler}, {Panuzzo}, {Riclet}, {Roux}, {Seabroke}, {Sordo}, {Th{\'e}venin}, {Gracia-Abril}, {Portell}, {Teyssier}, {Altmann}, {Andrae}, {Audard}, {Bellas-Velidis}, {Benson}, {Berthier}, {Blomme}, {Burgess}, {Busonero}, {Busso}, {C{\'a}novas}, {Carry}, {Cellino}, {Cheek}, {Clementini}, {Damerdji}, {Davidson}, {de Teodoro}, {Nu{\~n}ez Campos}, {Delchambre}, {Dell'Oro}, {Esquej},
  {Fern{\'a}ndez-Hern{\'a}ndez}, {Fraile}, {Garabato}, {Garc{\'\i}a-Lario}, {Gosset}, {Haigron}, {Halbwachs}, {Hambly}, {Harrison}, {Hern{\'a}ndez}, {Hestroffer}, {Hodgkin}, {Holl}, {Jan{\ss}en}, {Jevardat de Fombelle}, {Jordan}, {Krone-Martins}, {Lanzafame}, {L{\"o}ffler}, {Marchal}, {Marrese}, {Moitinho}, {Muinonen}, {Osborne}, {Pancino}, {Pauwels}, {Recio-Blanco}, {Reyl{\'e}}, {Riello}, {Rimoldini}, {Roegiers}, {Rybizki}, {Sarro}, {Siopis}, {Smith}, {Sozzetti}, {Utrilla}, {van Leeuwen}, {Abbas}, {{\'A}brah{\'a}m}, {Abreu Aramburu}, {Aerts}, {Aguado}, {Ajaj}, {Aldea-Montero}, {Altavilla}, {{\'A}lvarez}, {Alves}, {Anders}, {Anderson}, {Anglada Varela}, {Antoja}, {Baines}, {Baker}, {Balaguer-N{\'u}{\~n}ez}, {Balbinot}, {Balog}, {Barache}, {Barbato}, {Barros}, {Barstow}, {Bartolom{\'e}}, {Bassilana}, {Bauchet}, {Becciani}, {Bellazzini}, {Berihuete}, {Bernet}, {Bertone}, {Bianchi}, {Binnenfeld}, {Blanco-Cuaresma}, {Blazere}, {Boch}, {Bombrun}, {Bossini}, {Bouquillon}, {Bragaglia}, {Bramante}, {Breedt},
  {Bressan}, {Brouillet}, {Brugaletta}, {Bucciarelli}, {Burlacu}, {Butkevich}, {Buzzi}, {Caffau}, {Cancelliere}, {Cantat-Gaudin}, {Carballo}, {Carlucci}, {Carnerero}, {Carrasco}, {Casamiquela}, {Castellani}, {Castro-Ginard}, {Chaoul}, {Charlot}, {Chemin}, {Chiaramida}, {Chiavassa}, {Chornay}, {Comoretto}, {Contursi}, {Cooper}, {Cornez}, {Cowell}, {Crifo}, {Cropper}, {Crosta}, {Crowley}, {Dafonte}, {Dapergolas}, {David}, {David}, {de Laverny}, {De Luise}, \& {De March}}]{Gaia2023}
{Gaia Collaboration}, {Vallenari}, A., {Brown}, A.~G.~A., {et~al.} 2023, \aap, 674, A1

\bibitem[{{Gandolfi} {et~al.}(2026){Gandolfi}, {Rodighiero}, {Bisigello}, {Grazian}, {Finkelstein}, {Dickinson}, {Castellano}, {Merlin}, {Calabr{\`o}}, {Papovich}, {Bianchetti}, {Ba{\~n}ados}, {Benotto}, {Catone}, {Buitrago}, {Daddi}, {Girardi}, {Giulietti}, {Hirschmann}, {Holwerda}, {Arrabal Haro}, {Lapi}, {Lucas}, {Lyu}, {Massardi}, {Pacucci}, {P{\'e}rez-Gonz{\'a}lez}, {Ronconi}, {Tarrasse}, {Wilkins}, {Vulcani}, {Yung}, {Zavala}, {Backhaus}, {Bagley}, {Buat}, {Burgarella}, {Kartaltepe}, {Khusanova}, {Kirkpatrick}, {Kocevski}, {Koekemoer}, {Lambrides}, {Pirzkal}, \& {Yang}}]{gandolfi2025ultrahighredshiftcloserbydustobscured}
{Gandolfi}, G., {Rodighiero}, G., {Bisigello}, L., {et~al.} 2026, \aap, 708, A195

\bibitem[{{Genin} {et~al.}(2025){Genin}, {Shuntov}, {Brammer}, {Allen}, {Ito}, {Magdis}, {Matharu}, {Oesch}, {Toft}, \& {Valentino}}]{Genin_2025}
{Genin}, A., {Shuntov}, M., {Brammer}, G., {et~al.} 2025, \aap, 699, A343

\bibitem[{{Gottumukkala} {et~al.}(2024){Gottumukkala}, {Barrufet}, {Oesch}, {Weibel}, {Allen}, {Alcalde Pampliega}, {Nelson}, {Williams}, {Brammer}, {Fudamoto}, {Gonz{\'a}lez}, {Heintz}, {Illingworth}, {Magee}, {Naidu}, {Shuntov}, {Stefanon}, {Toft}, {Valentino}, \& {Xiao}}]{Gottumukkala2024}
{Gottumukkala}, R., {Barrufet}, L., {Oesch}, P.~A., {et~al.} 2024, \mnras, 530, 966

\bibitem[{{Grogin} {et~al.}(2011){Grogin}, {Kocevski}, {Faber}, {Ferguson}, {Koekemoer}, {Riess}, {Acquaviva}, {Alexander}, {Almaini}, {Ashby}, {Barden}, {Bell}, {Bournaud}, {Brown}, {Caputi}, {Casertano}, {Cassata}, {Castellano}, {Challis}, {Chary}, {Cheung}, {Cirasuolo}, {Conselice}, {Roshan Cooray}, {Croton}, {Daddi}, {Dahlen}, {Dav{\'e}}, {de Mello}, {Dekel}, {Dickinson}, {Dolch}, {Donley}, {Dunlop}, {Dutton}, {Elbaz}, {Fazio}, {Filippenko}, {Finkelstein}, {Fontana}, {Gardner}, {Garnavich}, {Gawiser}, {Giavalisco}, {Grazian}, {Guo}, {Hathi}, {H{\"a}ussler}, {Hopkins}, {Huang}, {Huang}, {Jha}, {Kartaltepe}, {Kirshner}, {Koo}, {Lai}, {Lee}, {Li}, {Lotz}, {Lucas}, {Madau}, {McCarthy}, {McGrath}, {McIntosh}, {McLure}, {Mobasher}, {Moustakas}, {Mozena}, {Nandra}, {Newman}, {Niemi}, {Noeske}, {Papovich}, {Pentericci}, {Pope}, {Primack}, {Rajan}, {Ravindranath}, {Reddy}, {Renzini}, {Rix}, {Robaina}, {Rodney}, {Rosario}, {Rosati}, {Salimbeni}, {Scarlata}, {Siana}, {Simard}, {Smidt}, {Somerville}, {Spinrad},
  {Straughn}, {Strolger}, {Telford}, {Teplitz}, {Trump}, {van der Wel}, {Villforth}, {Wechsler}, {Weiner}, {Wiklind}, {Wild}, {Wilson}, {Wuyts}, {Yan}, \& {Yun}}]{Grogin2011}
{Grogin}, N.~A., {Kocevski}, D.~D., {Faber}, S.~M., {et~al.} 2011, \apjs, 197, 35

\bibitem[{{Gwyn}(2012)}]{Gwyn2012CFHT}
{Gwyn}, S. D.~J. 2012, \aj, 143, 38

\bibitem[{{Hainline} {et~al.}(2024){Hainline}, {Johnson}, {Robertson}, {Tacchella}, {Helton}, {Sun}, {Eisenstein}, {Simmonds}, {Topping}, {Whitler}, {Willmer}, {Rieke}, {Suess}, {Hviding}, {Cameron}, {Alberts}, {Baker}, {Baum}, {Bhatawdekar}, {Bonaventura}, {Boyett}, {Bunker}, {Carniani}, {Charlot}, {Chevallard}, {Chen}, {Curti}, {Curtis-Lake}, {D'Eugenio}, {Egami}, {Endsley}, {Hausen}, {Ji}, {Looser}, {Lyu}, {Maiolino}, {Nelson}, {Pusk{\'a}s}, {Rawle}, {Sandles}, {Saxena}, {Smit}, {Stark}, {Williams}, {Willott}, \& {Witstok}}]{Hainline2024}
{Hainline}, K.~N., {Johnson}, B.~D., {Robertson}, B., {et~al.} 2024, \apj, 964, 71

\bibitem[{{Hassani} {et~al.}(2026){Hassani}, {Rosolowsky}, {Leroy}, {Sandstrom}, {Boquien}, {Thilker}, {Whitmore}, {Anand}, {Barnes}, {Cao}, {Chown}, {Congiu}, {Dale}, {Egorov}, {Gerasimov}, {Grasha}, {Indebetouw}, {Lee}, {Liang}, {Maschmann}, {Meidt}, {Oakes}, {Pessa}, {Pety}, {Querejeta}, {Ramambason}, {Rodr{\'\i}guez}, {Sarbadhicary}, {Sutter}, {{\'U}beda}, \& {Williams}}]{hassani2025hidden}
{Hassani}, H., {Rosolowsky}, E., {Leroy}, A.~K., {et~al.} 2026, \apjs, 284, 3

\bibitem[{{H{\"a}u{\ss}ler} {et~al.}(2022){H{\"a}u{\ss}ler}, {Vika}, {Bamford}, {Johnston}, {Brough}, {Casura}, {Holwerda}, {Kelvin}, \& {Popescu}}]{Haussler2022galapagos2}
{H{\"a}u{\ss}ler}, B., {Vika}, M., {Bamford}, S.~P., {et~al.} 2022, \aap, 664, A92

\bibitem[{{Ilbert} {et~al.}(2006){Ilbert}, {Arnouts}, {McCracken}, {Bolzonella}, {Bertin}, {Le F{\`e}vre}, {Mellier}, {Zamorani}, {Pell{\`o}}, {Iovino}, {Tresse}, {Le Brun}, {Bottini}, {Garilli}, {Maccagni}, {Picat}, {Scaramella}, {Scodeggio}, {Vettolani}, {Zanichelli}, {Adami}, {Bardelli}, {Cappi}, {Charlot}, {Ciliegi}, {Contini}, {Cucciati}, {Foucaud}, {Franzetti}, {Gavignaud}, {Guzzo}, {Marano}, {Marinoni}, {Mazure}, {Meneux}, {Merighi}, {Paltani}, {Pollo}, {Pozzetti}, {Radovich}, {Zucca}, {Bondi}, {Bongiorno}, {Busarello}, {de La Torre}, {Gregorini}, {Lamareille}, {Mathez}, {Merluzzi}, {Ripepi}, {Rizzo}, \& {Vergani}}]{Ilbert2006Lephare}
{Ilbert}, O., {Arnouts}, S., {McCracken}, H.~J., {et~al.} 2006, \aap, 457, 841

\bibitem[{{Ito} {et~al.}(2025){Ito}, {Tanaka}, {Shimasaku}, {Ando}, {Onoue}, {Tanaka}, {Matsui}, {Kakimoto}, \& {Valentino}}]{Ito2025}
{Ito}, K., {Tanaka}, T.~S., {Shimasaku}, K., {et~al.} 2025, \mnras, 538, 1501

\bibitem[{{Kirkpatrick} {et~al.}(2023){Kirkpatrick}, {Yang}, {Le Bail}, {Troiani}, {Bell}, {Cleri}, {Elbaz}, {Finkelstein}, {Hathi}, {Hirschmann}, {Holwerda}, {Kocevski}, {Lucas}, {McKinney}, {Papovich}, {P{\'e}rez-Gonz{\'a}lez}, {de la Vega}, {Bagley}, {Daddi}, {Dickinson}, {Ferguson}, {Fontana}, {Grazian}, {Grogin}, {Arrabal Haro}, {Kartaltepe}, {Kewley}, {Koekemoer}, {Lotz}, {Pentericci}, {Pirzkal}, {Ravindranath}, {Somerville}, {Trump}, {Wilkins}, \& {Yung}}]{Kirkpatrick2023}
{Kirkpatrick}, A., {Yang}, G., {Le Bail}, A., {et~al.} 2023, \apjl, 959, L7

\bibitem[{{Koekemoer} {et~al.}(2011){Koekemoer}, {Faber}, {Ferguson}, {Grogin}, {Kocevski}, {Koo}, {Lai}, {Lotz}, {Lucas}, {McGrath}, {Ogaz}, {Rajan}, {Riess}, {Rodney}, {Strolger}, {Casertano}, {Castellano}, {Dahlen}, {Dickinson}, {Dolch}, {Fontana}, {Giavalisco}, {Grazian}, {Guo}, {Hathi}, {Huang}, {van der Wel}, {Yan}, {Acquaviva}, {Alexander}, {Almaini}, {Ashby}, {Barden}, {Bell}, {Bournaud}, {Brown}, {Caputi}, {Cassata}, {Challis}, {Chary}, {Cheung}, {Cirasuolo}, {Conselice}, {Roshan Cooray}, {Croton}, {Daddi}, {Dav{\'e}}, {de Mello}, {de Ravel}, {Dekel}, {Donley}, {Dunlop}, {Dutton}, {Elbaz}, {Fazio}, {Filippenko}, {Finkelstein}, {Frazer}, {Gardner}, {Garnavich}, {Gawiser}, {Gruetzbauch}, {Hartley}, {H{\"a}ussler}, {Herrington}, {Hopkins}, {Huang}, {Jha}, {Johnson}, {Kartaltepe}, {Khostovan}, {Kirshner}, {Lani}, {Lee}, {Li}, {Madau}, {McCarthy}, {McIntosh}, {McLure}, {McPartland}, {Mobasher}, {Moreira}, {Mortlock}, {Moustakas}, {Mozena}, {Nandra}, {Newman}, {Nielsen}, {Niemi}, {Noeske}, {Papovich},
  {Pentericci}, {Pope}, {Primack}, {Ravindranath}, {Reddy}, {Renzini}, {Rix}, {Robaina}, {Rosario}, {Rosati}, {Salimbeni}, {Scarlata}, {Siana}, {Simard}, {Smidt}, {Snyder}, {Somerville}, {Spinrad}, {Straughn}, {Telford}, {Teplitz}, {Trump}, {Vargas}, {Villforth}, {Wagner}, {Wandro}, {Wechsler}, {Weiner}, {Wiklind}, {Wild}, {Wilson}, {Wuyts}, \& {Yun}}]{Koekemoer2011}
{Koekemoer}, A.~M., {Faber}, S.~M., {Ferguson}, H.~C., {et~al.} 2011, \apjs, 197, 36

\bibitem[{{Kriek} {et~al.}(2009){Kriek}, {van Dokkum}, {Labb{\'e}}, {Franx}, {Illingworth}, {Marchesini}, \& {Quadri}}]{Kriek2009FAST}
{Kriek}, M., {van Dokkum}, P.~G., {Labb{\'e}}, I., {et~al.} 2009, \apj, 700, 221

\bibitem[{{Kron}(1980)}]{Kron1980}
{Kron}, R.~G. 1980, \apjs, 43, 305

\bibitem[{{Laird} {et~al.}(2009){Laird}, {Nandra}, {Georgakakis}, {Aird}, {Barmby}, {Conselice}, {Coil}, {Davis}, {Faber}, {Fazio}, {Guhathakurta}, {Koo}, {Sarajedini}, \& {Willmer}}]{Laird_2009}
{Laird}, E.~S., {Nandra}, K., {Georgakakis}, A., {et~al.} 2009, \apjs, 180, 102

\bibitem[{{Larson} {et~al.}(2023){Larson}, {Hutchison}, {Bagley}, {Finkelstein}, {Yung}, {Somerville}, {Hirschmann}, {Brammer}, {Holwerda}, {Papovich}, {Morales}, \& {Wilkins}}]{Larson_2023}
{Larson}, R.~L., {Hutchison}, T.~A., {Bagley}, M., {et~al.} 2023, \apj, 958, 141

\bibitem[{{Li} {et~al.}(2023){Li}, {Ho}, {Shangguan}, {Zhuang}, \& {Li}}]{Liyang2023}
{Li}, Y.~A., {Ho}, L.~C., {Shangguan}, J., {Zhuang}, M.-Y., \& {Li}, R. 2023, \apjs, 267, 17

\bibitem[{{Lindegren} {et~al.}(2021){Lindegren}, {Klioner}, {Hern{\'a}ndez}, {Bombrun}, {Ramos-Lerate}, {Steidelm{\"u}ller}, {Bastian}, {Biermann}, {de Torres}, {Gerlach}, {Geyer}, {Hilger}, {Hobbs}, {Lammers}, {McMillan}, {Stephenson}, {Casta{\~n}eda}, {Davidson}, {Fabricius}, {Gracia-Abril}, {Portell}, {Rowell}, {Teyssier}, {Torra}, {Bartolom{\'e}}, {Clotet}, {Garralda}, {Gonz{\'a}lez-Vidal}, {Torra}, {Abbas}, {Altmann}, {Anglada Varela}, {Balaguer-N{\'u}{\~n}ez}, {Balog}, {Barache}, {Becciani}, {Bernet}, {Bertone}, {Bianchi}, {Bouquillon}, {Brown}, {Bucciarelli}, {Busonero}, {Butkevich}, {Buzzi}, {Cancelliere}, {Carlucci}, {Charlot}, {Cioni}, {Crosta}, {Crowley}, {del Peloso}, {del Pozo}, {Drimmel}, {Esquej}, {Fienga}, {Fraile}, {Gai}, {Garcia-Reinaldos}, {Guerra}, {Hambly}, {Hauser}, {Jan{\ss}en}, {Jordan}, {Kostrzewa-Rutkowska}, {Lattanzi}, {Liao}, {Licata}, {Lister}, {L{\"o}ffler}, {Marchant}, {Masip}, {Mignard}, {Mints}, {Molina}, {Mora}, {Morbidelli}, {Murphy}, {Pagani}, {Panuzzo}, {Pe{\~n}alosa
  Esteller}, {Poggio}, {Re Fiorentin}, {Riva}, {Sagrist{\`a} Sell{\'e}s}, {Sanchez Gimenez}, {Sarasso}, {Sciacca}, {Siddiqui}, {Smart}, {Souami}, {Spagna}, {Steele}, {Taris}, {Utrilla}, {van Reeven}, \& {Vecchiato}}]{GaiaEDR32021astrometry}
{Lindegren}, L., {Klioner}, S.~A., {Hern{\'a}ndez}, J., {et~al.} 2021, \aap, 649, A2

\bibitem[{{Ling} {et~al.}(2024){Ling}, {Goto}, {Kim}, {Wu}, {Hashimoto}, {Chien}, {Lin}, {Ho}, \& {Kilerci}}]{Ling2024}
{Ling}, C.-T., {Goto}, T., {Kim}, S.~J., {et~al.} 2024, \mnras, 528, 6025

\bibitem[{{McKay} {et~al.}(2025){McKay}, {Barger}, {Cowie}, \& {Nicandro Rosenthal}}]{Mckay2025}
{McKay}, S.~J., {Barger}, A.~J., {Cowie}, L.~L., \& {Nicandro Rosenthal}, M.~J. 2025, \apj, 988, 135

\bibitem[{{Merlin} {et~al.}(2016){Merlin}, {Bourne}, {Castellano}, {Ferguson}, {Wang}, {Derriere}, {Dunlop}, {Elbaz}, \& {Fontana}}]{Merlin2016tphot2}
{Merlin}, E., {Bourne}, N., {Castellano}, M., {et~al.} 2016, \aap, 595, A97

\bibitem[{{Merlin} {et~al.}(2015){Merlin}, {Fontana}, {Ferguson}, {Dunlop}, {Elbaz}, {Bourne}, {Bruce}, {Buitrago}, {Castellano}, {Schreiber}, {Grazian}, {McLure}, {Okumura}, {Shu}, {Wang}, {Amor{\'\i}n}, {Boutsia}, {Cappelluti}, {Comastri}, {Derriere}, {Faber}, \& {Santini}}]{Merlin2015tphot}
{Merlin}, E., {Fontana}, A., {Ferguson}, H.~C., {et~al.} 2015, \aap, 582, A15

\bibitem[{{Merlin} {et~al.}(2024){Merlin}, {Santini}, {Paris}, {Castellano}, {Fontana}, {Treu}, {Finkelstein}, {Dunlop}, {Arrabal Haro}, {Bagley}, {Boyett}, {Calabr{\`o}}, {Correnti}, {Davis}, {Dickinson}, {Donnan}, {Ferguson}, {Fortuni}, {Giavalisco}, {Glazebrook}, {Grazian}, {Grogin}, {Hathi}, {Hirschmann}, {Kartaltepe}, {Kewley}, {Kirkpatrick}, {Kocevski}, {Koekemoer}, {Leung}, {Lotz}, {Lucas}, {Magee}, {Marchesini}, {Mascia}, {McLeod}, {McLure}, {Nanayakkara}, {Napolitano}, {Nonino}, {Papovich}, {Pentericci}, {P{\'e}rez-Gonz{\'a}lez}, {Pirzkal}, {Ravindranath}, {Roberts-Borsani}, {Somerville}, {Trenti}, {Trump}, {Vulcani}, {Wang}, {Watson}, {Wilkins}, {Yang}, \& {Yung}}]{Merlin_2024}
{Merlin}, E., {Santini}, P., {Paris}, D., {et~al.} 2024, \aap, 691, A240

\bibitem[{{Momcheva} {et~al.}(2016){Momcheva}, {Brammer}, {van Dokkum}, {Skelton}, {Whitaker}, {Nelson}, {Fumagalli}, {Maseda}, {Leja}, {Franx}, {Rix}, {Bezanson}, {Da Cunha}, {Dickey}, {F{\"o}rster Schreiber}, {Illingworth}, {Kriek}, {Labb{\'e}}, {Ulf Lange}, {Lundgren}, {Magee}, {Marchesini}, {Oesch}, {Pacifici}, {Patel}, {Price}, {Tal}, {Wake}, {van der Wel}, \& {Wuyts}}]{Momcheva2016}
{Momcheva}, I.~G., {Brammer}, G.~B., {van Dokkum}, P.~G., {et~al.} 2016, \apjs, 225, 27

\bibitem[{{Nandra} {et~al.}(2005){Nandra}, {Laird}, {Adelberger}, {Gardner}, {Mushotzky}, {Rhodes}, {Steidel}, {Teplitz}, \& {Arnaud}}]{Nandra_2005}
{Nandra}, K., {Laird}, E.~S., {Adelberger}, K., {et~al.} 2005, \mnras, 356, 568

\bibitem[{{Nandra} {et~al.}(2015){Nandra}, {Laird}, {Aird}, {Salvato}, {Georgakakis}, {Barro}, {Perez-Gonzalez}, {Barmby}, {Chary}, {Coil}, {Cooper}, {Davis}, {Dickinson}, {Faber}, {Fazio}, {Guhathakurta}, {Gwyn}, {Hsu}, {Huang}, {Ivison}, {Koo}, {Newman}, {Rangel}, {Yamada}, \& {Willmer}}]{Nandra2015}
{Nandra}, K., {Laird}, E.~S., {Aird}, J.~A., {et~al.} 2015, \apjs, 220, 10

\bibitem[{{Nelson} {et~al.}(2024){Nelson}, {Brammer}, {Gim{\'e}nez-Arteaga}, {Oesch}, {Naidu}, {{\"U}bler}, {Matharu}, {Shapley}, {Whitaker}, {Wisnioski}, {F{\"o}rster Schreiber}, {Smit}, {van Dokkum}, {Chisholm}, {Endsley}, {Hartley}, {Gibson}, {Giovinazzo}, {Illingworth}, {Labbe}, {Maseda}, {Matthee}, {Covelo Paz}, {Price}, {Reddy}, {Shivaei}, {Weibel}, {Wuyts}, {Xiao}, {Alberts}, {Baker}, {Bunker}, {Cameron}, {Charlot}, {Eisenstein}, {de Graaff}, {Ji}, {Johnson}, {Jones}, {Maiolino}, {Robertson}, {Sandles}, {Suess}, {Tacchella}, {Williams}, \& {Witstok}}]{Nelson2024}
{Nelson}, E., {Brammer}, G., {Gim{\'e}nez-Arteaga}, C., {et~al.} 2024, \apjl, 976, L27

\bibitem[{{Newman} {et~al.}(2013){Newman}, {Cooper}, {Davis}, {Faber}, {Coil}, {Guhathakurta}, {Koo}, {Phillips}, {Conroy}, {Dutton}, {Finkbeiner}, {Gerke}, {Rosario}, {Weiner}, {Willmer}, {Yan}, {Harker}, {Kassin}, {Konidaris}, {Lai}, {Madgwick}, {Noeske}, {Wirth}, {Connolly}, {Kaiser}, {Kirby}, {Lemaux}, {Lin}, {Lotz}, {Luppino}, {Marinoni}, {Matthews}, {Metevier}, \& {Schiavon}}]{Newman2013}
{Newman}, J.~A., {Cooper}, M.~C., {Davis}, M., {et~al.} 2013, \apjs, 208, 5

\bibitem[{{Oke} \& {Gunn}(1983)}]{ABmag_oke1983secondary}
{Oke}, J.~B. \& {Gunn}, J.~E. 1983, \apj, 266, 713

\bibitem[{{Peng} {et~al.}(2002){Peng}, {Ho}, {Impey}, \& {Rix}}]{Peng2002galfit}
{Peng}, C.~Y., {Ho}, L.~C., {Impey}, C.~D., \& {Rix}, H.-W. 2002, \aj, 124, 266

\bibitem[{{Peng} {et~al.}(2010){Peng}, {Ho}, {Impey}, \& {Rix}}]{Peng2010galfit}
{Peng}, C.~Y., {Ho}, L.~C., {Impey}, C.~D., \& {Rix}, H.-W. 2010, \aj, 139, 2097

\bibitem[{{Revalski} {et~al.}(2023){Revalski}, {Rafelski}, {Fumagalli}, {Fossati}, {Pirzkal}, {Sunnquist}, {Prichard}, {Henry}, {Bagley}, {Dutta}, {Papini}, {Battaia}, {D'Odorico}, {Dayal}, {Estrada-Carpenter}, {Lofthouse}, {Lusso}, {Morris}, {Nedkova}, {Papovich}, \& {Peroux}}]{Revalski2023}
{Revalski}, M., {Rafelski}, M., {Fumagalli}, M., {et~al.} 2023, \apjs, 265, 40

\bibitem[{{Rutkowski} {et~al.}(2025){Rutkowski}, {Zabelle}, {Hagen}, {Alavi}, {Cohen}, {Conselice}, {Grogin}, {Guo}, {Hayes}, {Kaviraj}, {Koekemoer}, {Lucas}, {Mantha}, {Martin}, {Mehta}, {Mobasher}, {Hathi}, {Ji}, {Nedkova}, {O'Connell}, {Rafelski}, {Scarlata}, {Teplitz}, {Wang}, {Windhorst}, \& {Yung}}]{Rutkowski_2025}
{Rutkowski}, M.~J., {Zabelle}, B., {Hagen}, T., {et~al.} 2025, \apjl, 983, L32

\bibitem[{{Schreiber} {et~al.}(2017){Schreiber}, {Elbaz}, {Pannella}, {Merlin}, {Castellano}, {Fontana}, {Bourne}, {Boutsia}, {Cullen}, {Dunlop}, {Ferguson}, {Micha{\l}owski}, {Okumura}, {Santini}, {Shu}, {Wang}, \& {White}}]{Schreiber2017egg}
{Schreiber}, C., {Elbaz}, D., {Pannella}, M., {et~al.} 2017, \aap, 602, A96

\bibitem[{{Shapley} {et~al.}(2025){Shapley}, {Sanders}, {Topping}, {Reddy}, {Pahl}, {Oesch}, {Berg}, {Bouwens}, {Brammer}, {Carnall}, {Cullen}, {Dav{\'e}}, {Dunlop}, {Ellis}, {F{\"o}rster Schreiber}, {Furlanetto}, {Glazebrook}, {Illingworth}, {Jones}, {Kriek}, {McLeod}, {McLure}, {Narayanan}, {Pettini}, {Schaerer}, {Stark}, {Steidel}, {Tang}, {Clarke}, {Donnan}, \& {Kehoe}}]{Shapley2025}
{Shapley}, A.~E., {Sanders}, R.~L., {Topping}, M.~W., {et~al.} 2025, \apj, 981, 167

\bibitem[{{Shibuya} {et~al.}(2015){Shibuya}, {Ouchi}, \& {Harikane}}]{Shibuya2015}
{Shibuya}, T., {Ouchi}, M., \& {Harikane}, Y. 2015, \apjs, 219, 15

\bibitem[{{Skelton} {et~al.}(2014){Skelton}, {Whitaker}, {Momcheva}, {Brammer}, {van Dokkum}, {Labb{\'e}}, {Franx}, {van der Wel}, {Bezanson}, {Da Cunha}, {Fumagalli}, {F{\"o}rster Schreiber}, {Kriek}, {Leja}, {Lundgren}, {Magee}, {Marchesini}, {Maseda}, {Nelson}, {Oesch}, {Pacifici}, {Patel}, {Price}, {Rix}, {Tal}, {Wake}, \& {Wuyts}}]{Skelton20143dhst}
{Skelton}, R.~E., {Whitaker}, K.~E., {Momcheva}, I.~G., {et~al.} 2014, \apjs, 214, 24

\bibitem[{{Song} {et~al.}(2026){Song}, {Wang}, {Jia}, {Lyu}, {Chen}, {Wang}, {Li}, {Ding}, {Fang}, \& {Kong}}]{Song2025}
{Song}, J., {Wang}, E., {Jia}, C., {et~al.} 2026, \aap, 709, A205

\bibitem[{{Spergel} {et~al.}(2015){Spergel}, {Gehrels}, {Baltay}, {Bennett}, {Breckinridge}, {Donahue}, {Dressler}, {Gaudi}, {Greene}, {Guyon}, {Hirata}, {Kalirai}, {Kasdin}, {Macintosh}, {Moos}, {Perlmutter}, {Postman}, {Rauscher}, {Rhodes}, {Wang}, {Weinberg}, {Benford}, {Hudson}, {Jeong}, {Mellier}, {Traub}, {Yamada}, {Capak}, {Colbert}, {Masters}, {Penny}, {Savransky}, {Stern}, {Zimmerman}, {Barry}, {Bartusek}, {Carpenter}, {Cheng}, {Content}, {Dekens}, {Demers}, {Grady}, {Jackson}, {Kuan}, {Kruk}, {Melton}, {Nemati}, {Parvin}, {Poberezhskiy}, {Peddie}, {Ruffa}, {Wallace}, {Whipple}, {Wollack}, \& {Zhao}}]{Roman2015}
{Spergel}, D., {Gehrels}, N., {Baltay}, C., {et~al.} 2015, arXiv e-prints, arXiv:1503.03757

\bibitem[{{Stefanon} {et~al.}(2017){Stefanon}, {Yan}, {Mobasher}, {Barro}, {Donley}, {Fontana}, {Hemmati}, {Koekemoer}, {Lee}, {Lee}, {Nayyeri}, {Peth}, {Pforr}, {Salvato}, {Wiklind}, {Wuyts}, {Ashby}, {Castellano}, {Conselice}, {Cooper}, {Cooray}, {Dolch}, {Ferguson}, {Galametz}, {Giavalisco}, {Guo}, {Willner}, {Dickinson}, {Faber}, {Fazio}, {Gardner}, {Gawiser}, {Grazian}, {Grogin}, {Kocevski}, {Koo}, {Lee}, {Lucas}, {McGrath}, {Nandra}, {Newman}, \& {van der Wel}}]{stefanon2017}
{Stefanon}, M., {Yan}, H., {Mobasher}, B., {et~al.} 2017, \apjs, 229, 32

\bibitem[{{van der Wel} {et~al.}(2014){van der Wel}, {Franx}, {van Dokkum}, {Skelton}, {Momcheva}, {Whitaker}, {Brammer}, {Bell}, {Rix}, {Wuyts}, {Ferguson}, {Holden}, {Barro}, {Koekemoer}, {Chang}, {McGrath}, {H{\"a}ussler}, {Dekel}, {Behroozi}, {Fumagalli}, {Leja}, {Lundgren}, {Maseda}, {Nelson}, {Wake}, {Patel}, {Labb{\'e}}, {Faber}, {Grogin}, \& {Kocevski}}]{van_der_Wel_2014}
{van der Wel}, A., {Franx}, M., {van Dokkum}, P.~G., {et~al.} 2014, \apj, 788, 28

\bibitem[{{Wang} {et~al.}(2025){Wang}, {Sun}, {Zhou}, {Xu}, {Cheng}, {Li}, {Chen}, {Mo}, {Dekel}, {Yang}, {Wang}, {Chen}, {Zheng}, {Cai}, {Elbaz}, {Dai}, \& {Huang}}]{Wang_2025}
{Wang}, T., {Sun}, H., {Zhou}, L., {et~al.} 2025, \apjl, 988, L35

\bibitem[{{Williams} {et~al.}(2009){Williams}, {Quadri}, {Franx}, {van Dokkum}, \& {Labb{\'e}}}]{williams2009QG}
{Williams}, R.~J., {Quadri}, R.~F., {Franx}, M., {van Dokkum}, P., \& {Labb{\'e}}, I. 2009, \apj, 691, 1879

\bibitem[{{Willner} {et~al.}(2006){Willner}, {Coil}, {Goss}, {Ashby}, {Barmby}, {Huang}, {Ivison}, {Koo}, {Egami}, \& {Miyazaki}}]{Willner2006}
{Willner}, S.~P., {Coil}, A.~L., {Goss}, W.~M., {et~al.} 2006, \aj, 132, 2159

\bibitem[{{Wuyts} {et~al.}(2013){Wuyts}, {F{\"o}rster Schreiber}, {Nelson}, {van Dokkum}, {Brammer}, {Chang}, {Faber}, {Ferguson}, {Franx}, {Fumagalli}, {Genzel}, {Grogin}, {Kocevski}, {Koekemoer}, {Lundgren}, {Lutz}, {McGrath}, {Momcheva}, {Rosario}, {Skelton}, {Tacconi}, {van der Wel}, \& {Whitaker}}]{Wuyts2013}
{Wuyts}, S., {F{\"o}rster Schreiber}, N.~M., {Nelson}, E.~J., {et~al.} 2013, \apj, 779, 135

\bibitem[{{Xiao} {et~al.}(2025){Xiao}, {Williams}, {Oesch}, {Elbaz}, {Dessauges-Zavadsky}, {Marques-Chaves}, {Bing}, {Ji}, {Weibel}, {Bezanson}, {Brammer}, {Casey}, {Cloonan}, {Daddi}, {Dayal}, {Faisst}, {Franx}, {Glazebrook}, {Hutter}, {Kartaltepe}, {Labbe}, {Lagache}, {Lim}, {Magnelli}, {Martinez}, {Maseda}, {Nanayakkara}, {Schaerer}, \& {Whitaker}}]{Xiao2025}
{Xiao}, M., {Williams}, C.~C., {Oesch}, P.~A., {et~al.} 2025, \aap, 696, A156

\bibitem[{{Zhou} {et~al.}(2019){Zhou}, {Cooper}, {Newman}, {Ashby}, {Aird}, {Conselice}, {Davis}, {Dutton}, {Faber}, {Fang}, {Fazio}, {Guhathakurta}, {Kocevski}, {Koo}, {Nandra}, {Phillips}, {Rosario}, {Schlafly}, {Trump}, {Weiner}, {Willmer}, \& {Yan}}]{Zhou2019_DEEP3}
{Zhou}, R., {Cooper}, M.~C., {Newman}, J.~A., {et~al.} 2019, \mnras, 488, 4565

\bibitem[{{Zhuang} {et~al.}(2024){Zhuang}, {Li}, \& {Shen}}]{Zhuang2024psf}
{Zhuang}, M.-Y., {Li}, J., \& {Shen}, Y. 2024, \apj, 962, 93

\end{thebibliography}

\clearpage
\begin{appendix}
\nolinenumbers

\section{Mock image generation}\label{appendix:mock images}
While the completeness for point sources can be straightforwardly determined in magnitude bins, the detection of extended galaxies depends sensitively on surface brightness.
For a single \sersic model, this involves a combination of magnitude, effective radius, and \sersic index.
Consequently, any estimate of detection completeness at a given magnitude must account for the underlying distribution of galaxies in the mag-$R_\mathrm{e}$-$n$ parameter space. 
A key factor is that a higher fraction of low-surface-brightness objects at a fixed magnitude reduces the overall detection completeness, while a larger contribution from high-surface-brightness objects increases it.

To construct a single \sersic model, we compute the \sersic index and effective radius from the flux ratio of the two components~\citep{Euclid2023} as follows
\begin{equation}
    \begin{aligned}
    n &= n_\mathrm{b}\cdot b/t+n_\mathrm{d}\cdot (1-b/t)  \\
    R_{\mathrm{eff}} &= (b/t\cdot R_{\mathrm{bulge}})^a+((1-b/t) R_{\mathrm{disk}})^a
    \end{aligned}
\end{equation}
Here we adopt $n_\mathrm{d} = 1$ and $n_\mathrm{b} = 4$.
The exponent is set to 0.8 when $R_{\mathrm{bulge}} < R_{\mathrm{disk}}$, and to 2 otherwise.
We note that the resulting \sersic index $n$ is constrained to lie between 1 and 4 in our calculations.

We utilize \software{GALFIT} to create a single \sersic model that is matched to the zero-point and resolution of the F160W mosaic, as this model is used for source detection tests. 
The empirical PSF adopted in \software{GALFIT} is taken from Sect.~\ref{subsec:psf}.
The model image is set at $400\times400$ pixels, ensuring it is sufficiently large for normal extragalactic galaxy sizes in EGS F160W images.
To construct the background, we extract several flat regions, each $100\times100$ pixels, from various locations in the F160W mosaic. 
These small patches are then stitched together to build two background images: one of $400\times400$ pixels and a larger one of $6000\times6000$ pixels. 
For the larger background, the same small cutouts are reused to fill the area.
We note that pixel correlations introduced by the drizzling of multiple exposures can significantly affect source detection, as discussed in Sect.~\ref{sec:catalog_properties}.
Additionally, we generate a Gaussian background whose noise level matches that of the EGS background.
By combining the single \sersic model with the corresponding Poisson noise and the background image, we produce the final mock galaxies. 

\section{Source detection completeness tests}\label{appendix:completeness}

In addition to the test using a realistic distribution of galaxy morphologies with the HST background, we also examine how the \sersic index and background affect the detection completeness.

\subsection{Complete structural parameters}\label{appendix:completeness sersic index}
For the source detection completeness tests presented in Fig.~\ref{fig:completeness}, the \sersic index of each model galaxy is determined from a combination of disk ($n = 1$) and bulge ($n = 4$), resulting \sersic indices are restricted to the range 1.0--4.0 and therefore exclude more compact or more diffuse objects.
Here we test the detection completeness for galaxies with different \sersic indices ($n = 0.5$, $1.0$, $4.0$, $8.0$) using the same detection framework applied to the real observations.
For better comparison, we plot the 1-D completeness distribution of different \sersic index galaxies in Fig.~\ref{fig:1d_completeness_diff_n}.
We find no significant difference for galaxies fainter than 28 mag, as galaxies are all too faint to be detected, indicating that the detection limit is governed solely by magnitude.
For galaxies with $27 < m_\mathrm{H} <28$, the detection completeness remains below 50\% for all \sersic indices tested.
At a given magnitude in this range, the larger \sersic index implies that fewer pixel values exceed the detection threshold, which leads to lower completeness. 
However, for galaxies brighter than $\sim 27$ mag, the trend is reversed:  systems with higher \sersic indices exhibit higher completeness.
The reversal arises from background overestimation and stronger blending for higher \sersic indices.
We note that these differences primarily appear at large effective radii, which is not the parameter space occupied by the majority of galaxies in our catalog.
Therefore, for our sample, the detection completeness is not strongly sensitive to the \sersic index for bright galaxies.

\subsection{Background}\label{appendix:completeness background}
\begin{figure*}
    \centering
    \includegraphics[width=0.9\textwidth]{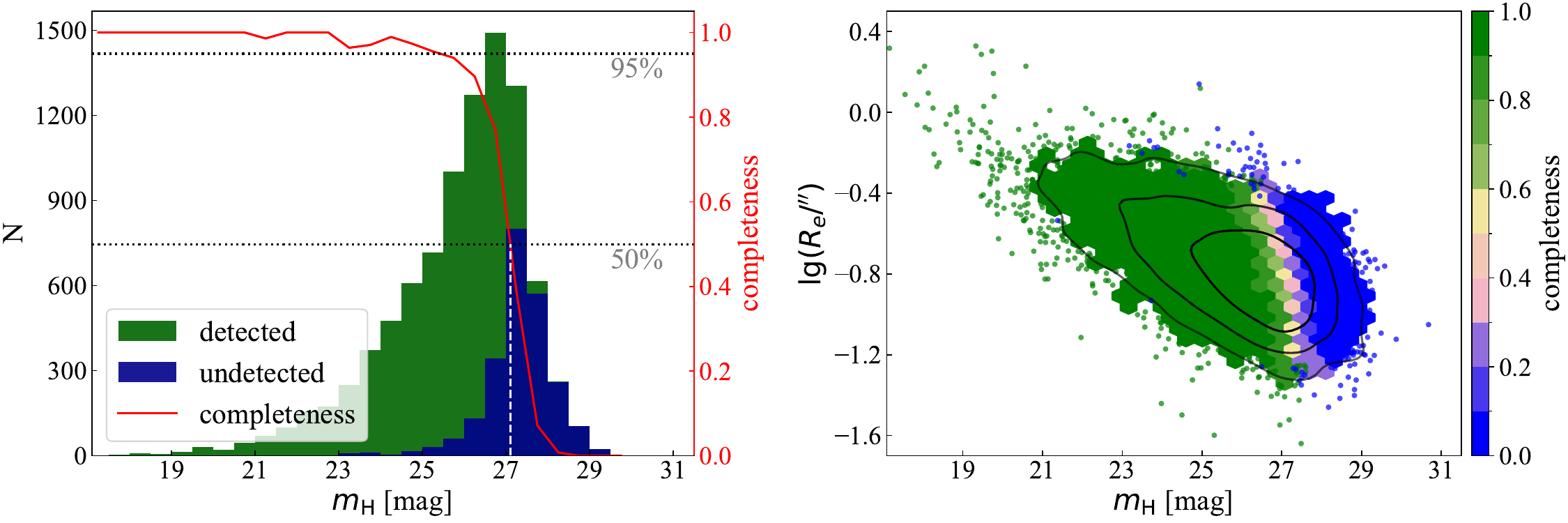}
    \caption{Same as Fig.~\ref{fig:completeness}, but showing the completeness distribution for galaxies with the Gaussian background. The results are similar to those obtained with the real background.}
    \label{fig:completeness_gauss}
\end{figure*}

The background is another factor that influences the detection completeness.
The pixel correlations introduced by drizzling in the EGS mosaic are nonnegligible.
In Fig.~\ref{fig:completeness_gauss}, we present the detection completeness measured with the Gaussian background.
Compared with Fig.~\ref{fig:completeness}, the completeness in the Gaussian background is consistent with that derived from the HST background.
However, the total detection number is lower in the Gaussian background, indicating a reduced false detection rate.

\section{Forced aperture photometry}\label{appenix:photometry}

\begin{figure*}[htpb]
    \centering
    \includegraphics[width=0.9\linewidth]{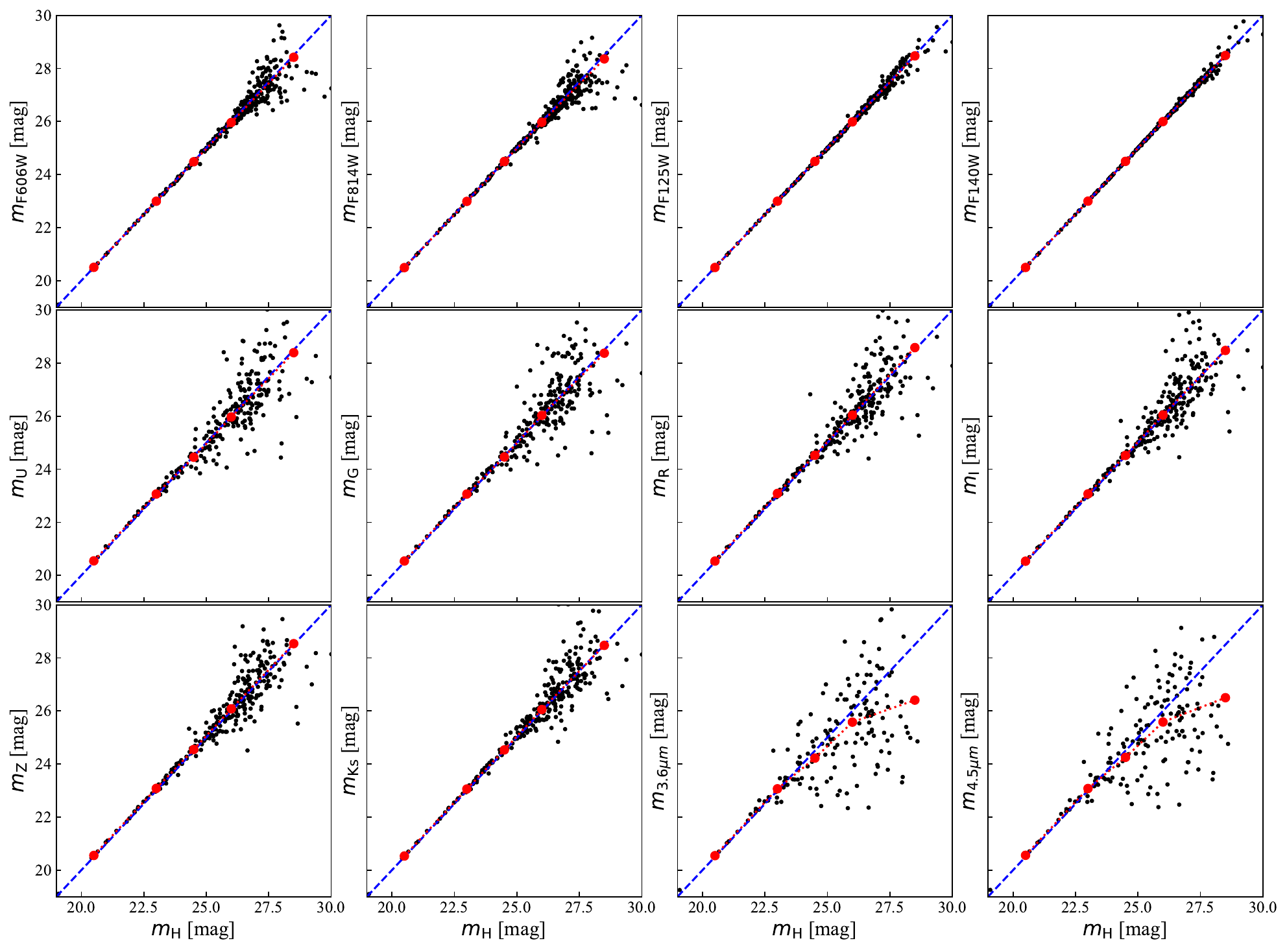}
    \caption{Results of the aperture-based PSF correction applied to HST, CFHT, and Spitzer images. The red dashed lines show the median value in each $m_\mathrm{H}$ bin.}
    \label{fig:check_aperture_psf_correction}
\end{figure*}

We convolve the images in different bands to match their PSF.
For high-resolution images, we bring their PSF to the resolution of the F160W band; for low-resolution images, we instead convolve the F160W image to match the PSFs of those bands.
This allows us to directly quantify the flux errors caused by uncorrected PSF mismatches and the challenges associated with lower resolution during deblending processes.
In Fig.~\ref{fig:check_aperture_psf_correction}, we compare the photometry of the same galaxies across images with different resolutions.

For bands with wavelength shorter than $3~\mu m$, the fluxes follow a tight one-to-one relation, as the image resolution is close to that of the F160W band.
Noticeable biases appear at the faint end ($m_\mathrm{H} > 25$ mag).
For IRAC/3.6$\mu m$ and 4.5$\mu m$, the magnitude bias and scatter are significantly larger than in other bands. 
This is expected because the PSF FWHM of these low-resolution bands is about ten times larger than that of the HST and JWST bands.

\section{Star galaxy separation}\label{appendix:star galaxy separation}

\begin{figure*}[htpb]
    \centering
    \includegraphics[width=0.75\textwidth]{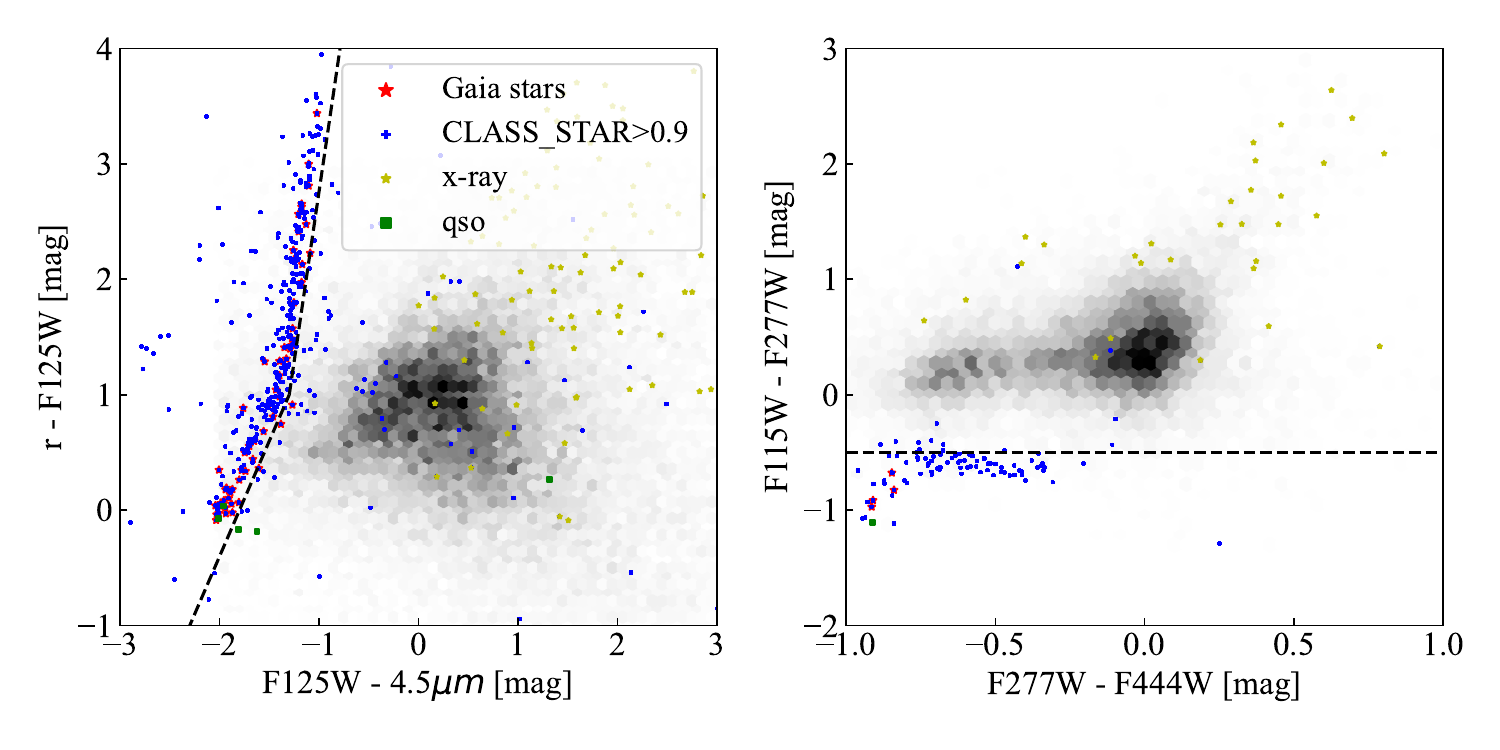}
    \caption{Color-color diagram used to separate stars and galaxies. Extended sources (galaxies) detected in the EGS catalog are shown as a grayscale density map, with deeper colors corresponding to higher source density. 
    Objects with CLASS\_STAR$>0.9$ that are identified as stars in the Gaia catalog are plotted as red star symbols, while those not identified as stars in the Gaia catalog are presented in blue points.
    AGNs are marked with green squares, and X-ray detected sources are presented in yellow dots.}
    \label{fig:color-color}
\end{figure*}

In source detection, a crucial step is separating the stars (point sources) from the galaxies (extended objects). 
Traditionally, point sources are identified using the CLASS\_STAR parameter from \software{SExtractor} (CLASS\_STAR$>0.9$).
Stars can also be distinguished from galaxies by their colors.
As shown in Fig.~\ref{fig:color-color}, the galaxies are shown in the gray hexbin distribution, while the stars from the Gaia catalog and the X-ray sources (AGNs) are represented by red stars and blue squares, respectively.
The two populations occupy distinct regions in color-color space, allowing effective separation.
However, due to the difference in typical SED of these objects and in image depth and sensitivity, not all the color--color diagrams can be used to cleanly separate these types.
For sources in the EGS field, we adopt the following criteria (shown as black dashed lines in Fig.~\ref{fig:color-color}):
\begin{itemize}
    \item CLASS\_STAR > 0.9
    \item $[F125W-4.5\mu \mathrm{m}] < 0.5 \times [R-F125W]-1.8 $, for [R-F125W]<1
    \item $[F125W-4.5\mu \mathrm{m}] < 0.17 \times [R-F125W]-1.47 $, for [R-F125W]>1
\end{itemize}
In the CEERS field, the separation of stars and galaxies is more straightforward.
We therefore adopt the following selection criteria:
\begin{itemize}
    \item CLASS\_STAR > 0.9
    \item $[F115W-F277W] < 0.5$
\end{itemize}

In the color-color diagram, objects with CLASS\_STAR$ >0.9$ are predominantly stars.
Note that most X-ray sources fall outside the stellar locus due to different SEDs.

\FloatBarrier

\section{Main SExtractor configuration parameters}\label{appendix:SE_params}

\begin{table}[ht!]
    \centering
    \caption{Main parameters in the configuration file of \software{SExtractor} for hot and cold modes}
    \begin{tabular}{l|cc|cc}
    \hline
    &\multicolumn{2}{c}{HST}&\multicolumn{2}{|c}{JWST}\\
    \hline
    Parameter Name&Cold Mode&Hot Mode&Cold Mode&Hot Mode\\
    \hline
    \texttt{DETECT\_MINAREA} & 4 & 8 & 4 & 8\\
    \texttt{DETECT\_THRESH} & 5.0 & 1.5 & 5.0 & 1.6\\
    \texttt{ANALYSIS\_THRESH}&5.0&1.5 & 5 &1.6\\
    \texttt{FILTER\_NAME}&default.conv&default.conv&default.conv&default.conv\\
    \texttt{DEBLEND\_NTHRESH}&32&64&32&64\\
    \texttt{DEBLEND\_MINCONT}&0.001&0.001&0.003&0.001\\
    \texttt{BACK\_SIZE}&256&64&128&64\\
    \texttt{BACK\_FILTERSIZE}&9&3&9&3\\
    \texttt{BACKPHOTO\_THICK}&100&24&100&25\\
    \texttt{CLEAN\_PARAM}&1&1&1&1\\
    \hline
    \end{tabular}
    \label{tab:SE_params}
\end{table}

\FloatBarrier
\clearpage

\onecolumn
\section{Content of the catalog}\label{appendix:source catalog}
Table~\ref{tab:source_catalog} lists the columns provided in the catalog, including source properties, photometry, photo-$z$, SED fitting results, and structure fitting results. 

\begin{longtable}{c l p{10cm}}

\caption{Multiwavelength photometric catalog entries} \\
\hline\hline
Col. & Name & Description \\
\hline
\endfirsthead

\hline
Col. & Name & Description \\
\hline
\endhead

    \stepcounter{colnum}{1} &ID&Sequential ID number in the F160W-based \software{SExtractor} catalog\\
    \coln &RA&Columns 2-3: R.A. and decl. (J2000) in the F160W mosaic.\\
    \coln &DEC&\multicolumn{1}{c}{\ldots}\\
    \coln&BKG&Local background.\\
    \coln&MAG\_AUTO&\\
    \coln&KRON\_RADIUS&\\
    \coln&FLUX\_RADIUS&\\
    \coln&ELONGATION&\\
    \coln&ELLIPTICITY& \\
    \coln&PA&\\
    \coln&ISO\_AREA&\\
    \coln&KRON\_FLUX&\\
    \coln&KRON\_FLUXERR&\\
    \coln&ISO\_FLUX&\\
    \coln&ISO\_FLUXERR&\\
    \coln&FWHM&\\
    \coln&CLASS\_STAR&\software{SExtractor} CLASS\_STAR parameter.\\
    \coln&AGN\_FLAG&=1 for those objects with a counterpart in the X-ray catalog, 0 otherwise.\\
    \coln&STAR\_FLAG&=1 for those objects with a counterpart in the Gaia catalog, 0 otherwise.\\
    \coln&FLAGS&Footprint flag. A value of 1 corresponds to sources falling in regions of low S/N as it can be at the borders of the mosaic; a value of 2 indicates that a source, as identified by its footprint in the segmentation map, falls close to a bright star or to its diffraction spikes; a value of 3 indicates that the source suffers from both a low S/N and contamination from bright stars. Sources free from any of the above effects have a flag of 0.\\
    \\
    \coln&ACS\_F606W\_FLUX&Columns 21-56: Flux densities and associated uncertainties measured\\
    \coln&ACS\_F606W\_FLUXERR&from forced aperture photometry, expressed in $\mu Jy$. A value of $-99$ has\\
    \coln&ACS\_F814W\_FLUX&been set to the flux and associated uncertainty for those objects falling\\
    \coln&ACS\_F814W\_FLUXERR&outside the coverage of the mosaic in a specific band or when bad\\
    \coln&WFC3\_F125W\_FLUX&pixels within the segmentation map contaminate the flux measurement.\\
    \coln&WFC3\_F125W\_FLUXERR&\multicolumn{1}{c}{\ldots}\\
    \coln&WFC3\_F140W\_FLUX&\multicolumn{1}{c}{\ldots}\\
    \coln&WFC3\_F140W\_FLUXERR&\multicolumn{1}{c}{\ldots}\\
    \coln&WFC3\_F160W\_FLUX&\multicolumn{1}{c}{\ldots}\\
    \coln&WFC3\_F160W\_FLUXERR&\multicolumn{1}{c}{\ldots}\\
    \coln&NIRCAM\_F115W\_FLUX&\multicolumn{1}{c}{\ldots}\\
    \coln&NIRCAM\_F115W\_FLUXERR&\multicolumn{1}{c}{\ldots}\\
    \coln&NIRCAM\_F150W\_FLUX&\multicolumn{1}{c}{\ldots}\\
    \coln&NIRCAM\_F150W\_FLUXERR&\multicolumn{1}{c}{\ldots}\\
    \coln&NIRCAM\_F200W\_FLUX&\multicolumn{1}{c}{\ldots}\\
    \coln&NIRCAM\_F200W\_FLUXERR&\multicolumn{1}{c}{\ldots}\\
    \coln&NIRCAM\_F277W\_FLUX&\multicolumn{1}{c}{\ldots}\\
    \coln&NIRCAM\_F277W\_FLUXERR&\multicolumn{1}{c}{\ldots}\\
    \coln&NIRCAM\_F356W\_FLUX&\multicolumn{1}{c}{\ldots}\\
    \coln&NIRCAM\_F356W\_FLUXERR&\multicolumn{1}{c}{\ldots}\\
    \coln&NIRCAM\_F410M\_FLUX&\multicolumn{1}{c}{\ldots}\\
    \coln&NIRCAM\_F410M\_FLUXERR&\multicolumn{1}{c}{\ldots}\\
    \coln&NIRCAM\_F444W\_FLUX&\multicolumn{1}{c}{\ldots}\\
    \coln&NIRCAM\_F444W\_FLUXERR&\multicolumn{1}{c}{\ldots}\\
    \coln&CFHT\_u\_FLUX&\multicolumn{1}{c}{\ldots}\\
    \coln&CFHT\_u\_FLUXERR&\multicolumn{1}{c}{\ldots}\\
    \coln&CFHT\_g\_FLUX&\multicolumn{1}{c}{\ldots}\\
    \coln&CFHT\_g\_FLUXERR&\multicolumn{1}{c}{\ldots}\\
    \coln&CFHT\_r\_FLUX&\multicolumn{1}{c}{\ldots}\\
    \coln&CFHT\_r\_FLUXERR&\multicolumn{1}{c}{\ldots}\\
    \coln&CFHT\_i\_FLUX&\multicolumn{1}{c}{\ldots}\\
    \coln&CFHT\_i\_FLUXERR&\multicolumn{1}{c}{\ldots}\\
    \coln&CFHT\_z\_FLUX&\multicolumn{1}{c}{\ldots}\\
    \coln&CFHT\_z\_FLUXERR&\multicolumn{1}{c}{\ldots}\\
    \coln&WIRCAM\_Ks\_FLUX&\multicolumn{1}{c}{\ldots}\\
    \coln&WIRCAM\_Ks\_FLUXERR&\multicolumn{1}{c}{\ldots}\\
    \coln&IRAC\_CH1\_FLUX&Columns 57-60: Flux densities and associated uncertainties measured\\
    \coln&IRAC\_CH1\_FLUXERR&from \software{TPHOT} template fitting, expressed in $\mu Jy$.\\
    \coln&IRAC\_CH2\_FLUX&\multicolumn{1}{c}{\ldots}\\
    \coln&IRAC\_CH2\_FLUXERR&\multicolumn{1}{c}{\ldots}\\
    \coln&MAG\_F160W&Columns 61–71: Single \sersic fitting results ($mag$, $R_\mathrm{e}$, $n$, $q$, and $PA$) \\
    \coln&RE\_F160W&using \software{GALFITM} on WFC3/F125W and F160W.\\
    \coln&N\_F160W&\multicolumn{1}{c}{\ldots}\\
    \coln&Q\_F160W&\multicolumn{1}{c}{\ldots}\\
    \coln&PA\_F160W&\multicolumn{1}{c}{\ldots}\\
    \coln&MAG\_F125W&\multicolumn{1}{c}{\ldots}\\
    \coln&RE\_F125W&\multicolumn{1}{c}{\ldots}\\
    \coln&N\_F125W&\multicolumn{1}{c}{\ldots}\\
    \coln&Q\_F125W&\multicolumn{1}{c}{\ldots}\\
    \coln&PA\_F125W&\multicolumn{1}{c}{\ldots}\\
    \coln&GALFIT\_CHI2&\multicolumn{1}{c}{\ldots}\\
    \coln&C&Concentration parameter $C$ from \textsc{statmorph}.\\
    \coln&A&Asymmetry parameter $A$ from \textsc{statmorph}.\\
    \coln&S&Smoothness $S$ from \textsc{statmorph}.\\
    \coln&Gini&Gini parameter $Gini$ from \textsc{statmorph}.\\
    \coln&M20&$M20$ parameter from \textsc{statmorph}.\\
    \coln&n\_statmorph&\sersic index $n$ from \textsc{statmorph}.\\
    \coln&SPEC\_Z&Spectroscopic redshift from the DEEP2, DEEP3 and RUBIES catalog.\\
    \coln&PHOT\_Z&Photometric redshift from \software{EAZY}. \\
    \coln&PHOT\_Z\_CHI2&\\
    \coln&REST\_FRAME\_U&\\
    \coln&REST\_FRAME\_V&\\
    \coln&REST\_FRAME\_J&\\
    \coln&M\_star&Stellar mass measurement from \textsc{Cigale}.\\
    \coln&age&Age measurement from \textsc{Cigale}.\\
    \coln&tau&$\tau$ of star formation history from \textsc{Cigale}.\\
    \coln&E\_BV&Stellar continuum color excess $E(B-V)$ from \textsc{Cigale}.\\
    \coln&SFR&Star formation rate measurement from \textsc{Cigale}.\\
    \coln&CIGALE\_CHI2&SED fitting $\chi^2$ from \textsc{Cigale}.\\
\hline
\label{tab:source_catalog}
\end{longtable}

\twocolumn

\end{appendix}

\end{document}